\documentclass[11pt,a4paper,useAMS]{emulateapj}

\usepackage{longtable}
\usepackage{natbib}
\usepackage{amsmath}
\usepackage{graphicx}
\newcommand{\rbracket}{]}
\newcommand{\lbracket}{[}

\begin{document}

\title{\vspace{-5mm}LOFAR Low-band antenna observations of the 3C\,295 and Bo\"otes fields: source counts and Ultra-Steep Spectrum sources}
\interfootnotelinepenalty=10000

\author{
R.~J.~van~Weeren\altaffilmark{1$\star$},  W.~L.~Williams\altaffilmark{2,3}, C. Tasse\altaffilmark{4,5,6}, H.~J.~A.~R\"ottgering\altaffilmark{2}, D. A. Rafferty\altaffilmark{2}, S.~van~der~Tol\altaffilmark{2,3}, G. Heald\altaffilmark{3}, G.~J. White\altaffilmark{15,16}, A.~Shulevski\altaffilmark{7},  P.~Best\altaffilmark{8}
% ...
}

\affil{\altaffilmark{}}

\affil{\altaffilmark{1}Harvard-Smithsonian Center for Astrophysics, 60 Garden Street, Cambridge, MA 02138, USA}
\affil{\altaffilmark{2}Leiden Observatory, Leiden University, P.O. Box 9513, NL-2300 RA Leiden, The Netherlands}
\affil{\altaffilmark{3}Netherlands Institute for Radio Astronomy (ASTRON), PO Box 2, 7990AA Dwingeloo, The Netherlands}
\affil{\altaffilmark{4}Department of Physics \& Electronics, Rhodes University, PO Box 94, Grahamstown, 6140, South Africa}
\affil{\altaffilmark{5}SKA South Africa, 3rd Floor, The Park, Park Road, Pinelands, 7405, South Africa}
\affil{\altaffilmark{6}GEPI, Observatoire de Paris, CNRS, Universit{\'e} Paris Diderot, 5 place Jules Janssen, 92190 Meudon, France}
\affil{\altaffilmark{7}Kapteyn Astronomical Institute, PO Box 800, 9700 AV Groningen, The Netherlands}
\affil{\altaffilmark{8}Institute for Astronomy, University of Edinburgh, Royal Observatory of Edinburgh, Blackford Hill, Edinburgh EH9 3HJ, UK}
\affil{\altaffilmark{9}National Radio Astronomy Observatory, Socorro, NM 87801, USA}
\affil{\altaffilmark{10}Department of Astrophysics, Institute for Mathematics, Astrophysics and Particle Physics (IMAPP), Radboud University Nijmegen, P.O. Box 9010, 6500 GL Nijmegen, The Netherlands}
\affil{\altaffilmark{11}Max Planck Institute for Astrophysics, Karl-Schwarzschildstr.1, 85748 Garching, Germany}
\affil{\altaffilmark{12}Hamburger Sternwarte, University of Hamburg, Gojenbergsweg 112, 21029 Hamburg, Germany}
\affil{\altaffilmark{13}Astronomical Observatory, Jagiellonian University, ul. Orla 171, 30-244, Krak\'ow, Poland}
\affil{\altaffilmark{14}INAF - Istituto di Radioastronomia, Via Gobetti 101, I-40129 Bologna, Italy}
\affil{\altaffilmark{15}Department of Physics \& Astronomy, The Open University, UK}
\affil{\altaffilmark{16}Space Science Department, Rutherford Appleton Laboratory, Chilton, UK}
\affil{\altaffilmark{17}Max-Planck-Institut f\"ur Radioastronomie, Auf dem H\"ugel 69, 53121 Bonn, Germany}
\affil{\altaffilmark{18}Leibniz-Institut f\"ur Astrophysik Potsdam (AIP), An der Sternwarte 16, D-14482 Potsdam, Germany}
\affil{\altaffilmark{19}Astrophysics, Department of Physics, Keble Road, Oxford, OX1 3RH, UK}
\affil{\altaffilmark{20}Centre for Astrophysics, Science \& Technology Research Institute, University of Hertfordshire, Hatfield, Herts, AL10 9AB, UK}
\affil{\altaffilmark{21}Physics Department, University of the Western Cape, Private Bag X17, Bellville 7535, South Africa}
\affil{\altaffilmark{22}Astronomical Institute `Anton Pannekoek', University of Amsterdam, Postbus 94249, 1090 GE Amsterdam, The Netherlands}
\affil{\altaffilmark{23}Laboratoire Lagrange, UMR7293, Universit\'e de Nice Sophia-Antipolis, CNRS, Observatoire de la C\^ote d'Azur, 06300 Nice, France}
\affil{\altaffilmark{24}School of Physics, Astronomy and Mathematics, University of Hertfordshire, College Lane, Hatfield, Hertfordshire AL10 9AB, UK}
\affil{\altaffilmark{25}Th\"uringer Landessternwarte, Sternwarte 5, D-07778 Tautenburg, Germany}
\affil{\altaffilmark{26}Toru\'n Centre for Astronomy, Faculty of Physics, Astronomy and Informatics, NCU, Grudziacka 5, 87-100 Toru\'n, Poland}
\affil{\altaffilmark{27}Department of Earth and Space Sciences, Chalmers University of Technology, Onsala Space Observatory, SE-439 92 Onsala, Sweden}
\affil{\altaffilmark{28}National Centre for Radio Astrophysics, TIFR, Pune University Campus, Post Bag 3, Pune 411 007, India}
\affil{\altaffilmark{29}School of Physics \& Astronomy, University of Southampton, Highfield, Southampton, SO17 1BJ, UK}
\affil{\altaffilmark{30}Max-Planck Institut f\"ur Extraterrestrische Physik, Giessenbachstr., 85748, Garching, Germany}
\affil{\altaffilmark{32}Universit\'e de Lyon, Lyon, F-69003, France}
\affil{\altaffilmark{33}Centre de Recherche Astrophysique de Lyon, Observatoire de Lyon, 9 av Charles Andr\'{e}, 69561 Saint Genis Laval Cedex, France}
\affil{\altaffilmark{34}Ecole Normale Sup\'erieure de Lyon, Lyon, F-69007, France}
\affil{\altaffilmark{35}Space Telescope Science Institute, 3700 San Martin Drive, Baltimore, MD 21218, USA} 
\affil{\altaffilmark{36}SRON Netherlands Institute for Space Research, PO Box 800, 9700 AV Groningen, The Netherlands}
\affil{\altaffilmark{37}ARC Centre of Excellence for All-sky astrophysics (CAASTRO), Sydney Institute of Astronomy, University of Sydney Australia} 
\affil{\altaffilmark{38}University of Twente,  Postbus 217, 7500 AE Enschede, The Netherlands}
\affil{\altaffilmark{39}Research School of Astronomy and Astrophysics, Australian National University, Mt Stromlo Obs., via Cotter Road, Weston, A.C.T. 2611, Australia}
\affil{\altaffilmark{40}SmarterVision BV, Oostersingel 5, 9401 JX Assen, The Netherlands}
\affil{\altaffilmark{41}LPC2E - Universite d'Orleans/CNRS, France}
\affil{\altaffilmark{42}Station de Radioastronomie de Nancay, Observatoire de Paris - CNRS/INSU, USR 704 - Univ. Orleans, OSUC , Route de Souesmes, 18330 Nancay, France}
\affil{\altaffilmark{43}Jodrell Bank Center for Astrophysics, School of Physics and Astronomy, The University of Manchester, Manchester M13 9PL, UK}
\affil{\altaffilmark{44}Astronomisches Institut der Ruhr-Universit\"{a}t Bochum, Universitaetsstrasse 150, 44780 Bochum, Germany}
\affil{\altaffilmark{45}Astro Space Center of the Lebedev Physical Institute, Profsoyuznaya str. 84/32, Moscow 117997, Russia}
\affil{\altaffilmark{46}Sodankyl\"{a} Geophysical Observatory, University of Oulu, T\"{a}htel\"{a}ntie 62, 99600 Sodankyl\"{a}, Finland}
\affil{\altaffilmark{47}STFC Rutherford Appleton Laboratory,  Harwell Science and Innovation Campus,  Didcot  OX11 0QX, UK}
\affil{\altaffilmark{48}Center for Information Technology (CIT), University of Groningen, The Netherlands}
\affil{\altaffilmark{49}Fakult\"{a}t f\"ur Physik, Universit\"{a}t Bielefeld, Postfach 100131, D-33501, Bielefeld, Germany}
\affil{\altaffilmark{50}LESIA, UMR CNRS 8109, Observatoire de Paris, 92195 Meudon, France}
\email{E-mail: rvanweeren@cfa.harvard.edu}

\altaffiltext{$\star$}{Einstein Fellow}

\shorttitle{LOFAR observations of the Bo\"otes and 3C\,295 fields}
\shortauthors{van Weeren et al.}

\vspace{0.5cm}
\begin{abstract}
We present LOFAR Low Band observations of the Bo\"otes and 3C\,295 fields. Our images made at 34, 46, and 62~MHz reach noise levels of 12, 8, and 5 mJy~beam$^{-1}$, making them the deepest images ever obtained in this frequency range. In total, we detect between 300 and 400 sources in each of these images, covering an area of 17 to 52 deg$^{2}$. From the observations we derive Euclidean-normalized differential source counts. The 62~MHz source counts agree with previous GMRT 153~MHz and VLA 74~MHz differential source counts, scaling with a spectral index of $-0.7$. {We find that a spectral index scaling of $-0.5$ is required to match up the LOFAR 34~MHz source counts. This result is also in agreement with source counts from the 38~MHz 8C survey, indicating that the average spectral index of radio sources flattens towards lower frequencies.} {We also find evidence for spectral flattening using the individual flux measurements of sources between 34 and 1400~MHz and by calculating the  spectral index averaged over the source population.} To select ultra-steep spectrum ($\alpha < -1.1$) radio sources, that could be associated with massive high redshift radio galaxies, we compute spectral indices between 62~MHz, 153~MHz and 1.4~GHz for  sources in the Bo\"otes field. We cross-correlate these radio sources with optical and infrared catalogues and fit the spectral energy distribution to obtain photometric redshifts. We find that most of these ultra-steep spectrum sources are located in the $ 0.7 \lesssim  z \lesssim 2.5$ range. 
\\ \\
\end{abstract}





\section{Introduction}
Low-frequency surveys of the sky are an important tool to address various open questions in astrophysics ranging from the evolution of galaxies, active galactic nuclei (AGN), galaxy clusters, to pulsars. The half power beam width (HPBW) of radio telescopes scales with wavelength, making low-frequency radio observations ($\lesssim 300$~MHz) an efficient way to carry out large-area surveys. In addition, these observations take advantage of the steep synchrotron spectra ($F_{\nu} \propto \nu^{\alpha}$, with $\alpha$ the spectral index) of many extragalactic radio sources, with the flux densities  increasing towards lower frequencies.

Low-frequency observations are particularly important to locate distant high-redshift radio galaxies (HzRG). Empirically it has been found that the radio spectral index correlates with the redshift of host galaxies, with the steepest spectra corresponding to the highest redshifts.  Therefore massive high-redshift galaxies can be found by selecting radio sources with  ultra-steep radio spectra (USS), especially in combination with an optical or near-IR magnitude cut  \citep[e.g.,][]{2000A&AS..143..303D,2008A&ARv..15...67M,2012MNRAS.420.2644K}. However, USS sources are rare so large surveys are needed to find them. The fraction of  USS sources with $\alpha^{1400}_{\sim350}<-1.3$ is about 0.5\% \citep{2000A&AS..143..303D}. Deep observations at $\lesssim 150$~MHz have the potential to detect sources with $\alpha \lesssim -2$, because these sources become too faint to be detected in sensitive high-frequency observations.

Radio sources in the last stages of the AGN evolution (both short and long-lived) are also most efficiently selected at low-frequencies.  These relic or dying radio sources have steep and curved radio spectra due to synchrotron and inverse Compton losses as the central energy supply has been switched off \citep[e.g.,][]{2007A&A...470..875P,2010MNRAS.408.2261K,2011A&A...526A.148M}.

Recently, most deep low-frequency surveys have been carried out with the GMRT at around 150 MHz \citep[e.g.,][]{2007ASPC..380..237I, 2009MNRAS.392.1403S, 2010MNRAS.405..436I, 2011A&A...535A..38I,2013A&A...549A..55W}. These surveys reach a rms noise level of the order of a mJy per beam. Below 100~MHz, there are no radio surveys that reach a similar depth. \cite{2004ApJS..150..417C} carried out a 165~deg$^{2}$  74~MHz survey with a central noise of 24 mJy beam$^{-1}$ at a resolution of 25\arcsec. 
{\cite{2006A&A...456..791T} surveyed the XMM-LSS field at 74~MHz with a resolution of 30\arcsec, covering an area of 132~deg$^{2}$. The median rms noise over the field was 32~mJy~beam$^{-1}$.} Larger, but shallower surveys below 100 MHz, are the 74 MHz VLSS \citep{2007AJ....134.1245C,2012RaSc...47.....L} and 38~MHz \citep{1995MNRAS.274..447H, 1990MNRAS.244..233R} surveys.

The LOw Frequency ARray (LOFAR) is a new generation radio telescope operating at 10--240~MHz \citep{2013A&A...556A...2V}. With its multi-beaming capabilities, high-spatial resolution, and large fractional bandwidth, it is an ideal instrument to carry out large surveys. Here we report on the first LOFAR Low Band Antenna commissioning observations of the Bo\"otes and the 3C\,295 fields (which includes the Groth Strip). Both the Bo\"otes field and the Groth Strip have been extensively studied at higher radio frequencies and other parts of the electromagnetic spectrum. For the Bo\"otes field, observations have been carried out at 153~MHz \citep{2011A&A...535A..38I,2013A&A...549A..55W}, 325~MHz \citep{2008AJ....135.1793C},  1.4~GHz \citep{2002AJ....123.1784D,2005ApJ...626...58H}, and 3.1~GHz \citep{2013ApJ...762...93C}. The Groth strip has been observed at 1.4~GHz \citep{2007ApJ...660L..77I}. 

The outline of this paper is as follows. The observations and data reduction are described in Sect.~\ref{sec:obs}. The results and analysis are presented in Sects.~\ref{sec:results} and \ref{sec:analysis}. This is followed by the conclusions in Sect. \ref{sec:conclusions}. All coordinates and images use the J2000 coordinate system.

\begin{figure}
\begin{center}
\includegraphics[angle =0, trim =0cm 0cm 0cm 0cm,width=0.45\textwidth]{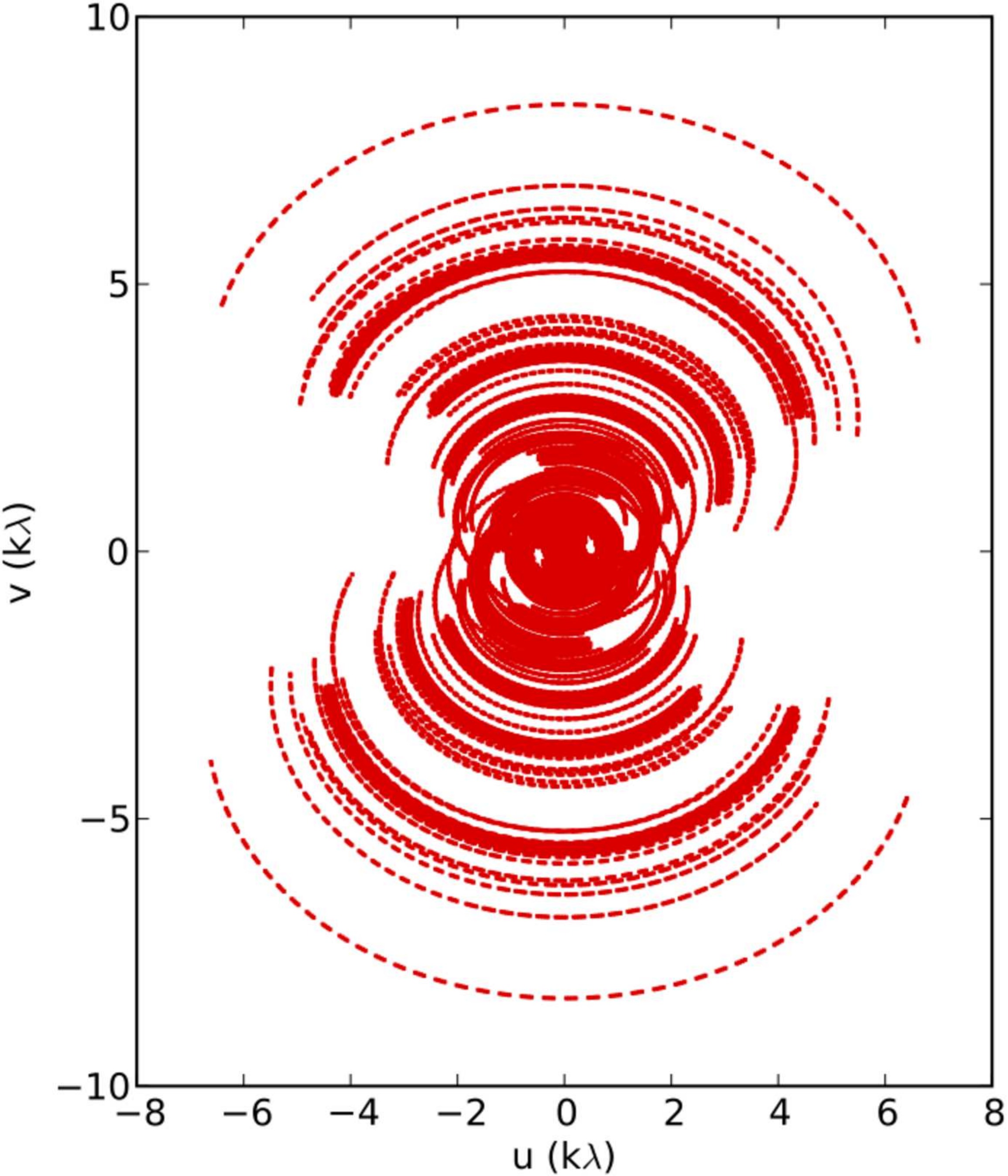}
\end{center}
\caption{UV-coverage of the 3C\,295 observations. 
The relatively large fractional bandwidth fills the uv-plane radially (not shown in the figure). }

\label{fig:uv}
\end{figure}

\section{Observations \& data reduction}
\label{sec:obs}

The Bo\"otes and 3C\,295 fields 
 were simultaneously observed on April 12, 2012 as part of a multi-beam observation with the LOFAR Low Band Antenna (LBA) stations. The idea behind the multi-beam setup is that we use the 3C\,295 observations as a calibrator field to transfer the gain amplitudes to the (target) Bo\"otes field.  
The total integration time on both fields was 10.25~hr. 
  An overview of the observations is given in Table~\ref{tab:observations}. Complete frequency coverage was obtained between 54 and 70~MHz for both fields, while non-contiguous frequency coverage was obtained between 30 and 54 MHz for the 3C\,295 only.
    All four correlation products were recorded. By default, the frequency band was divided into subbands, each 195.3125~kHz wide. Each subband was further divided in 64 channels and the integration time was 1~s.

  Nine Dutch remote stations were used, and 22 core stations, giving baselines that range between 90 m and 80~km. The resulting uv-coverage is displayed in Fig.~\ref{fig:uv}. The LBA\_OUTER configuration was used.  In the LBA\_OUTER configuration 48 LBA antennas are used, located mostly in the outer part of the stations (which have diameters of about 81~m). This increases the sidelobe levels for the station beams, but reduces the field of view (FoV)  with respect to other station antenna configurations available. 
The HPBW is about 3.6\degr, 4.8\degr, and 7.2\degr~at 60, 45, and 30 MHz, respectively. It should be noted though that the station beams are complex-valued, time and direction dependent, and differ  from station to station.

\begin{table}
\begin{center}
\caption{LBA Observations}
\begin{tabular}{lllll}
\hline
\hline
\hline
Observations ID   &  L56691 \\
Pointing center 3C\,295    & 14$^{\rm{h}}$11$^{\rm{m}}$20.9$^{\rm{s}}$, +52\degr13{\arcmin}55\arcsec  \\ 
Pointing center Bo\"otes   & 14$^{\rm{h}}$32$^{\rm{m}}$03.0$^{\rm{s}}$, +34\degr16{\arcmin}33\arcsec  \\ 
Integration time     & 1~s \\
Observation date & 12 April, 2012 \\ 
Total on-source time		&  10.25 hr\\
Correlations           & XX, XY, YX, YY \\
Frequency setup (a)               & 54--70~MHz full coverage\\
Frequency setup (b)               & 40--54~MHz  25 subbands$^{*}$\\ 
Frequency setup (c)              & 30--40~MHz  21  subbands$^{*}$\\ 
Bandwidth (a, b, c)   & 16 MHz,  4.9 MHz,  4.1 MHz\\
Bandwidth per subband		  & 195.3125~kHz \\
Channels per subband                & 64\\
\hline
\hline
\end{tabular}
\label{tab:observations}
\end{center}
(a) 54--70 MHz~Bo\"otes and 3C\,295 fields\\
(b) 40--54~MHz 3C\,295 field\\
(c) 30--44~MHz 3C\,295 field\\
$^{*}$ subbands are more or less evenly distributed within this frequency range, the total bandwidth is reported in Table~\ref{tab:lbaimages}\\

\end{table}

\subsection{Data reduction}
Our data reduction broadly consists of the following steps: (1) flagging, (2) bright off-axis source removal, (3) averaging, (4) solving for the 3C\,295 complex gains (in a circular basis to deal with differential Faraday Rotation), (5) transfer of the amplitude solutions from 3C\,295 to the Bo\"otes  field, and (6) phase-only calibration of the Bootes field against a GMRT model, and (7) imaging of the 3C\,295 field and Bo\"otes  fields. All calibration steps are performed with the {\tt BlackBoard Selfcal} ({\tt BBS}) software system \citep{2009ASPC..407..384P}. Below these steps are explained in more detail.

\subsubsection{Flagging, bright off-axis source removal, and averaging}
The first step in the reduction consisted of the automatic flagging of radio frequency interference (RFI) using the {\tt AOFlagger} \citep{2010MNRAS.405..155O, 2012A&A...539A..95O}. The first and last three channels of each subband were also flagged. Typically about 2\% of the data was flagged as RFI in the 50 to 70 MHz range. Between 30 and 40 MHz this percentage increases by a factor of $\sim$~2--3 \citep[see][for an overview of the LOFAR RFI environment]{2013A&A...549A..11O}. About a dozen subbands were lost due to failures of the data storage system.

\begin{figure*}
\begin{center}
\includegraphics[trim =0cm 0cm 0cm 0cm,angle=180, width=0.49\textwidth]{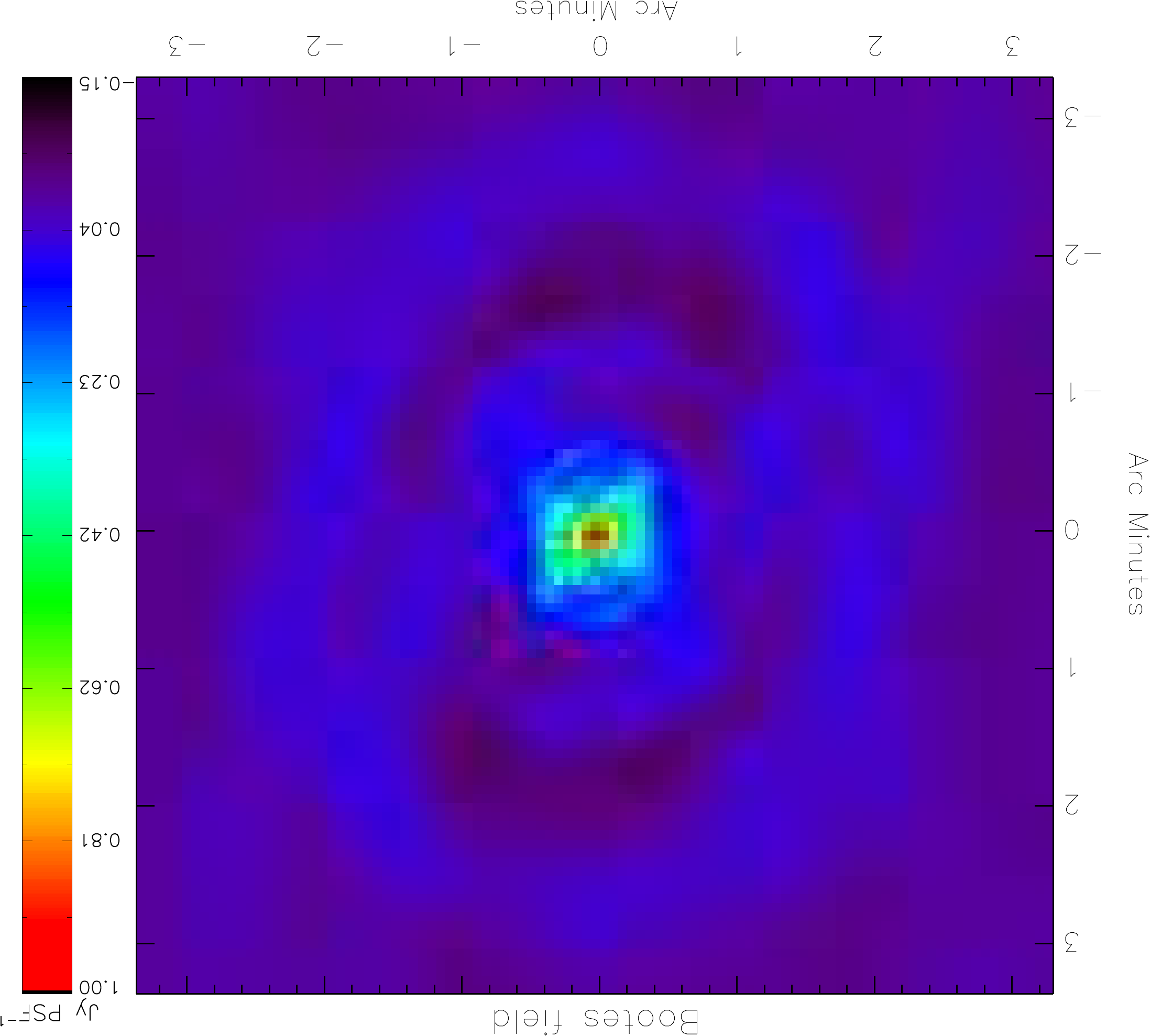}
\includegraphics[trim =0cm 0cm 0cm 0cm,angle=180, width=0.49\textwidth]{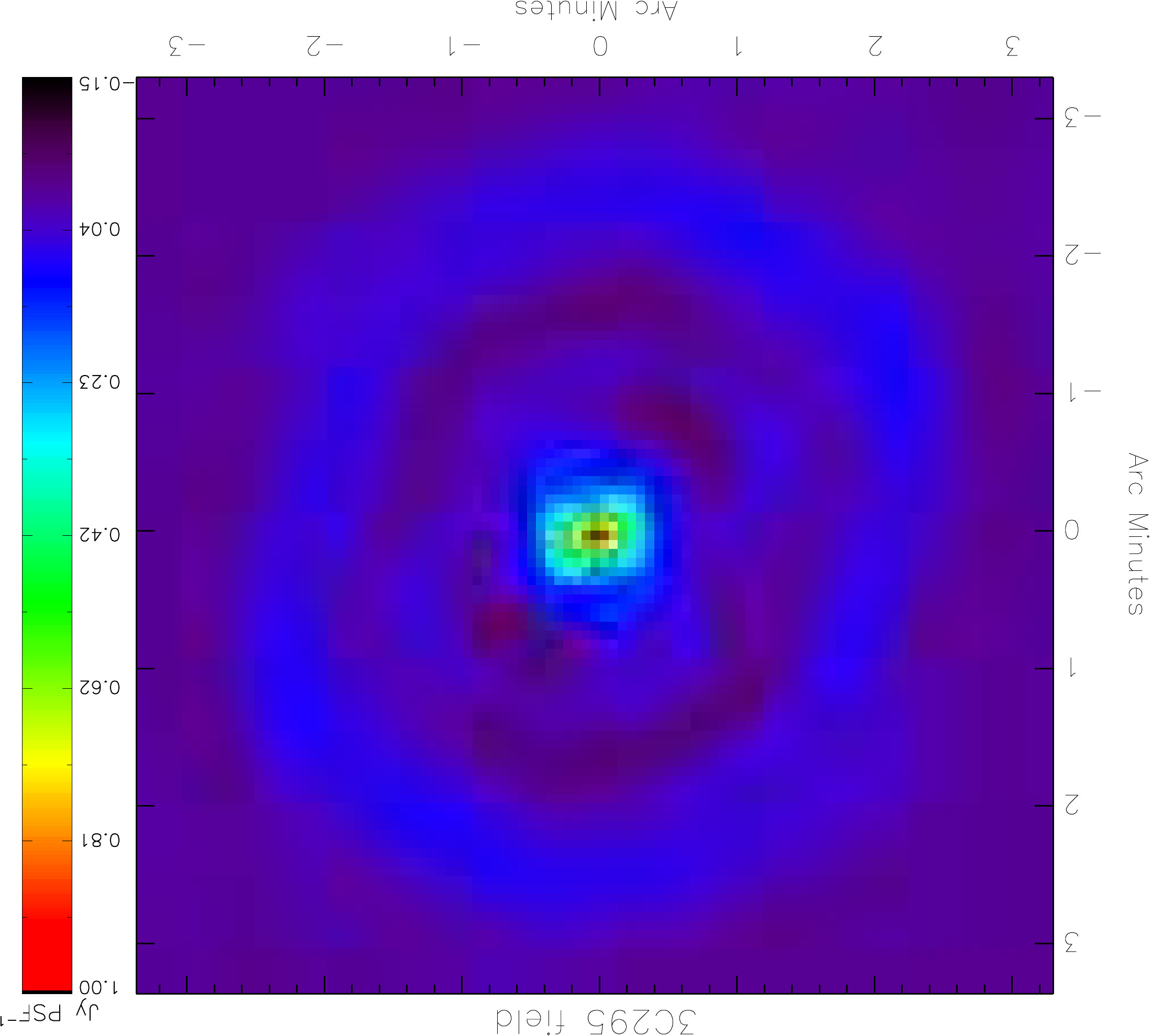}
\end{center}
\caption{Point Spread Functions for the Bo\"otes (\textit{left}) and 3C\,295 (\textit{right}) images covering the 54--70~MHz frequency range.}
\label{fig:psf}
\end{figure*}

A next step consisted of the removal of the bright ``A-team'' radio sources Cas~A and Cyg~A.  These sources have integrated flux densities of 18 and 17~kJy at 74~MHz, respectively.  Although they are located outside the main FoV, they are sufficiently bright to prevent proper calibration and imaging of sources in the central part of the FoV if detected in the secondary lobes of the beam. The amplitudes of these off-axis sources are strongly modulated as they move in and out of the station beam sidelobes. These sources were removed with the ``demixing'' method described by \cite{2007ITSP...55.4497V} and which is part of the standard LOFAR pre-processing pipeline \citep{2010iska.meetE..57H}. For the models of Cas~A and Cyg~A we took the clean component models at 74~MHz from Very Large Array (VLA) A-array\footnote{http://lwa.nrl.navy.mil/tutorial/} observations \citep{2007ApJS..172..686K} with a resolution of 25\arcsec. After flagging and subtracting out Cas~A and Cyg~A, we averaged the data in time to 5~s and one channel per subband. The time resolution is set by the requirement to avoid decorrelation due to rapid ionospheric phase variations. At large radial distances from the field center there is some bandwidth smearing. At the HPBW, the source width increases by a factor of $\sim1.2$ at 62~MHz and a factor of $\sim1.9$ at 34 MHz, due to this effect.
 
 \begin{figure*}
\begin{center}
\includegraphics[angle =180, trim =0cm 0cm 0cm 0cm,width=0.49\textwidth]{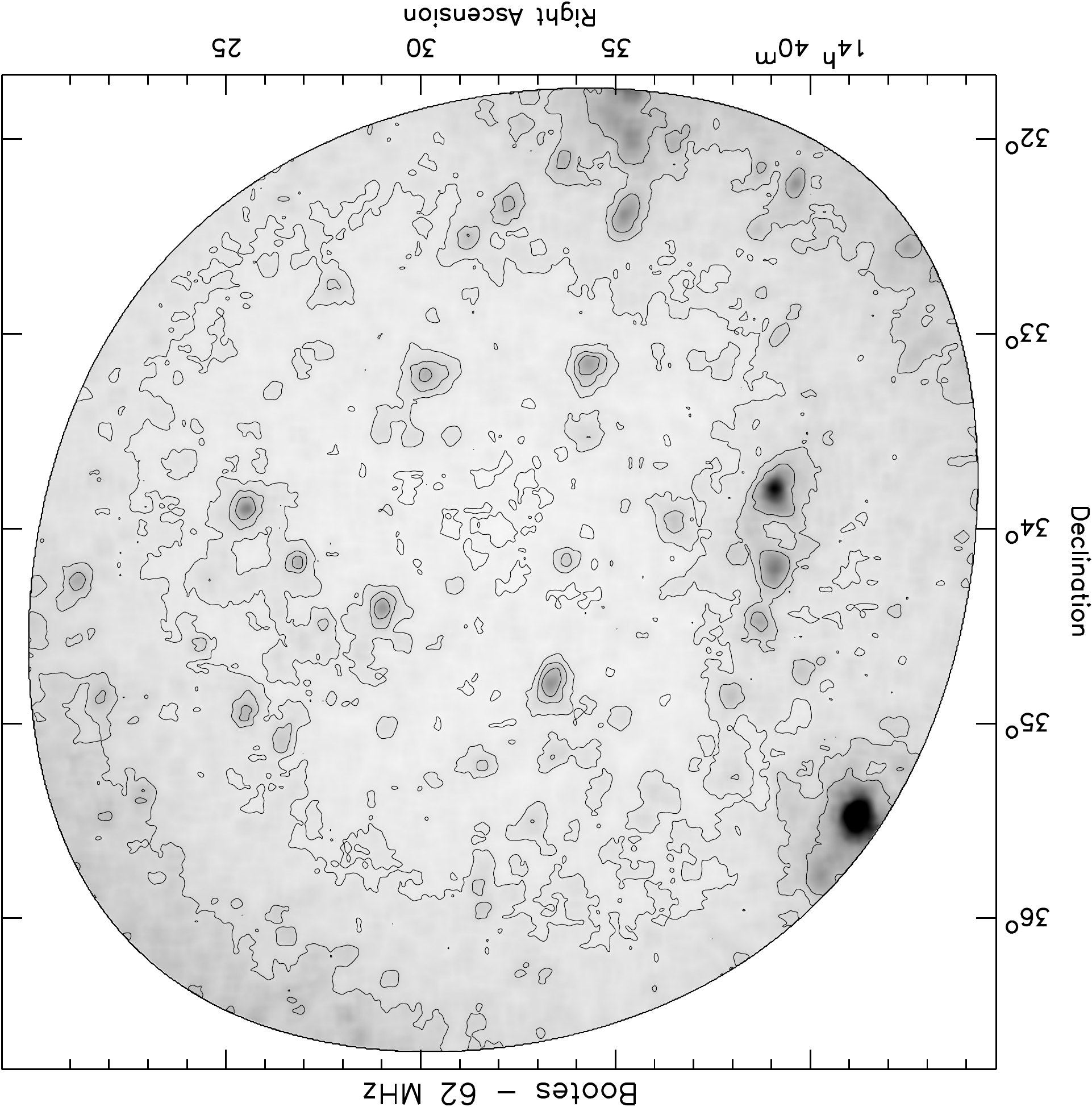}
\includegraphics[angle =180, trim =0cm 0cm 0cm 0cm, width=0.49\textwidth]{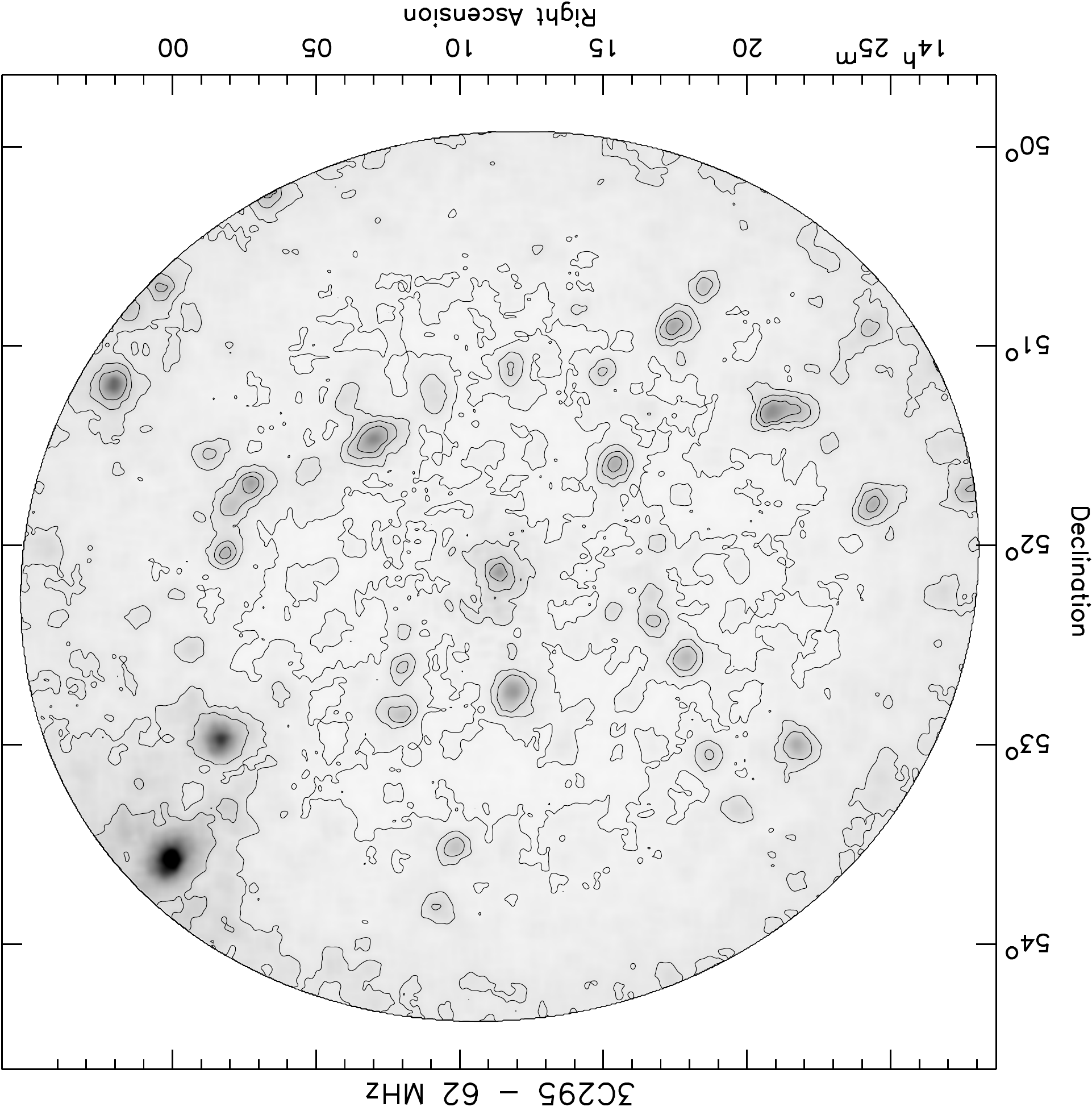}
\includegraphics[angle =180, trim =0cm 0cm 0cm 0cm, width=0.49\textwidth]{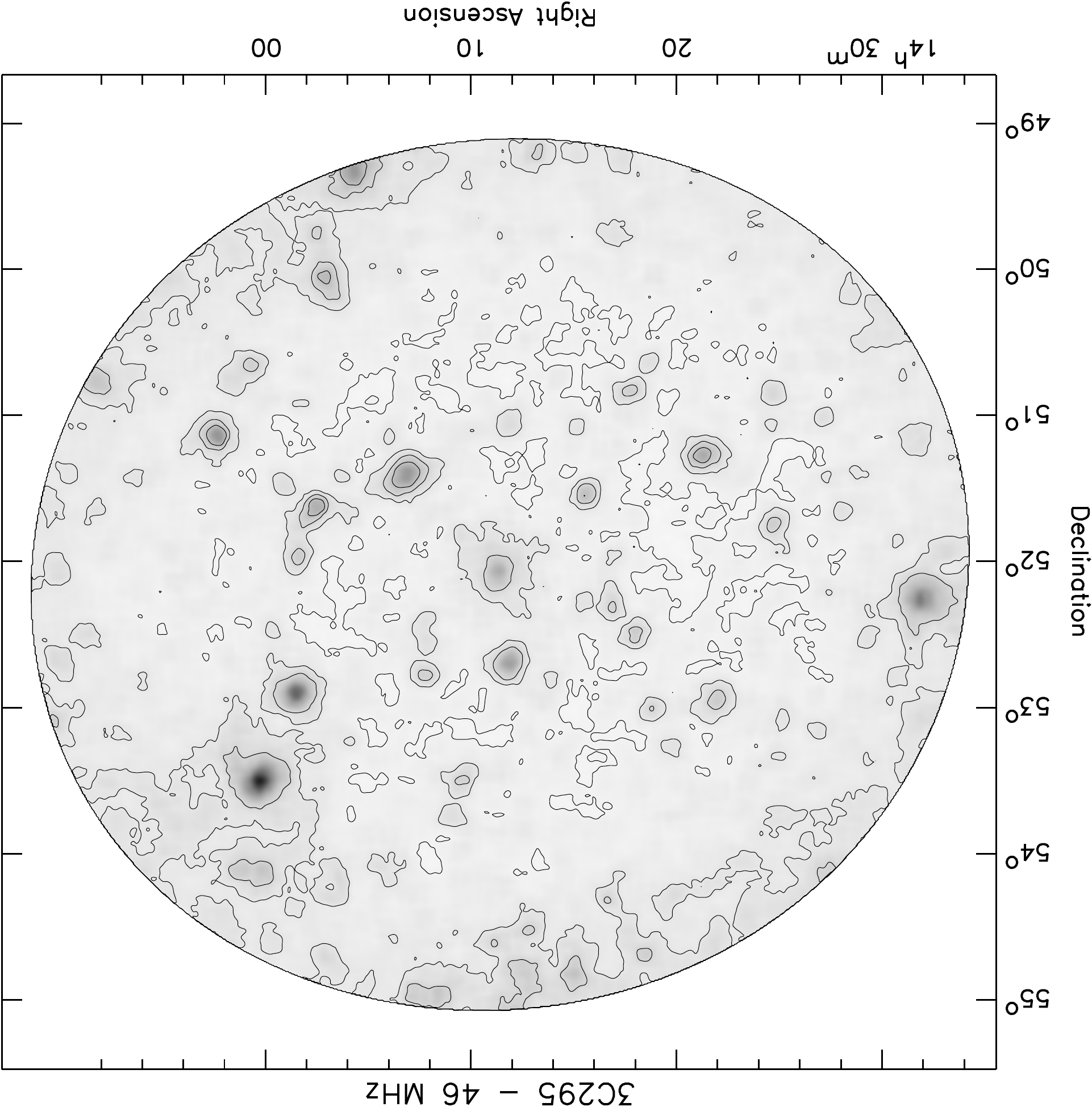}
\includegraphics[angle =180, trim =0cm 0cm 0cm 0cm, width=0.49\textwidth]{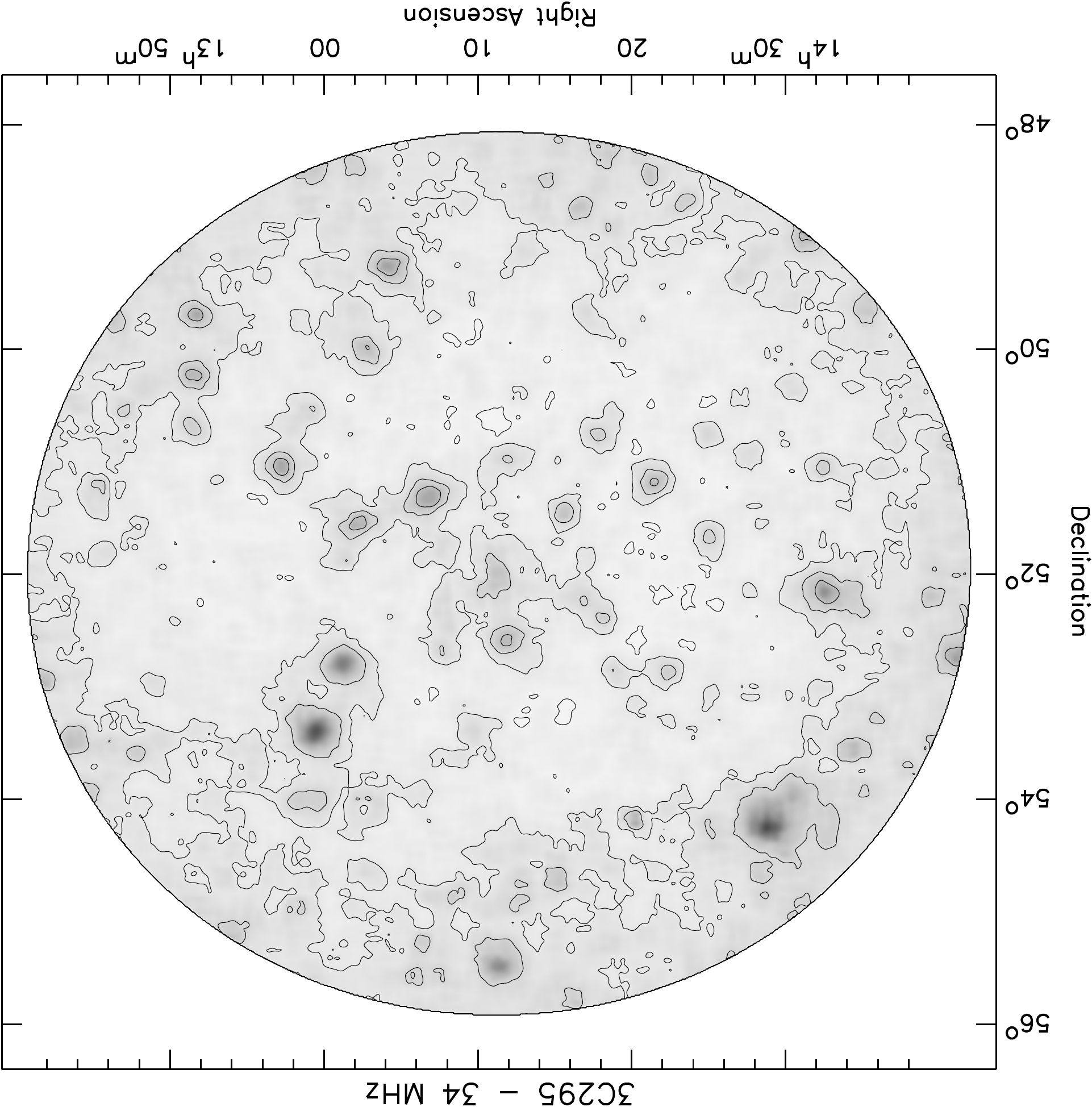}
\end{center}
\caption{Figures showing the local rms noise in the images. A box size of $80\times80$ pixels was used in computing the local rms noise. The pixel sizes are listed in Table~\ref{tab:lbaimages}.
The gray scales display the range from $0.5\sigma_{\rm{rms}}$ to $10\sigma_{\rm{rms}}$, with the $\sigma_{\rm{rms}}$ values  taken from Table~\ref{tab:lbaimages}. 
Contour levels are drawn at $\left(\sqrt{2}\right)^n  \times \sigma_{\rm{rms}}$, with $n$ ranging from $-1$ to 3. The local noise variations are correlated with the position of strong sources. }
\label{fig:rmsmaps}
\end{figure*}

\subsubsection{3C\,295 field}
The primary calibrator 3C\,295 has a sufficiently high flux density that it dominates the total flux in the main FoV. 3C\,295 consists of two main source components \citep[e.g.,][]{1991AJ....101.1623P} and has an angular size of only $\sim 5\arcsec$. Given that our longest baseline is $\sim80$~km (corresponding to a resolution of about 10\arcsec), we used a simple two clean component model for the source. The integrated flux density of the source is given by the model of \cite{2012MNRAS.423L..30S}.

Before calibrating, we converted the linear XX, XY, YX, YY correlations ($V_{\rm{XY}}$) to circular RR, RL, LR, LL correlations $V_{\rm{RL}}$ using the transformation described by \cite{1996A&AS..117..161H}
\begin{equation}
V_{\rm{RL}} = C_A V_{\rm{XY}} C^{*}_{A}   \mbox{ ,}
\end{equation}
with
\begin{eqnarray}
C_A = \frac{1}{\sqrt{2}}  \begin{pmatrix} 1 & \mbox{ }\mbox{ }i \\ 1 & -i \end{pmatrix}  \mbox{ .}
\end{eqnarray}

This transformation was done via a {\tt Python} script. The effects of the station beams\footnote{{The station beam model is derived using the dipole beam model based on the interpolation of electromagnetic simulations of the LBA dipole beam response, and the dipole locations within a station \citep{hamaker}.}} were taken out as well in the direction of 3C\,295 with {\tt BBS}.  This is needed because  the LBA stations do not record ``true'' linear correlation products due to the fixed orientation of the  dipole antennas on the ground. 

The (only) reason for converting to circular correlations is that differential Faraday Rotation, which is important in the LBA frequency range, only affects the RR and LL phases, while in linear correlations, flux from XX and YY leaks into the cross-hand XY and YX correlations. 
Therefore by converting to circular correlations the calibration is simplified, since we only have to solve for the RR and LL phases to remove the effects of differential Faraday Rotation \citep[e.g.,][]{2011A&A...527A.107S}. {The conversion from linear to circular correlations depends on the accuracy of the beam models. It is also possible to solve for differential Faraday Rotation in a more direct way using the observed linear correlations but this requires solving for an extra free parameter.}

After converting to circular correlations we obtained amplitude and phase solutions for the RR and LL correlations for each subband using the 3C\,295 model (with {\tt BBS}) . We used a solution interval of 5~s. This takes care of the frequency dependence of ionospheric phase variations, differential Faraday Rotation,  clock errors, and the overall LBA bandpass (with a single complex gain correction for each time interval per subband). Good quality solutions were obtained over the entire time and frequency range, except for time periods affected by RFI. We then subtracted 3C\,295 from the data using these gain solutions. {This avoids many clean cycles and clean dynamic range limitations such as described in \cite{2008A&A...490..455C}.}
After the 3C\,295 gain calibration we converted back the calibrated visibilities from circular to linear correlations because of limitations in the imaging software.
 
\subsubsection{Bo\"otes field}
We transferred the amplitude solutions from the corresponding frequencies of the 3C\,295 observation to the Bo\"otes field dataset. The Bo\"otes field does not contain any bright dominating sources. This means that there is not enough signal available per subband for a phase-only calibration on a timescale of 5~sec. To increase the signal to noise, all subbands were combined into a new measurement set consisting of 81~channels covering the entire 54-70~MHz range, with each channel corresponding to one individual subband. We then performed a phase-only calibration for groups of 27~channels each to obtain sufficient signal to noise to calibrate the distant remote stations against the GMRT 153~MHz model. 

For the Bo\"otes field, the calibration model is derived from a deep GMRT 153~MHz image \citep{2013A&A...549A..55W} using the  {\tt PyBDSM} source detection software\footnote{http://dl.dropboxusercontent.com/u/1948170/html/index.html}. 
\subsection{Imaging and cleaning}

 Imaging and cleaning was carried out with {\tt awimager} \citep{2013A&A...553A.105T}, which incorporates the complex valued, time varying and frequency dependent individual station beams using A-Projection \citep{2008A&A...487..419B}. For LOFAR, all $4\times4$ Mueller terms have to be taken into account in the A-Projection. 
 For {\tt awimager} a hybrid AW-projection algorithm was developed to apply the time, frequency, baseline, and direction dependent effects in
full-polarization in an efficient way. Also, a new parallel gridding technique is used, which differs from the {\tt casapy}\footnote{http://casa.nrao.edu} gridder. 

For the imaging, we combined all available 54--70 MHz subbands for the Bo\"otes and 3C\,295 fields to improve the uv-coverage with multi-frequency synthesis (MFS). We did not correct for the spectral index of individual sources \citep{2011A&A...532A..71R} because such an algorithm is not yet implemented for {\tt awimager}. For the 3C\,295 field, we made two additional images from the subbands in the ranges 30--40~MHz and 40--54 MHz. We employed various robust weighting schemes  \citep{Briggs_phd} to find that a robust parameter of about $0.0$ typically gave the lowest rms noise level. All final images have sizes of $8192^2$ pixels and were made with a robust value of $0.0$ and all baselines were included. The Point Spread Functions for the 54--70~MHz images are shown in Fig.~\ref{fig:psf}. An overview of the image properties is given in Table~\ref{tab:lbaimages}.

We used clean masks during the final imaging step to minimize clean bias effects \citep[e.g.,][]{1998AJ....115.1693C, 1997ApJ...475..479W}. The mask was derived from a previous imaging run without any mask. The clean mask was  generated with {\tt PyBDSM}, detecting islands of emission with a $3\sigma_{\rm{rms}}$ island threshold, a pixel threshold of $5\sigma_{\rm{rms}}$, and a locally varying rms box with  a size of $80\times80$ pixels to take into account artifacts around strong sources. {The 80-pixels approximately correspond to the spatial scale over which the local rms noise changes in the presence of strong sources.}  Maps of the local rms noise are shown in Fig.~\ref{fig:rmsmaps}.  

\section{Results}
\label{sec:results}
\begin{figure*}
\begin{center}
\includegraphics[angle =180, trim =0cm 0cm 0cm 0cm,width=1.0\textwidth]{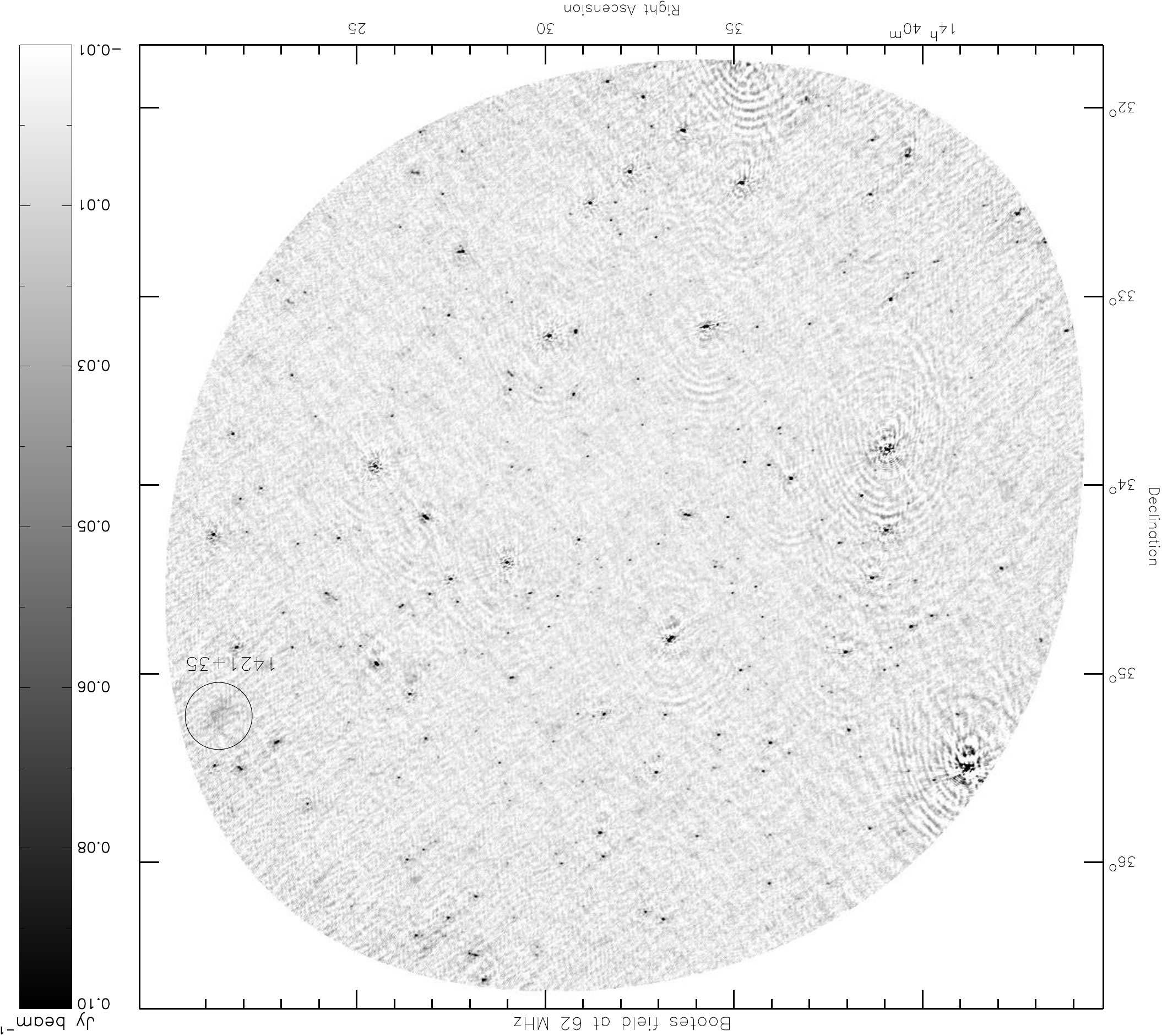}
\end{center}
\caption{Primary beam corrected Bo\"otes field 62~MHz image. The image is blanked beyond a primary beam attenuation factor of 0.4. The position of the diffuse source 1421+35 is indicated.}
\label{fig:60mhzfullbootes}
\end{figure*}

\begin{figure*}
\begin{center}
\includegraphics[trim =0cm 0cm 0cm 0cm,angle=180, width=1.0\textwidth]{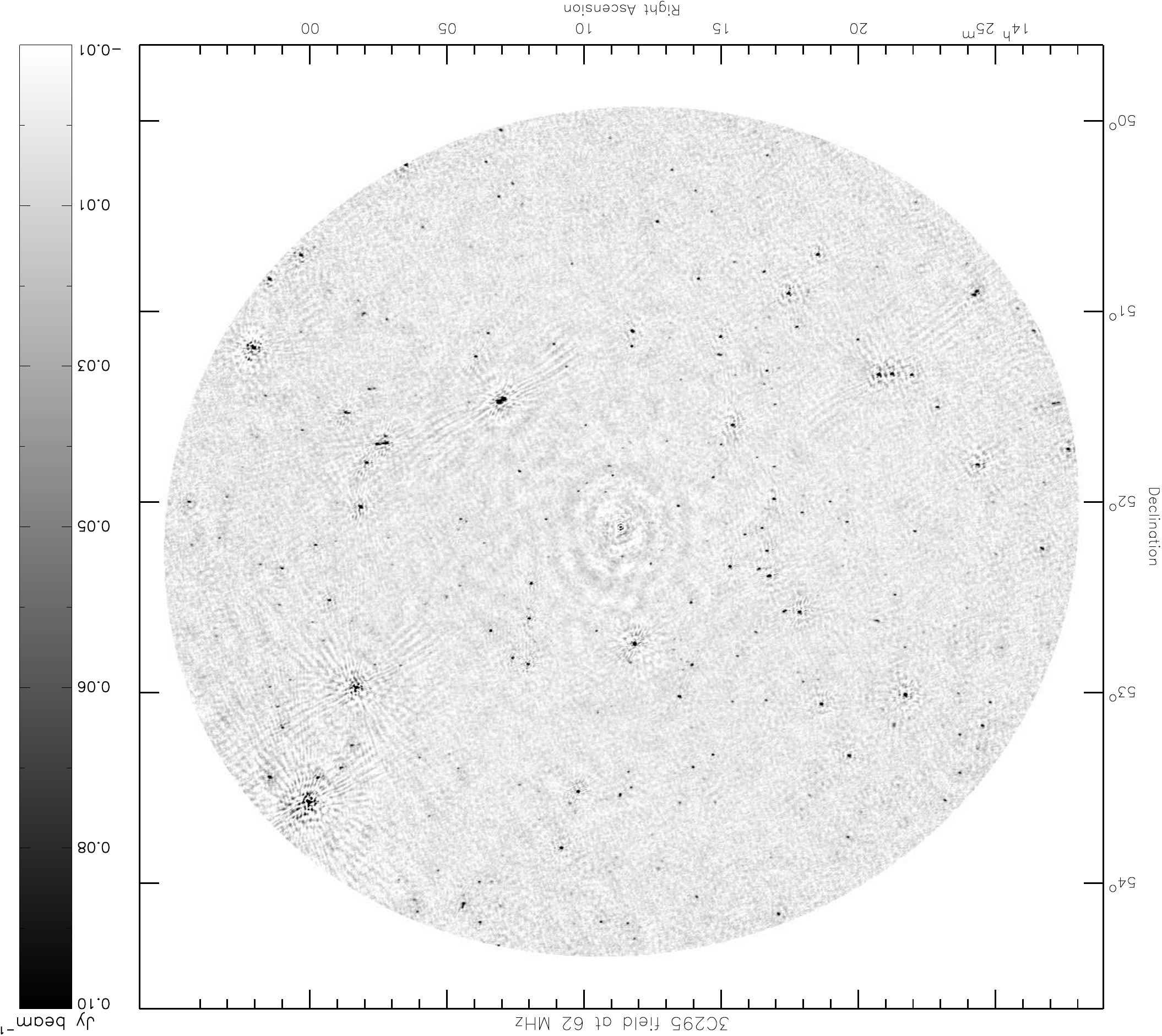}
\end{center}
\caption{Primary beam corrected 3C\,295 field 62~MHz image. The image is blanked beyond a primary beam attenuation factor of 0.4.}
\label{fig:60mhzfull}
\end{figure*}

\begin{figure*}
\begin{center}
\includegraphics[trim =0cm 0cm 0cm 0cm,angle=180, width=1.0\textwidth]{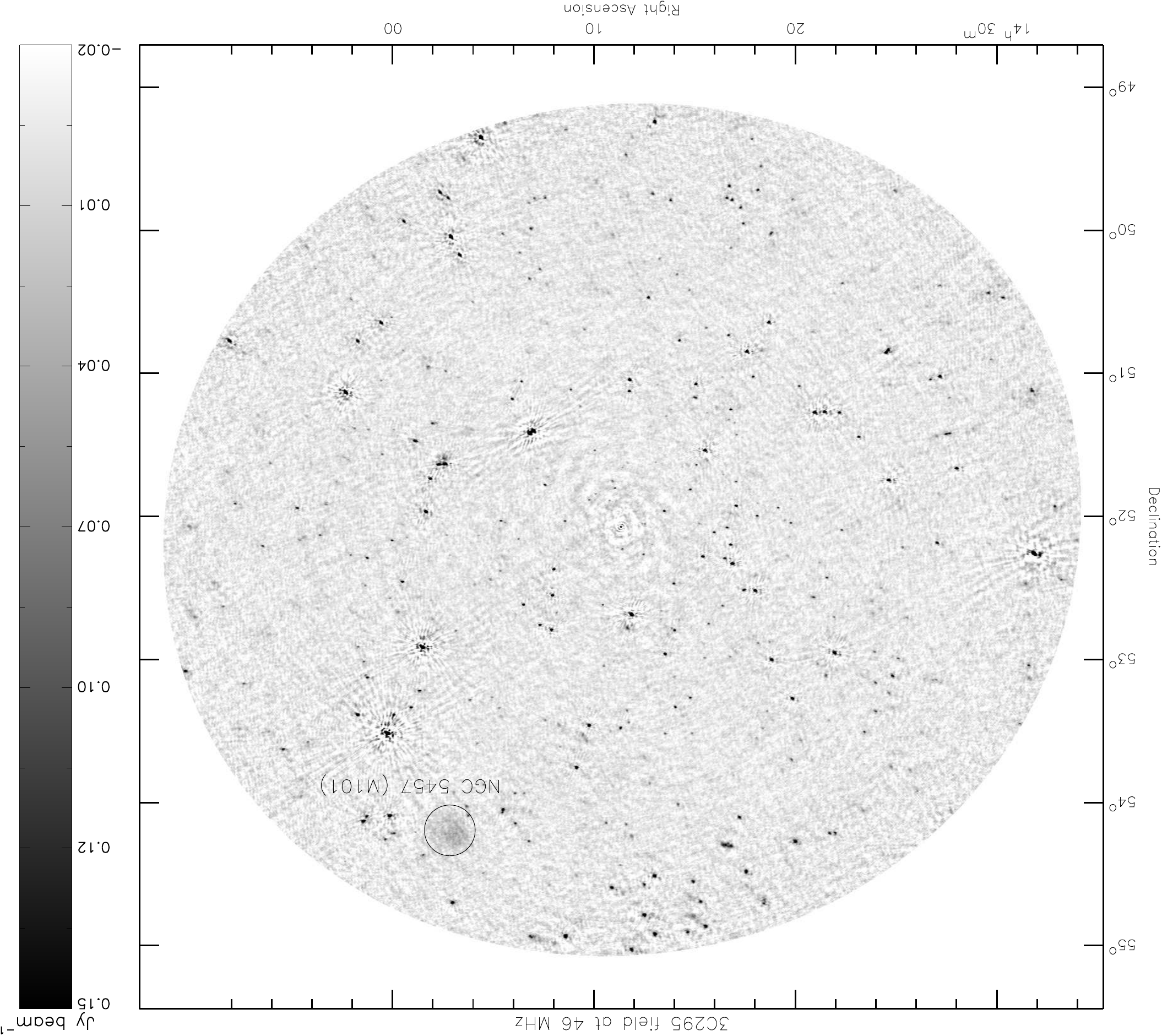}
\end{center}
\caption{Primary beam corrected 3C\,295 field 46~MHz image. The image is blanked beyond a primary beam attenuation factor of 0.4. The position of NGC\,5457 is indicated.}
\label{fig:46mhzfull}
\end{figure*}

\begin{figure*}
\begin{center}
\includegraphics[ trim =0cm 0cm 0cm 0cm,angle=180, width=1.0\textwidth]{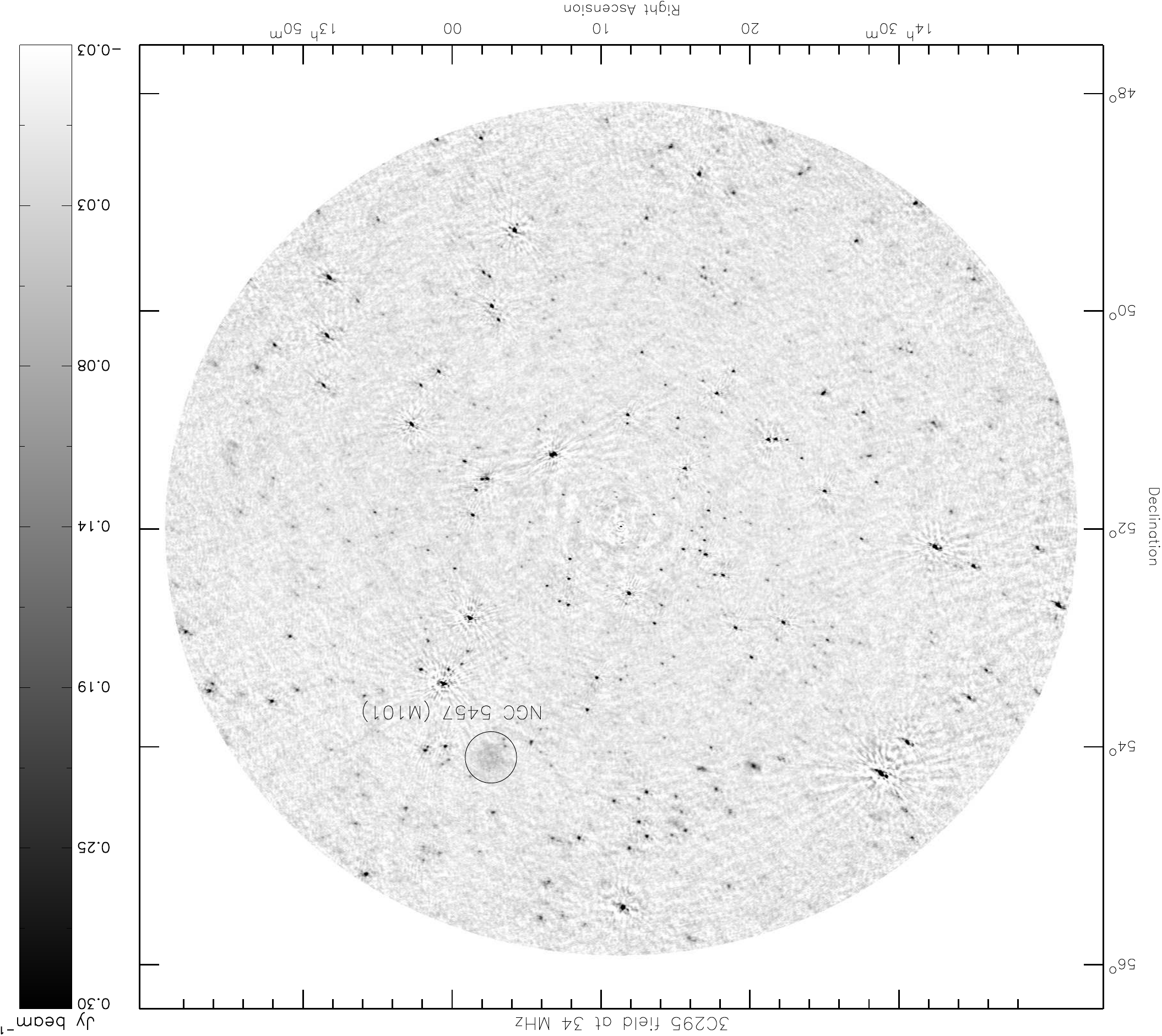}
\end{center}
\caption{Primary beam corrected 3C\,295 field 34~MHz image. The image is blanked beyond a primary beam attenuation factor of 0.4. The position of NGC\,5457 is indicated.}
\label{fig:34mhzfull}
\end{figure*}

An overview of the resulting images, resolution, FoV,  and noise levels obtained is given Table~\ref{tab:lbaimages}. The primary beam corrected images are displayed in Figs.~\ref{fig:60mhzfullbootes} to \ref{fig:34mhzfull}. The artifacts  visible around the brighter sources in the fields are due to imperfect calibration and errors in the station beam model. These artifacts also give rise to the increased noise around bright sources (Fig.~\ref{fig:rmsmaps}).  The ``spoke''-like patterns are likely caused by direction dependent ionospheric phase errors. The spokes are not visible at the position of 3C\,295 because the ionospheric phase variations in this direction were properly taken into account (phase calibration was performed in the 3C\,295 direction).

The ``smudge'' visible in the 3C\,295 field (labeled with a circle in Figs.~\ref{fig:46mhzfull} and \ref{fig:34mhzfull}) at 14$^{\rm{h}}$03$^{\rm{m}}$ +54\degr21\arcmin~is the galaxy \object{NGC\,5457} (M101). In the Bo\"otes field, faint diffuse emission is found at  14$^{\rm{h}}$21.5$^{\rm{m}}$ +35\degr12\arcmin, labeled with a circle in Fig.~\ref{fig:60mhzfullbootes}. This source (1421+35) was previously also noted by \cite{2006AN....327..561D} and \cite{2013A&A...549A..55W}. A more detailed study of the source was performed by \cite{2014MNRAS.440.1542D}. They conclude that the extended radio emission is the remnant of a past AGN activity cycle of \object{NGC 5590} at $z=0.0107$.

\begin{table*}
\begin{center}
\caption{LOFAR LBA image characteristics}
\begin{tabular}{lllllll}
\hline
\hline
field &frequency & field of view$^a$ & bandwidth &  rms noise ($\sigma_{\rm{rms}}$) & synthesized beam  & pixel size\\
          &  MHz         &    deg$^2$    &      MHz                    & mJy~beam$^{-1}$    & arcsec    & arcsec         \\
\hline
Bo\"otes  & 62   &   19.4       & 16       & 4.8  &  $31\times19$  & 4.0  \\
3C295  & 62   &    17.0       & 16      &  5.3   & $29\times18$  & 4.0    \\
3C295  & 46   &     30.5       & 4.9     &  8.2  &   $40\times24$& 6.0\\
3C295  & 34   &      52.3      & 4.1     &  12  & $56\times30$  & 8.0  \\
\hline
\hline
\end{tabular}
\label{tab:lbaimages}
\end{center}
$^{a}$ with a primary beam correction factor $< 0.4$\\
\end{table*}

\subsection{Source detection}
\label{sec:pybdsm}
We used the {\tt PyBDSM} software for source detection. {\tt PyBDSM} works by identifying islands of contiguous pixels above a certain detection threshold and fitting each island with Gaussians. For detecting islands, we took a threshold of $3\sigma_{\rm{rms}}$ and a pixel threshold of $5\sigma_{\rm{rms}}$, meaning that at least one pixel in each island needs to be above $5\sigma_{\rm{rms}}$. We used a locally varying rms noise with a sliding box size of $80\times80$ pixels to take into account the rms noise increase around the bright sources. We manually inspected the output source catalogues to remove about a dozen false detections. {These false detections were associated to side-lobes near bright sources.} No sources beyond a primary beam attenuation factor of 0.4 were included.

Because the sources are distorted and smeared, and this distortion varies across FoV due to the ionosphere, the fitted major and minor axes for the Gaussian components cannot be simply used to determine whether a source is resolved or not.  To first order, the derived integrated flux densities for the sources should not be affected by the smearing. We carried out a visual inspection for actual resolved sources, images of these sources are given in Appendix~\ref{sec:extim}. {In Fig.~\ref{fig:ion} we plot histograms of the fitted major and minor axes for the sources in the 3C\,295 and  Bo\"otes fields. The decrease of the effective resolution towards lower frequencies can be seen by the broadening of the distribution of fitted major and minor axes.} 
The final source list at 62~MHz contains 329 sources for both the Bo\"otes and 3C\,295 fields. At 46 and 34~MHz, the lists contain 367 and 392 sources  for the 3C\,295 field, respectively. Our LBA images reach a similar depth as the 325~MHz WENSS survey (scaling with a spectral index of $-0.7$). Because of the ionospheric distortions, we do not classify resolved or unresolved sources. The uncertainties for the  measured flux densities and positions are discussed in Sects.~\ref{sec:astrometric} and \ref{sec:fluxuncertainties}. {An example of the source catalogues is shown in Table~\ref{tab:catalog}. For each source we list the source name, the flux weighted coordinates and uncertainties, and the integrated flux densities and uncertainties. }

\begin{table*}[h!]
 \centering
 \begin{center}
 \caption{Sample of the LOFAR $62$\ MHz Bo\"otes field source catalogue}
 \label{tab:catalog}
\begin{tabular}{rccccr@{\,$\pm$\,}lr@{}lr@{\,$\pm$\,}lr@{\,$\pm$\,}lr@{\,$\pm$\,}lcc}
\hline
 Source ID & RA & $\sigma_{\rm{RA}}$ & DEC &  $\sigma_{\rm{DEC}}$  &  $S$ &   $\sigma_{S}$&   \\
  & \multicolumn{1}{c}{[deg]} & \multicolumn{1}{c}{[$\arcsec$]} & \multicolumn{1}{c}{[deg]} & \multicolumn{1}{c}{[$\arcsec$]} & \multicolumn{2}{c}{[mJy]} \\
\hline
    J143859.5+345312 & 219.74800 &  2.2 & 34.88676 &  2.1 &    237 &     40 \\
    J143856.4+343310 & 219.73524 &  3.2 & 34.55297 &  2.5 &    145 &     29 \\
    J143849.0+335015 & 219.70420 &  1.8 & 33.83753 &  1.9 &   6469 &   1008 \\
    J143849.3+341553 & 219.70580 &  1.8 & 34.26481 &  1.9 &   2816 &    439 \\
    J143850.4+350020 & 219.71027 &  2.0 & 35.00571 &  2.0 &    298 &     49 \\
    J143831.3+335652 & 219.63057 &  4.4 & 33.94792 &  3.2 &    174 &     36 \\
    J143828.9+343107 & 219.62059 &  1.8 & 34.51874 &  1.9 &   2035 &    317 \\
    J143819.1+321149 & 219.57987 &  2.1 & 32.19706 &  2.1 &    566 &     91 \\
    J143831.6+355053 & 219.63205 &  2.1 & 35.84832 &  2.1 &    516 &     81 \\
    J143817.1+322905 & 219.57145 &  1.8 & 32.48483 &  2.0 &   1251 &    196 \\
    J143821.6+344000 & 219.59040 &  2.6 & 34.66683 &  2.2 &    194 &     34 \\
    J143814.2+342010 & 219.55944 &  2.8 & 34.33632 &  2.2 &    148 &     27 \\
    J143810.9+340500 & 219.54543 &  1.8 & 34.08339 &  1.9 &    797 &    125 \\
    J143814.8+352807 & 219.56189 &  2.6 & 35.46880 &  2.3 &    186 &     32 \\
    J143750.2+345451 & 219.45921 &  1.8 & 34.91425 &  2.1 &   2106 &    328 \\
 \hline
\end{tabular}
 \end{center}
\end{table*}

\begin{figure*}
\begin{center}
\includegraphics[trim =0cm 0cm 0cm 0cm,angle=180, width=0.49\textwidth]{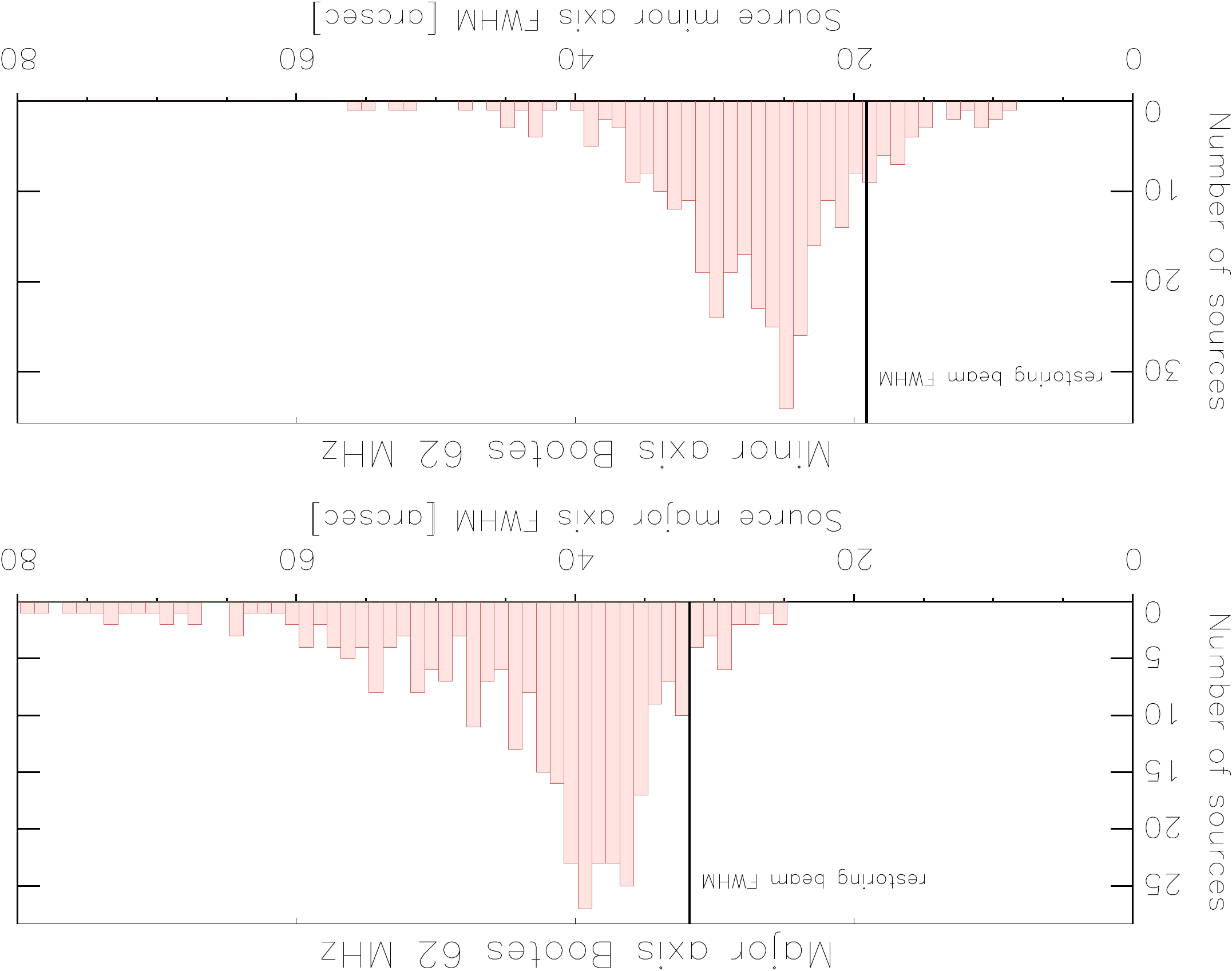}
\includegraphics[trim =0cm 0cm 0cm 0cm,angle=180, width=0.49\textwidth]{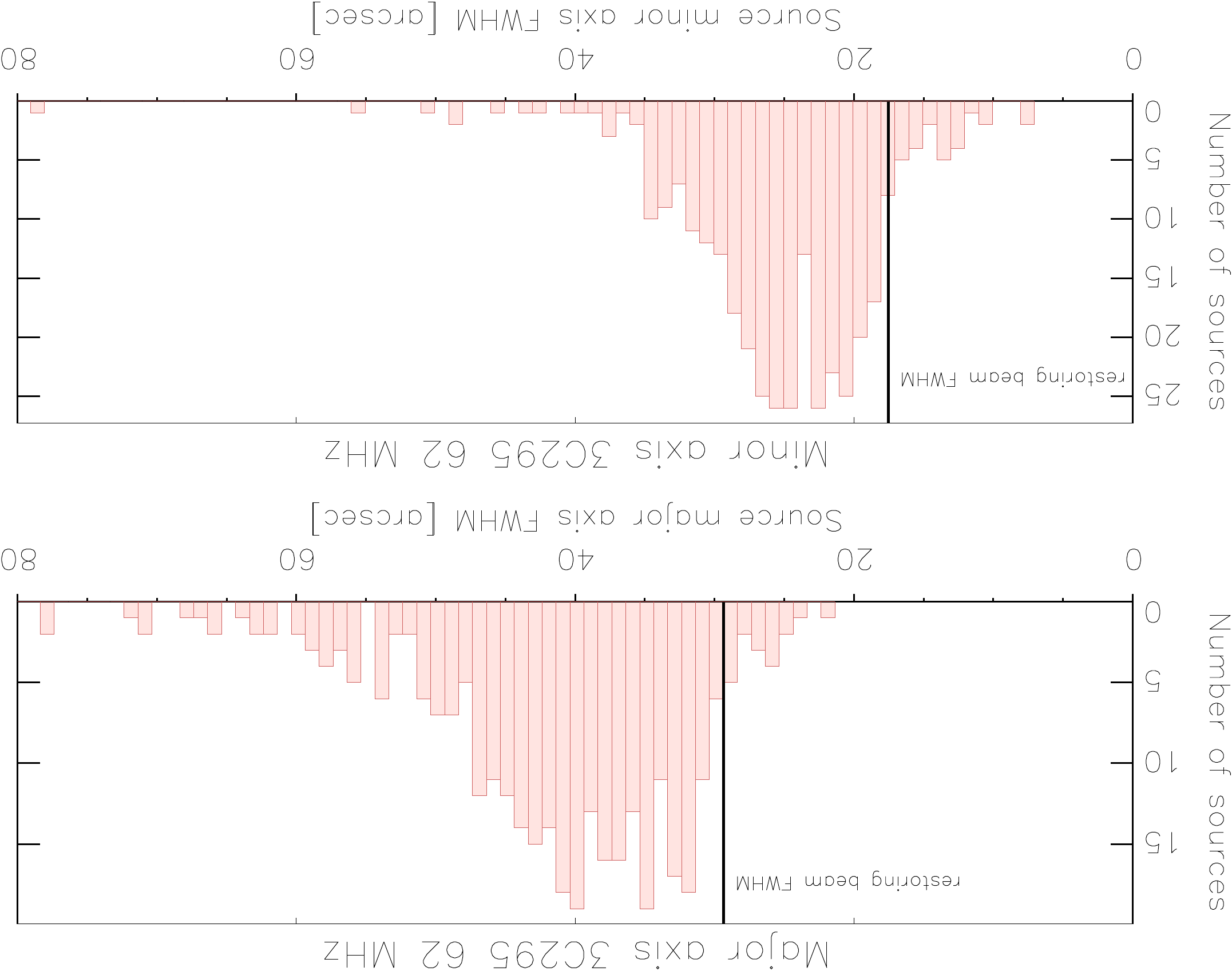}
\includegraphics[trim =0cm 0cm 0cm 0cm,angle=180, width=0.49\textwidth]{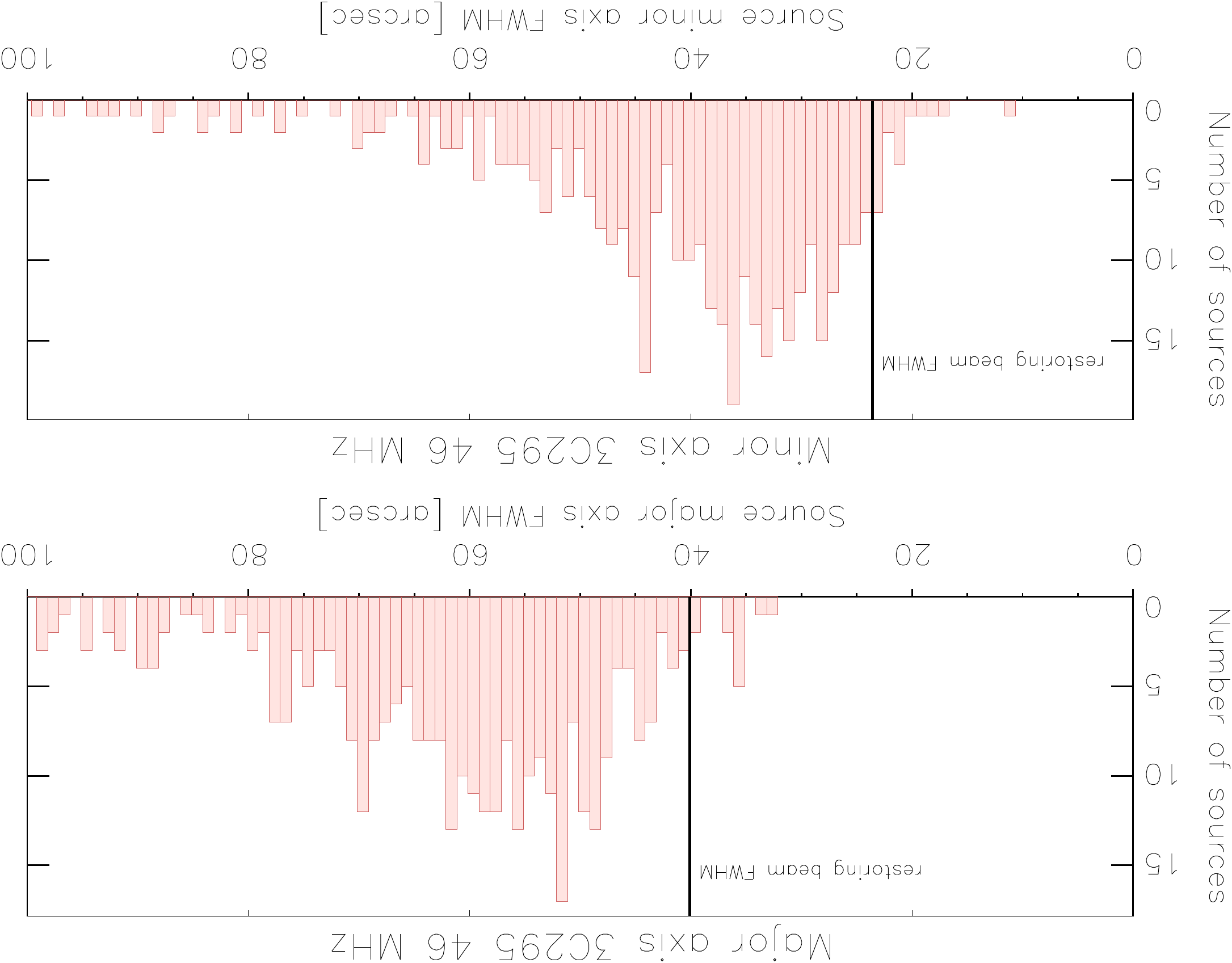}
\includegraphics[trim =0cm 0cm 0cm 0cm,angle=180, width=0.49\textwidth]{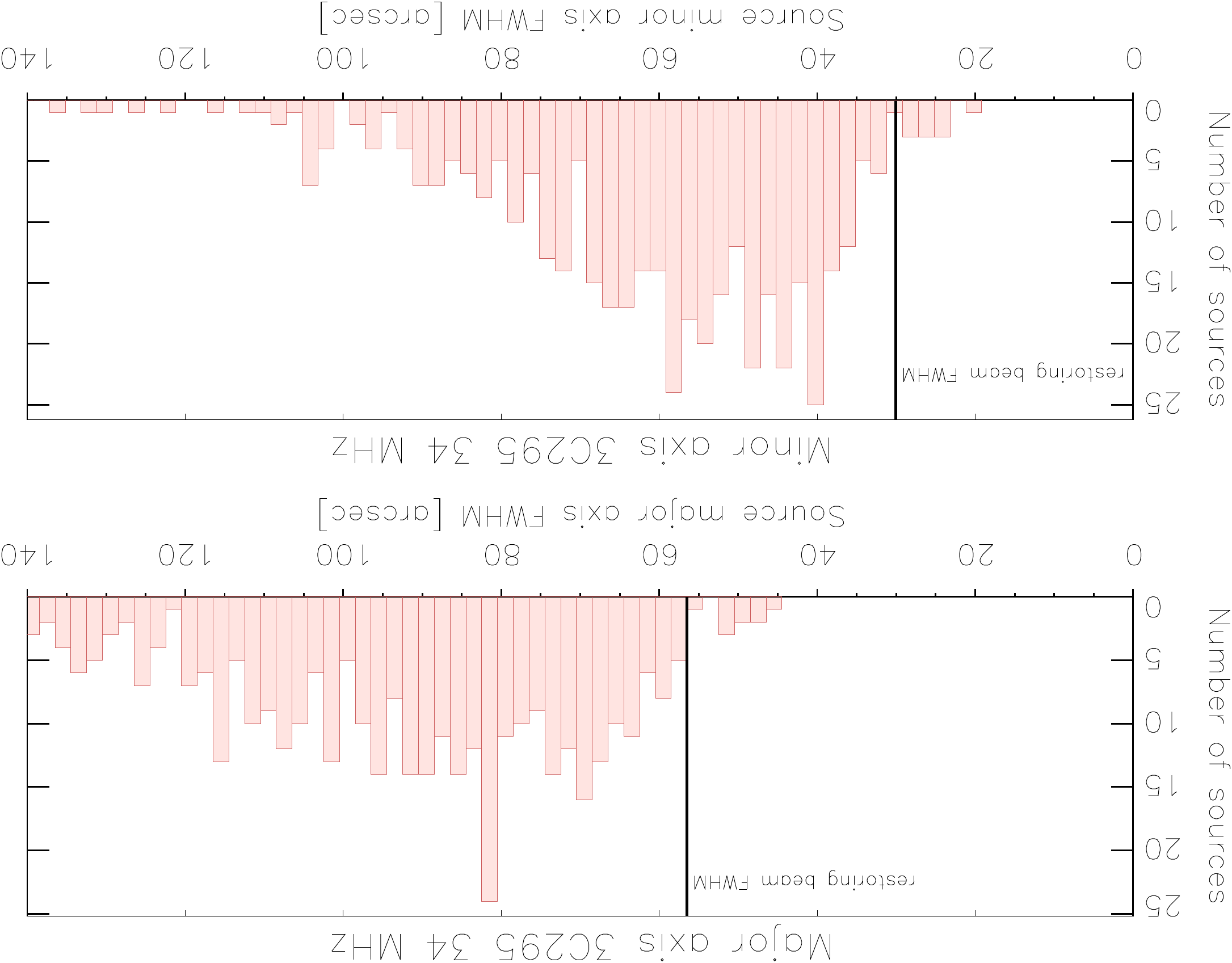}
\end{center}
\caption{Distribution of fitted major and minor axes (full width half maximum) for the Bo\"otes and 3C\,295 fields. These values are not deconvolved by the beam. Solid black line indicates the fitted restoring beam based on the uv coordinates. Uncorrected ionospheric phase variations causes ``smearing'' of sources and decreases the effective resolution. This effect increases towards lower frequencies.}
\label{fig:ion}
\end{figure*}

\subsection{Astrometric uncertainties}
\label{sec:astrometric}


Ionospheric phase distortions and residual calibration errors can have an effect on the source positions. To assess the accuracy of the LBA source positions we compared them to the source positions from the 325~MHz WENSS survey \citep{1997A&AS..124..259R}. The positional accuracy of the WENSS survey is reported to be $5-10\arcsec$~for the faintest sources and increases to 1.5\arcsec~for the brighter sources.

\begin{figure*}
\begin{center}
\includegraphics[ trim =0cm 0cm 0cm 0cm,angle=180, width=0.49\textwidth]{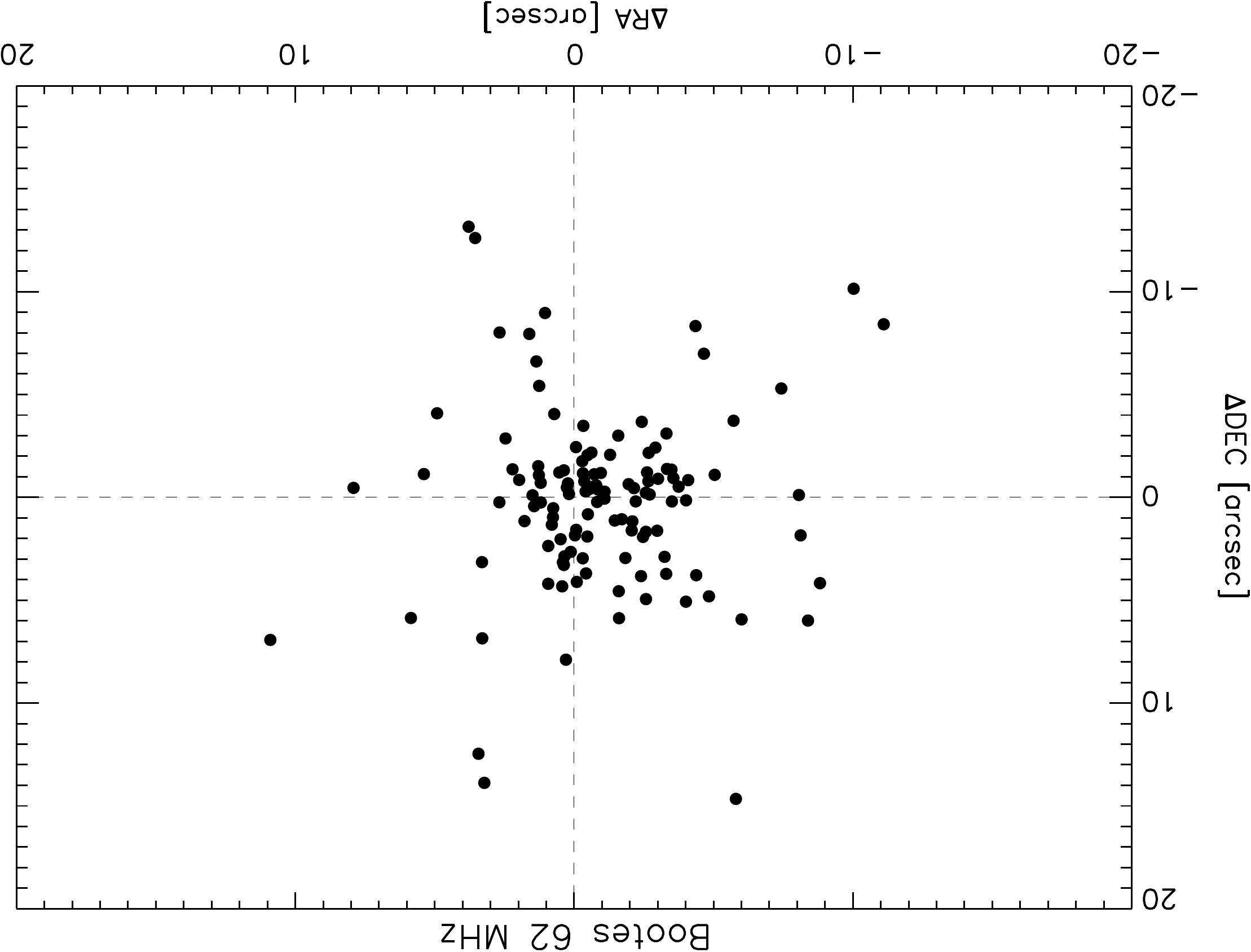}
\includegraphics[ trim =0cm 0cm 0cm 0cm,angle=180, width=0.49\textwidth]{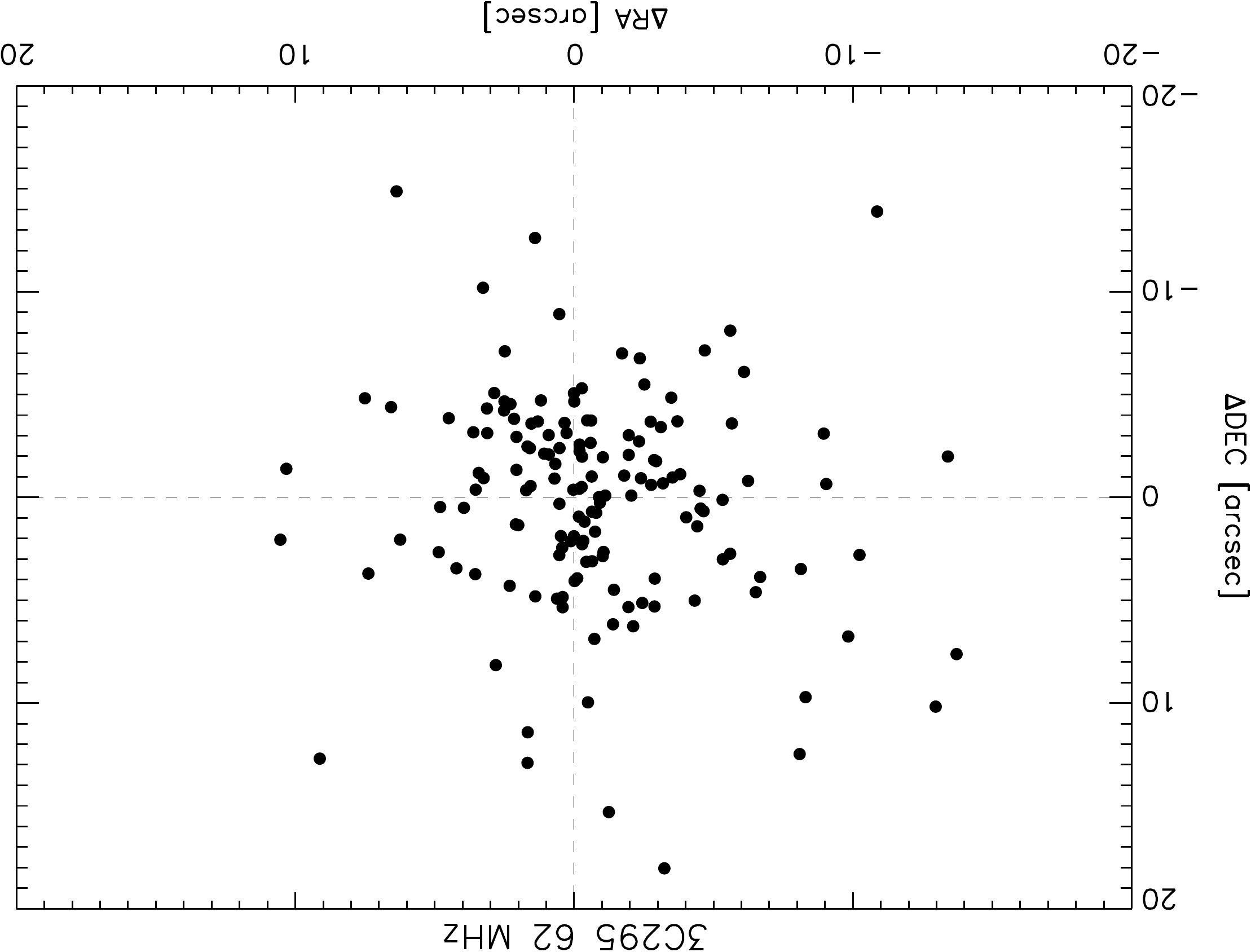}
\includegraphics[ trim =0cm 0cm 0cm 0cm,angle=180, width=0.49\textwidth]{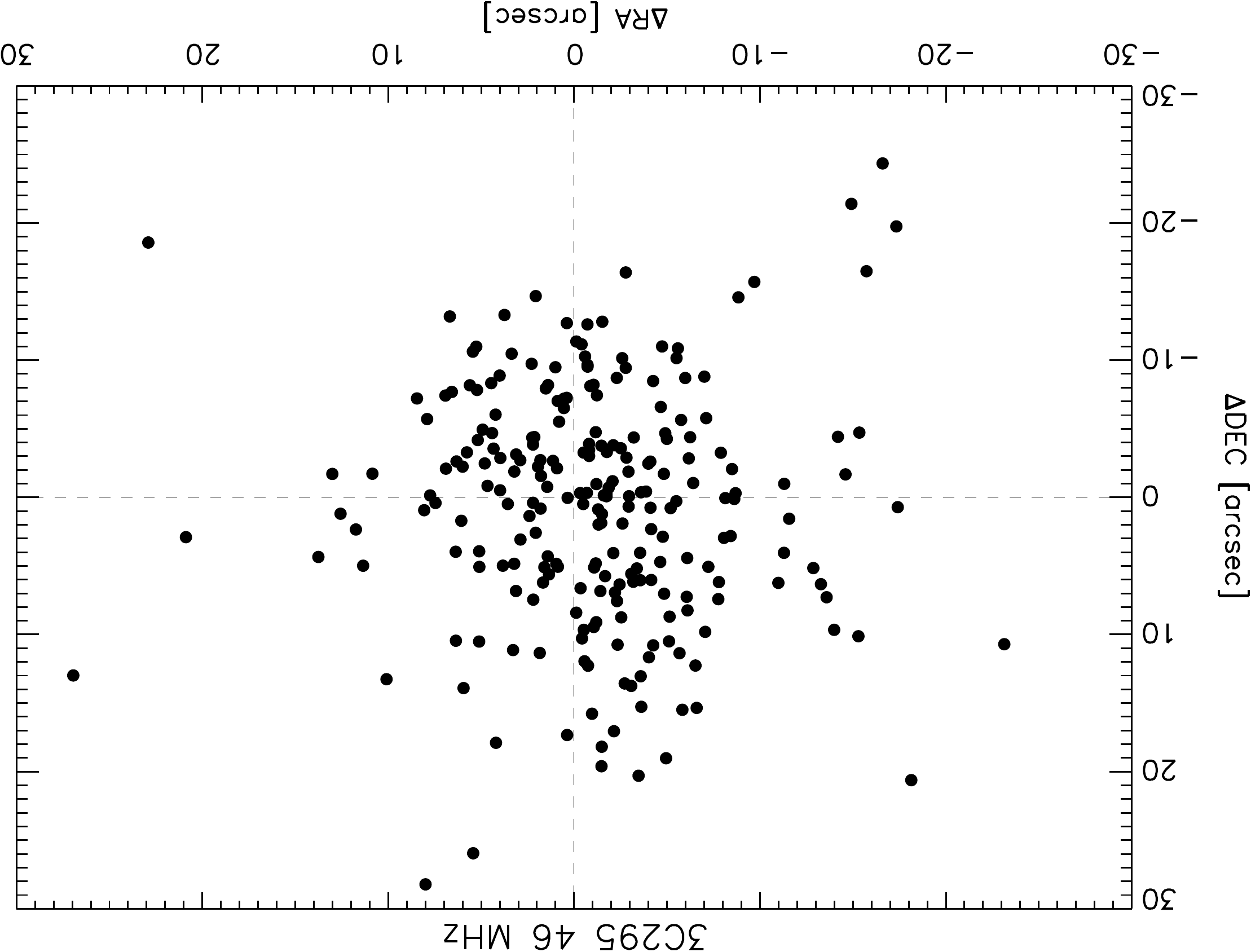}
\includegraphics[ trim =0cm 0cm 0cm 0cm,angle=180, width=0.49\textwidth]{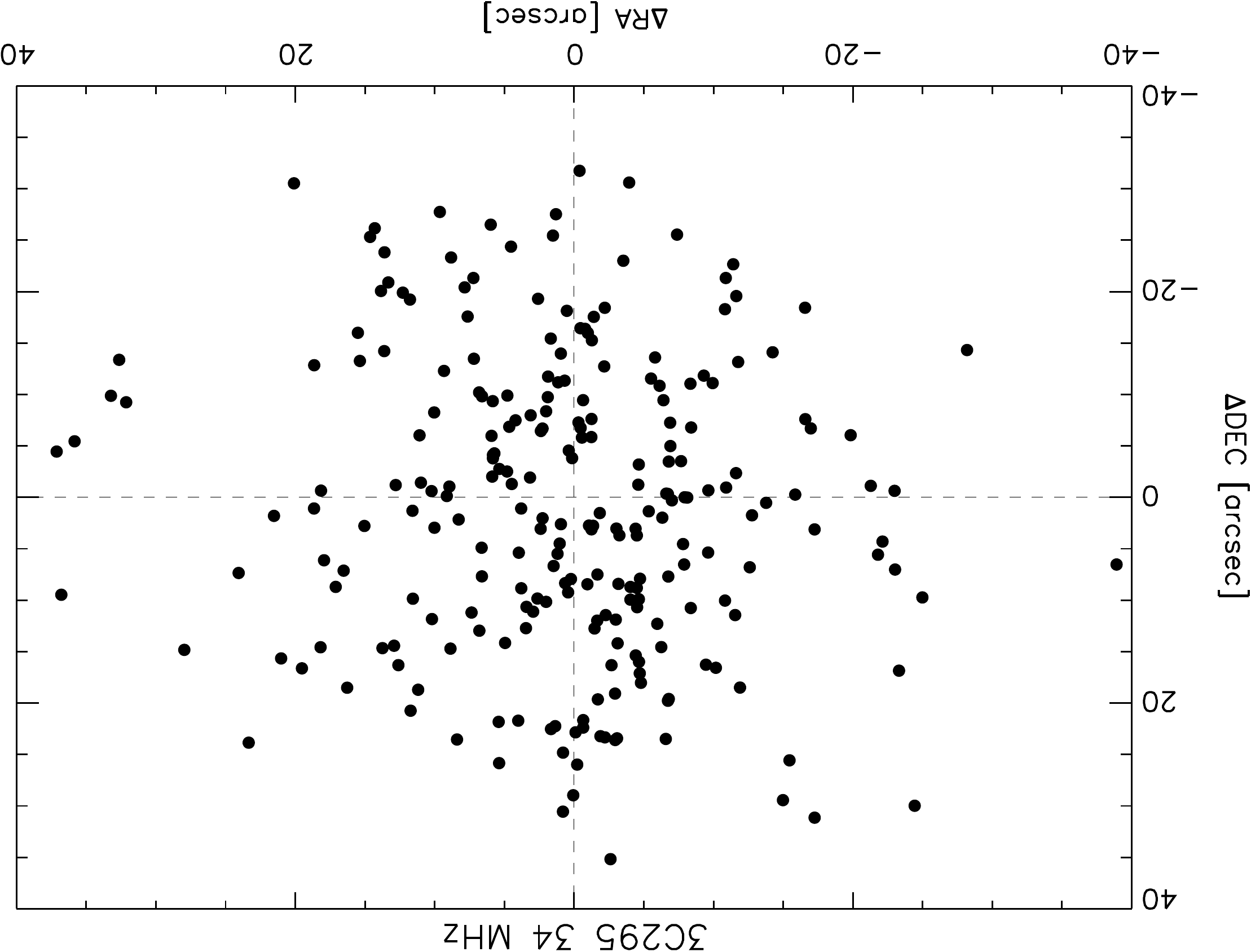}
\end{center}
\caption{The position offsets between the LOFAR LBA and the 325~MHz WENSS sources.}
\label{fig:coord}
\end{figure*}

For all our sources detected in the LOFAR images, we searched for the closest counterpart in the WENSS survey. The difference between the LOFAR positions and WENSS positions are displayed in Fig.~\ref{fig:coord}. The positional offsets ($\Delta\alpha, \Delta\delta$) are a combination of imperfect calibration, noise dependent offsets from position determination using Gaussian fitting by {\tt PyBDSM}, and offsets due to differences in source structure between 325 and 34--62~MHz, related to spectral index variations across the sources and/or differences in resolution between the WENSS and LOFAR images.

The median source position offsets between LOFAR and WENSS are smaller than 1/10th of the beam size for all frequencies, and therefore we do not correct our lists for systematic position offsets. To reduce the effect of the noise dependent term in the position offsets, we re-calculated the offsets taking only sources that are detected with a signal to noise ratio larger than 20 in the LOFAR images. Using only these bright sources, we find a scatter of  
$(\sigma_\alpha, \sigma_\delta=1.8\arcsec, 1.9\arcsec)_{\rm{Bootes, 62 MHz}}$, 
$(\sigma_\alpha, \sigma_\delta=2.1\arcsec, 3.1\arcsec)_{\rm{3C\,295, 62 MHz}}$,  
$(\sigma_\alpha, \sigma_\delta=3.7\arcsec, 5.6\arcsec)_{\rm{3C\,295, 46 MHz}}$, and
$(\sigma_\alpha, \sigma_\delta=6.5\arcsec, 10.2\arcsec)_{\rm{3C\,295, 34 MHz}}$ between LOFAR and WENSS.
We added these values in quadrature to the position uncertainties determined from the Gaussian fitting. The strong increase in the scatter towards the lower frequencies suggests that this is the result of residual ionospheric phase errors. The Bo\"otes field has the smallest spread in position offsets. 

\subsection{Flux density uncertainties}
\label{sec:fluxuncertainties}

For our absolute flux calibration (bootstrapping) we took the scale from \cite{2012MNRAS.423L..30S} for 3C\,295. \citeauthor{2012MNRAS.423L..30S} report an uncertainty in the 3C\,295 flux-scale of about 8\% at 34~MHz, 6\% at 46~MHz, and 4\% at 62~MHz.

{We  performed a check on the accuracy of the beam model and bootstrapping of the flux-scale. We did this by checking for flux density variations within the FoV, and by looking for an overall scaling factor (which applies to all sources within a field). 
For this, we compared the measured LBA flux densities to predicted flux densities from external surveys. These predicted fluxes are based on the NVSS \citep{1998AJ....115.1693C}, WENSS \citep{1997A&AS..124..259R}, GMRT 153 MHz \citep[][in the case of the Bo\"otes field only]{2013A&A...549A..55W}, and VLSS Redux \citep[VLSSr,][]{2012RaSc...47.....L} surveys. We fitted second order polynomials to these flux measurements in $\log{(S)}-\log{(\nu)}$ space. We use these polynomial fits to predict the flux densities at the relevant LBA frequencies.  To obtain reliable predictions, we only included LOFAR sources that were detected in all external surveys.}

\begin{figure}
\begin{center}
\includegraphics[trim =0cm 0cm 0cm 0cm,angle=180, width=0.49\textwidth]{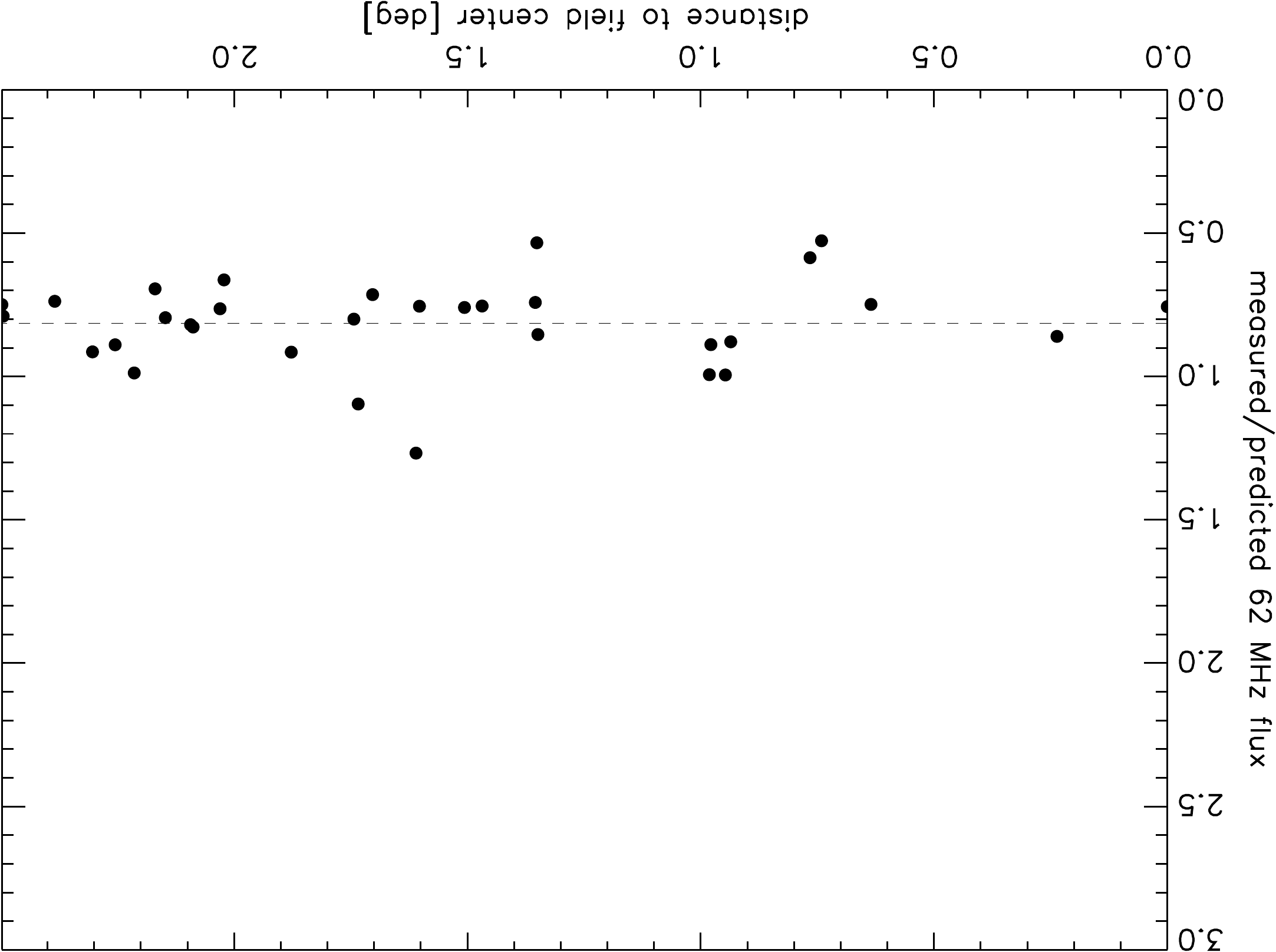}
\end{center}
\caption{Measured flux densities over the predicted flux densities as function of radial distance to the Bo\"otes field center. Dashed line shows the robust mean of the plotted data points.}
\label{fig:radial}
\end{figure}

{
For the Bo\"otes field, we find a scatter of 15\% between the measured and predicted 62~MHz fluxes and a mean flux ratio of 0.8 (measured flux divided by predicted flux), see Fig.~\ref{fig:radial}. If we use the polynomial fits from NVSS, WENSS, and GMRT 153~MHz to predict the VLSSr fluxes we find a scatter of 5\%. Therefore, some of the scatter can be attributed to the intrinsic uncertainties in the predicted LOFAR LBA fluxes due to measurement errors. For the 3C\,295 field, we find a scatter of 18\%,  29\%, 48\%, at 62, 46 and 34~MHz, respectively. No clear trends with radial distance from the field center are found. The increase in the scatter for the 3C\,295 field at 62~MHz, compared to the Bo\"otes field, is not unexpected since we do not have high-quality GMRT 153~MHz flux density measurements available which help to predict the LBA fluxes. In addition, the predicted 46 and 34~MHz flux densities are considerably more uncertain as we extrapolate from higher frequency data. We therefore argue that the  Bo\"otes field 62~MHz fluxes are best suited to determine the flux-scale accuracy across the FoV.}

{
The average measured to predicted flux ratios for the 3C\,295 field are 1.0, 1.0, and 1.05, at 62, 46 and 34~MHz, respectively. 
The mean flux density ratios for the 3C\,295 field are consistent with the uncertainty in the adopted flux-scale for 3C\,295 itself, reported by \cite{2012MNRAS.423L..30S}. The mean flux ratio of 0.8 for the Bo\"otes field likely resulted from the amplitude transfer from 3C\,295 to the Bo\"otes field. This transfer relies on the accuracy of the global beam model. At the time of our observations there were issues with the remote station processing (RSP) boards which could have affected the beam shapes and sensitivities of some stations, resulting in errors when transferring the flux-scale from one pointing to another. The RSP boards were fixed about half a year after our observations.}

{
From the above results, we conclude that the relative uncertainties in the flux-scale within a single FoV due to uncertainties in the beam model, are likely less than 15\%. {We note that this 15\% refers to the averaged beam model of all stations over the entire period of the observations.}  
This result is similar to the $\sim10\%$ we found for LBA observations of Abell 2256 \citep{2012A&A...543A..43V}. The transfer of the flux-scale from one field to the other (i.e., from calibrator to target) seems to be more uncertain, in our case we find a mean ratio of 0.8 (Fig.~\ref{fig:radial}). To bring the Bo\"otes field flux densities to the same scale as the 3C\,295 field, we multiplied them by a factor of 1.25. }

The  integrated flux density errors ($\sigma_{S}$, Eq.~\ref{eq:fluxerror}) are thus a combination of the uncertainties from 3C\,295 flux-scale, the uncertainties from the Gaussian fitting ($\sigma_{{gauss}}$), and a conservative 15\% uncertainty to account for the beam model used during the imaging process:

\begin{equation}
{\sigma^{2}_{S}} = \left\{ 
  \begin{array}{l l}
      \left({0.04S}\right)^2 + \sigma^{2}_{{gauss}}   +  \left({0.15S}\right)^2 & \quad \text{62 MHz}\\
      \left({0.06S}\right)^2 + \sigma^{2}_{{gauss}}   +  \left({0.15S}\right)^2  & \quad \text{46 MHz}\\
     \left({0.08S}\right)^2 + \sigma^{2}_{{gauss}}    +  \left({0.15S}\right)^2 & \quad \text{34 MHz}
  \end{array} \right.
    \label{eq:fluxerror}
\end{equation}

In addition, averaging over a wide frequency range leads to an additional flux density error that depends on spectral index of the source. In this work we neglect this error as it is smaller than 1\% for a source with $\alpha=-1$.

\subsection{Completeness and reliability}
\label{sec:completeness}
To quantify the completeness and reliability of the source lists, we performed a Monte-Carlo (MC) simulation in which we generated $25$ random fields corresponding to each LOFAR image. Each field contains $\sim\!1200$~randomly positioned point sources with peak flux densities between $2.5$\,mJy and $6.3$\,Jy (the catalogue range) for the Bo\"otes $62$\,MHz field, $2.3$\,mJy and $6.6$\,Jy for the 3C\,295 $62$\,MHz field, $3.9$\,mJy and $8.3$\,Jy for the 3C\,295 $46$\,MHz field, and $8.9$\,mJy and $8.0$\,Jy for the 3C\,295 $34$\,MHz field. The source flux densities are drawn randomly from the source count distribution, $dN/dS \propto S^{-1.6}$ \citep{2013A&A...549A..55W}. We neglect the deviation of the true source counts from a power law slope at high flux densities because there are very few sources at these flux densities. {The effect of the beam is naturally taken into account by inserting sources in the noise-inhomogeneous maps. We also deal with non-Gaussian noise (calibration errors) in this way.} Our MC simulation also accounts for the strong ionospheric and bandwidth smearing in the real LOFAR images by scaling the size of the point sources with radial distance from the centre of the field. {The radial scaling factor is determined by the median value within radial distance bins of the ratio of the measured fitted major-axes to the beam major axis in each field.} For comparison we also ran the MC simulation without any smearing. Simulated sources were inserted into the residual images resulting after source detection with {\tt PyBDSM}. Source detection was performed for each randomly simulated field in the same manner as described in Sect.~\ref{sec:pybdsm}. Only $\sim\!300-400$ sources in each field satisfy the detection criterion of peak flux density $>5\sigma$. 

\begin{figure*}[p]
 \centering
\includegraphics[trim =0cm 0cm 0cm 0cm,angle=180, width=0.35\textwidth]{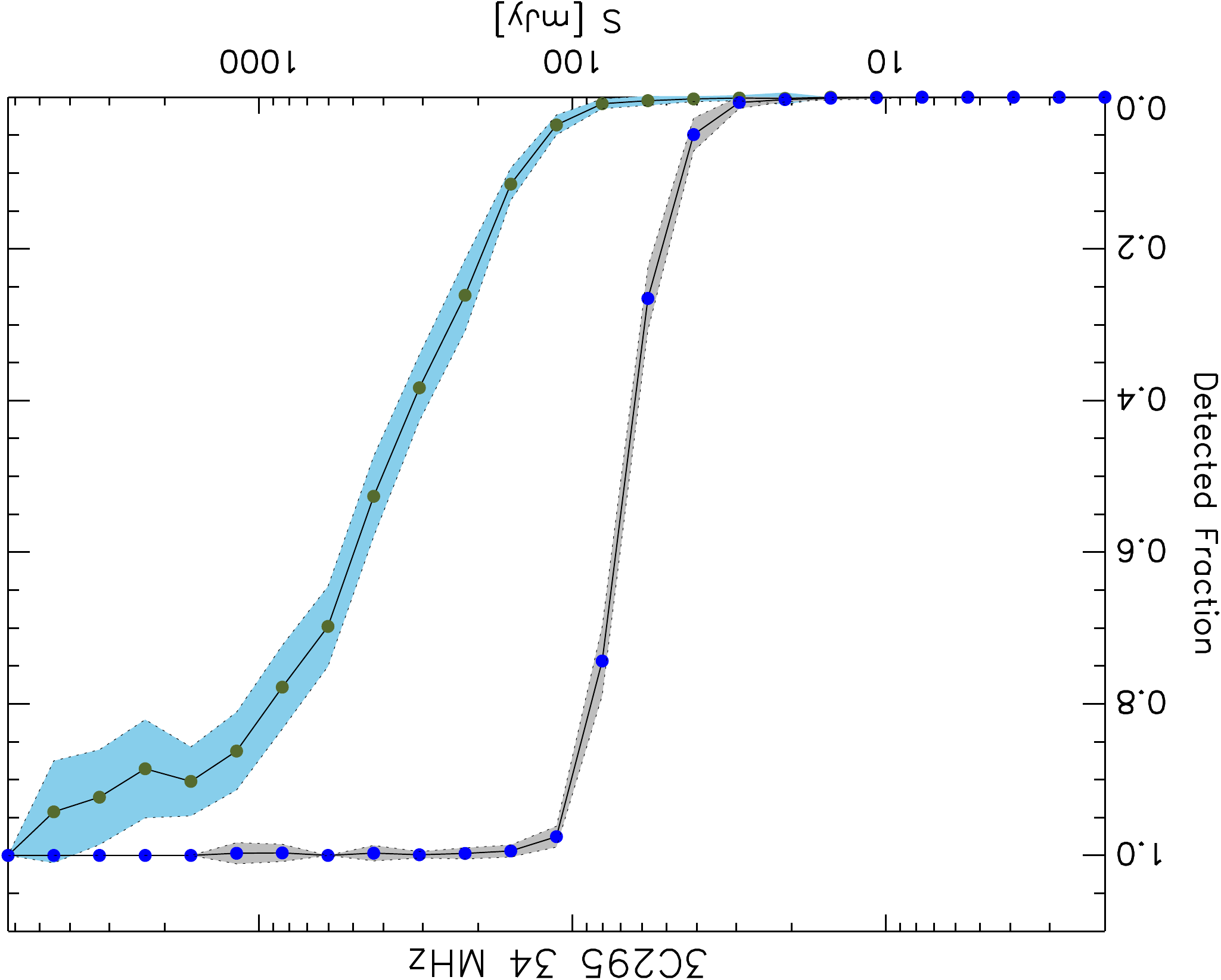}
\includegraphics[trim =0cm 0cm 0cm 0cm,angle=180, width=0.35\textwidth]{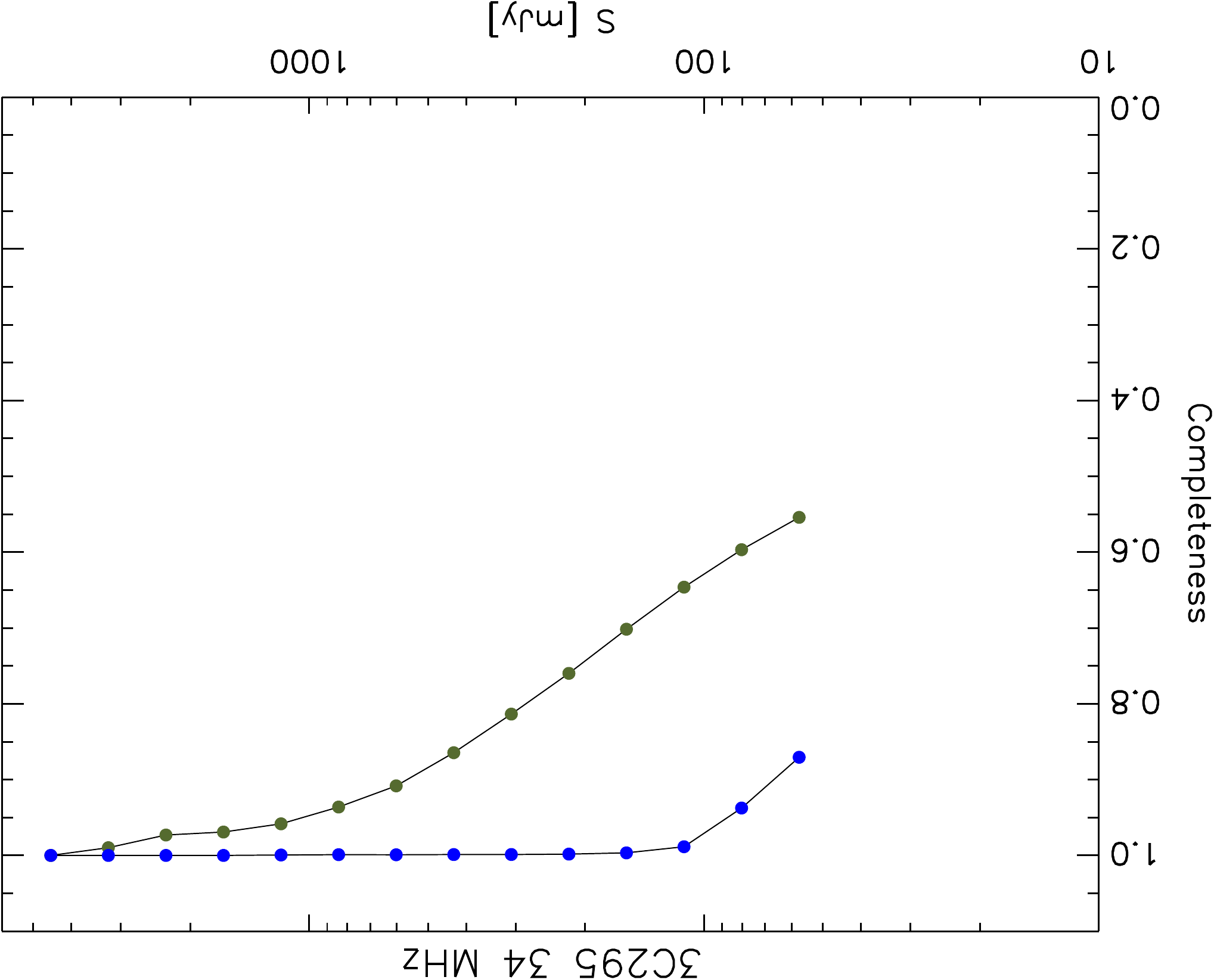}
\includegraphics[trim =0cm 0cm 0cm 0cm,angle=180, width=0.35\textwidth]{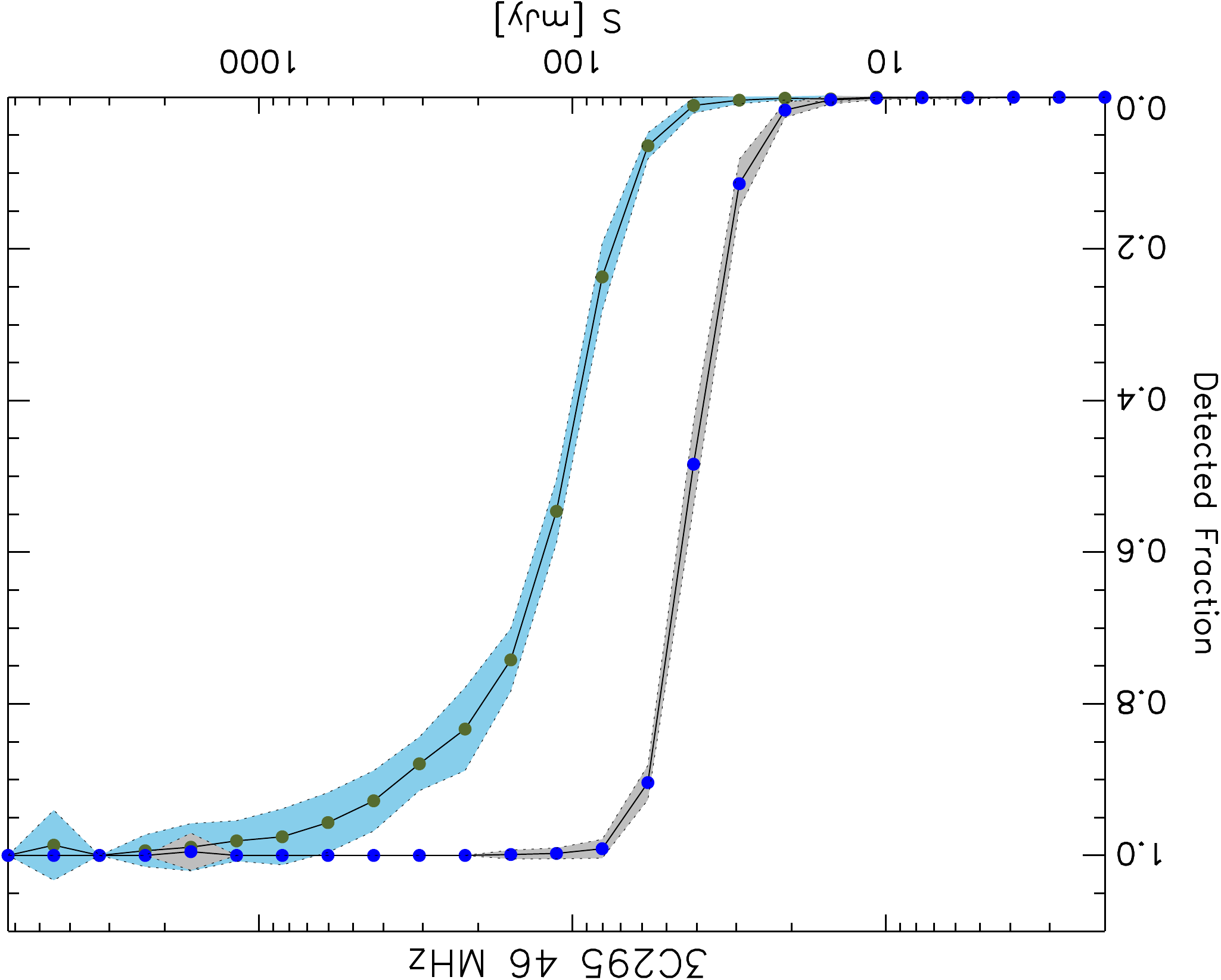}
\includegraphics[trim =0cm 0cm 0cm 0cm,angle=180, width=0.35\textwidth]{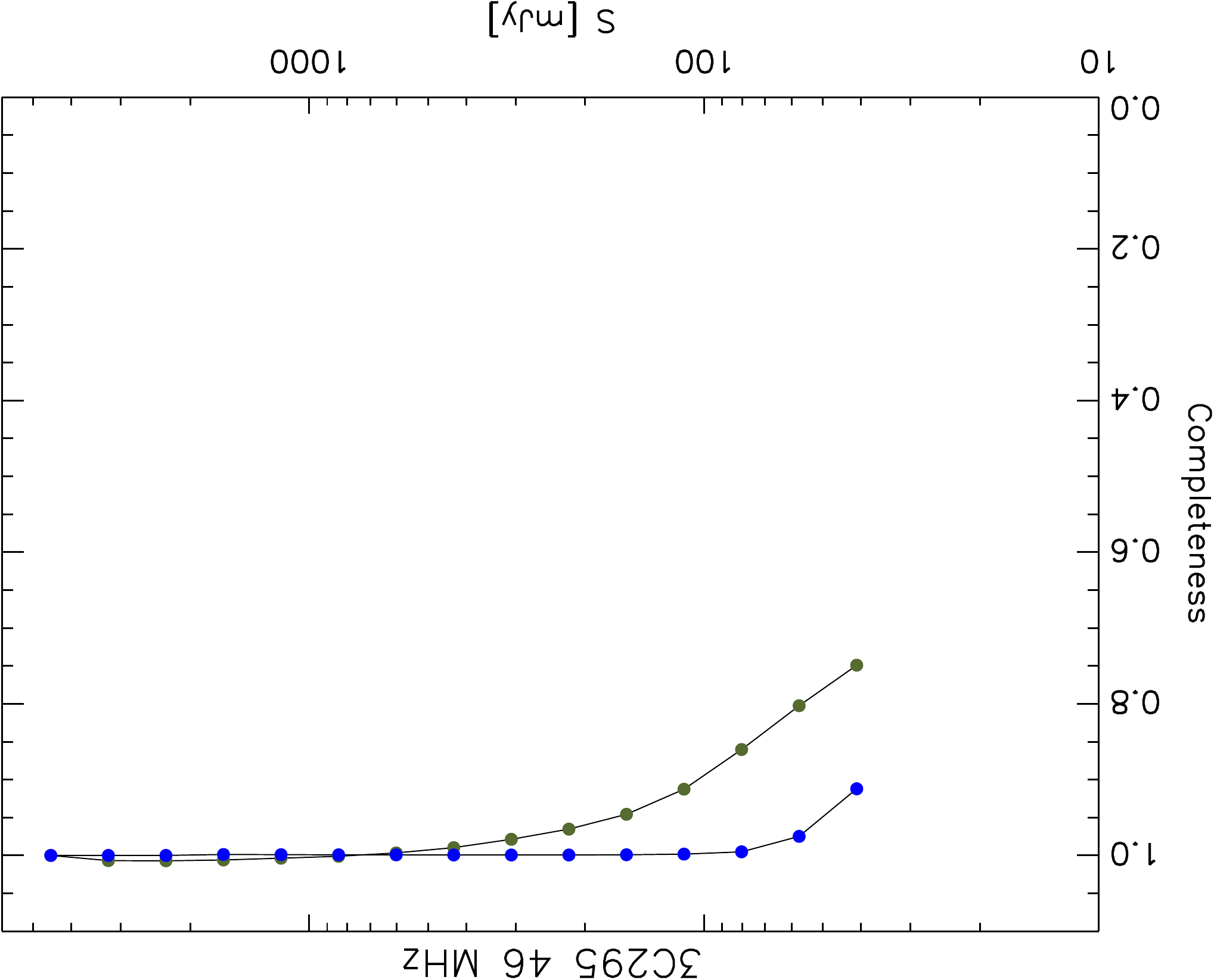}
\includegraphics[trim =0cm 0cm 0cm 0cm,angle=180, width=0.35\textwidth]{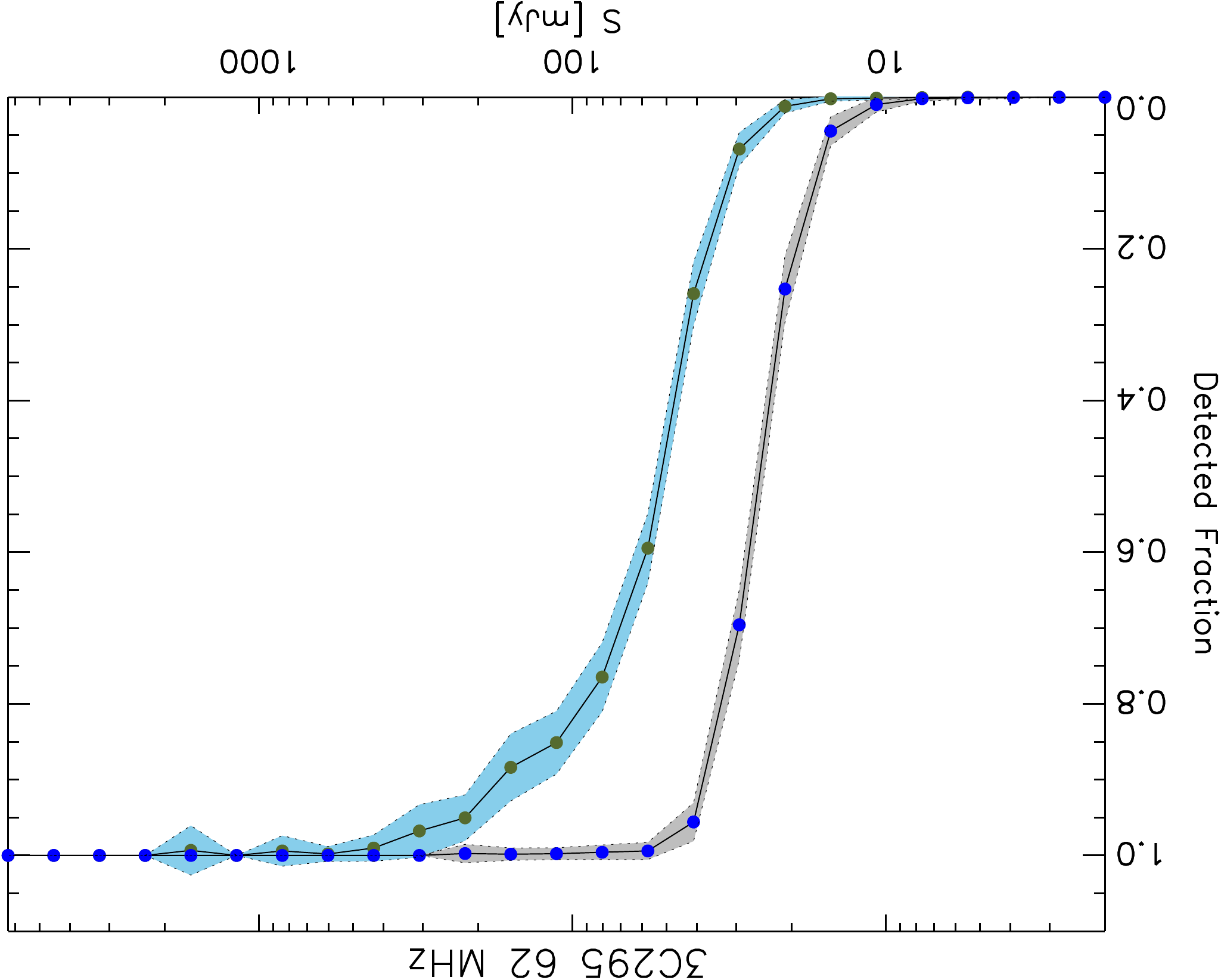}
\includegraphics[trim =0cm 0cm 0cm 0cm,angle=180, width=0.35\textwidth]{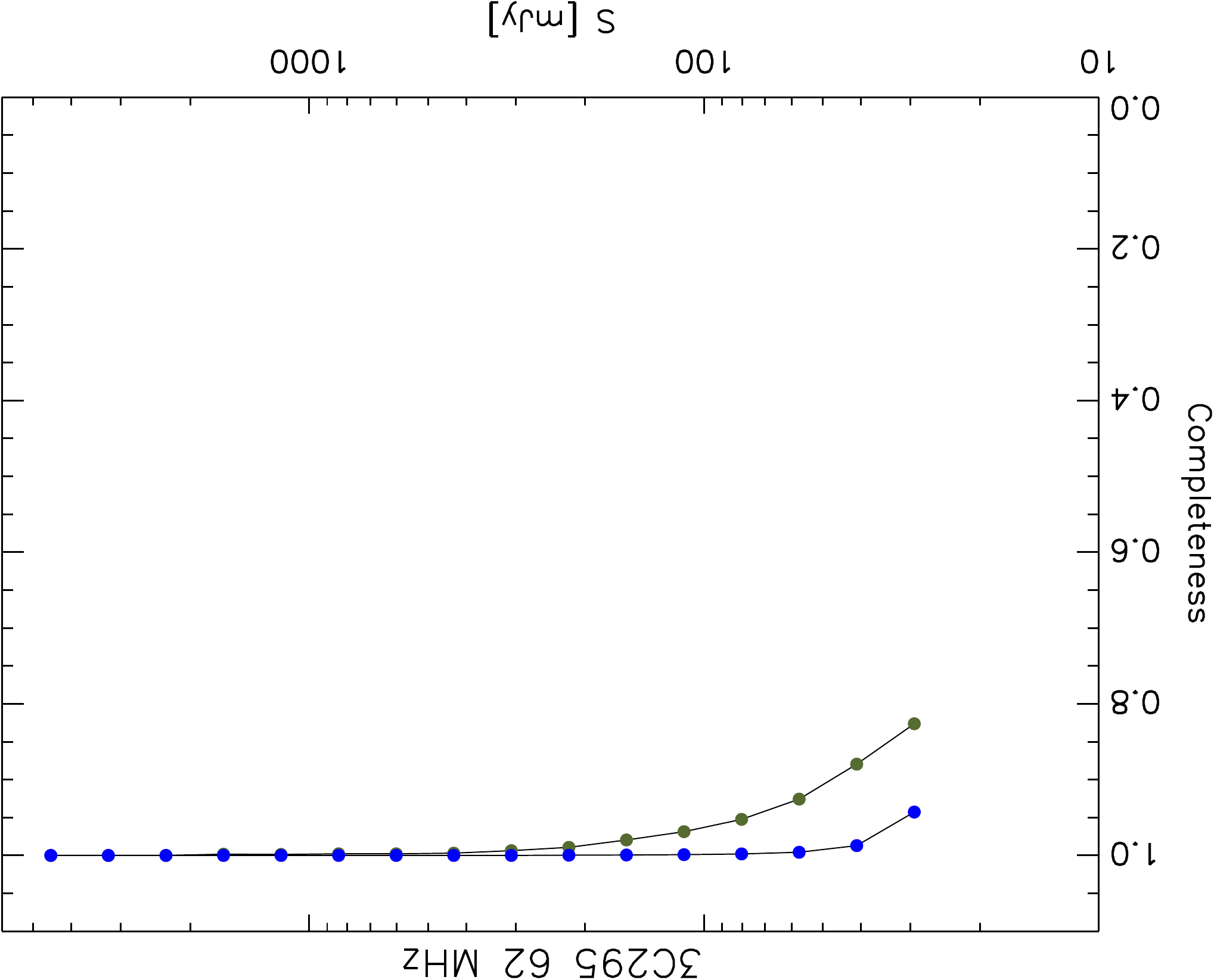}
\includegraphics[trim =0cm 0cm 0cm 0cm,angle=180, width=0.35\textwidth]{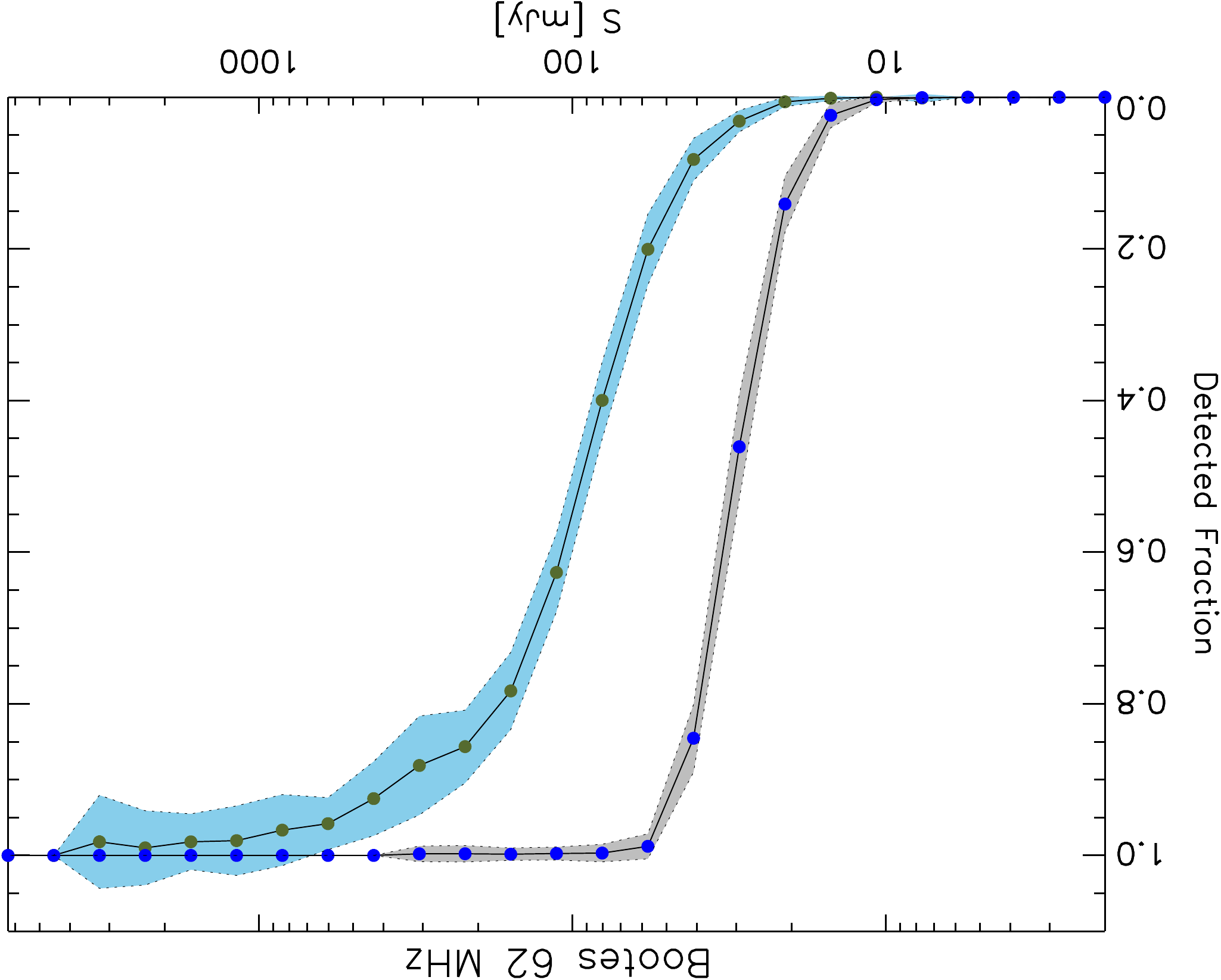}
\includegraphics[trim =0cm 0cm 0cm 0cm,angle=180, width=0.35\textwidth]{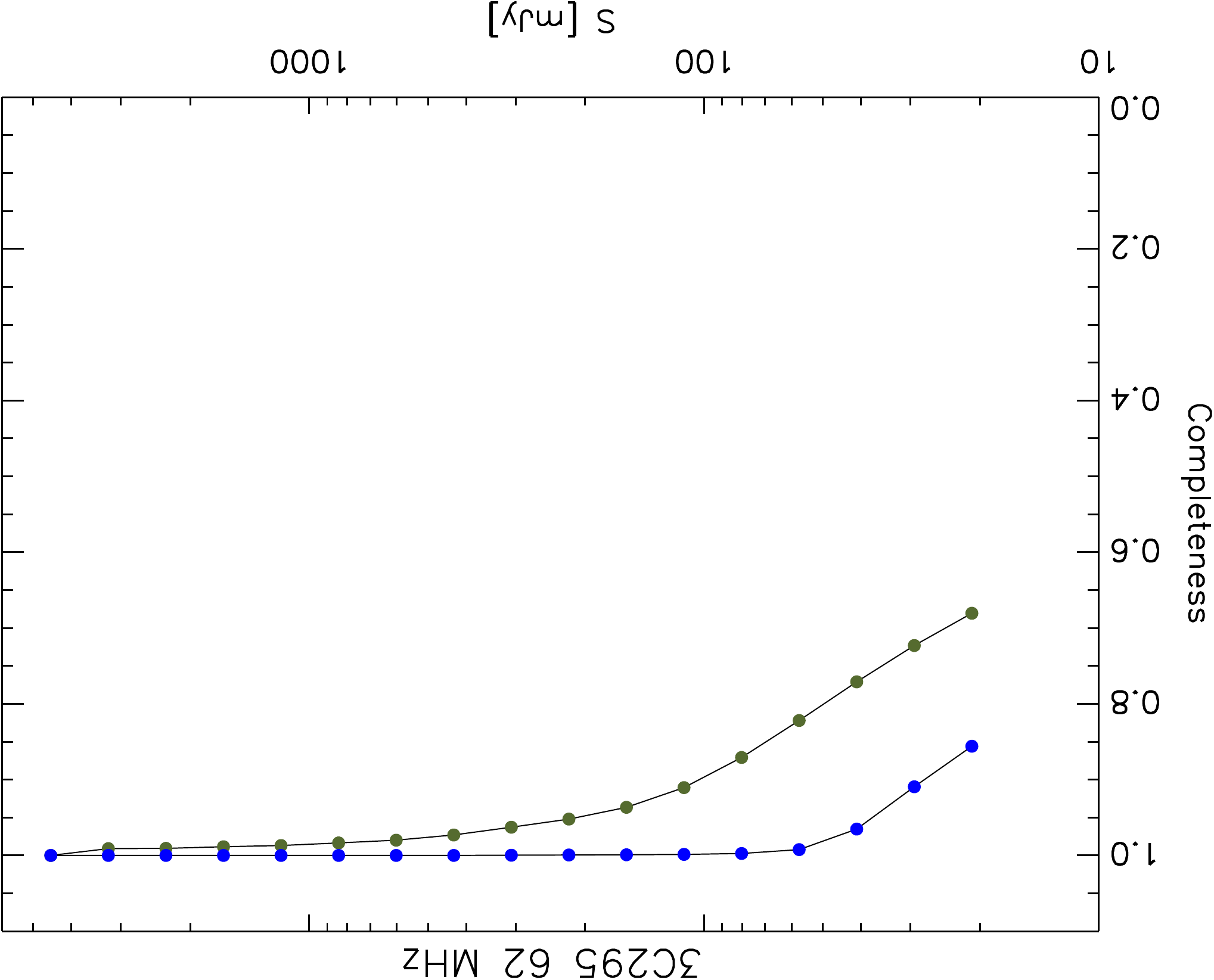}
\caption{Detection fraction and completeness. From \textit{top} to \textit{bottom:} 3C\,295 34, 46 and 62 MHz and Bo\"{o}tes 62 MHz.  \textit{Left:} Fraction of sources detected as a function of integrated flux density calculated from $25$ Monte-Carlo simulations. The solid line shows the mean of all $25$ randomly generated fields and the shaded areas show the $1\sigma$ uncertainty. The blue shaded areas and olive points include source smearing in the MC simulations (see the main text of Sect.~\ref{sec:completeness}). The grey shaded areas and blue points do not include source smearing. \textit{Right:} Estimated completeness of the catalogue as a function of integrated flux density limit accounting for the varying sensitivity across the field of view. The olive points include source smearing, the blue points do not.}
\label{fig:complete}
\end{figure*}

We have estimated the catalogue completeness by plotting the fraction of detected sources {in our MC simulation} as a function of integrated flux density {(\textit{left} panel of Fig.~\ref{fig:complete})}, i.e.,  the fraction of input sources that have a catalogued flux density {using the same detection parameters}. The completeness at a given flux density is determined by integrating the detected fraction upwards from a given flux density limit and is plotted as a function of integrated flux density in {the \textit{right} panel of} Fig.~\ref{fig:complete}. Due to the variation in the rms noise across the image, the detection fraction has first been multiplied by the fraction of the total area in which the source can be detected. We thus estimate that the catalogue is $95$~per~cent complete above a peak flux density of $37$~mJy (Bo\"otes $62$\,MHz) and $88$~mJy, $51$~mJy, and $30$~mJy (3C\,295 $34$, $46$ and $62$~MHz respectively).

\begin{figure*}[p]
 \centering
\includegraphics[trim =0cm 0cm 0cm 0cm,angle=180, width=0.35\textwidth]{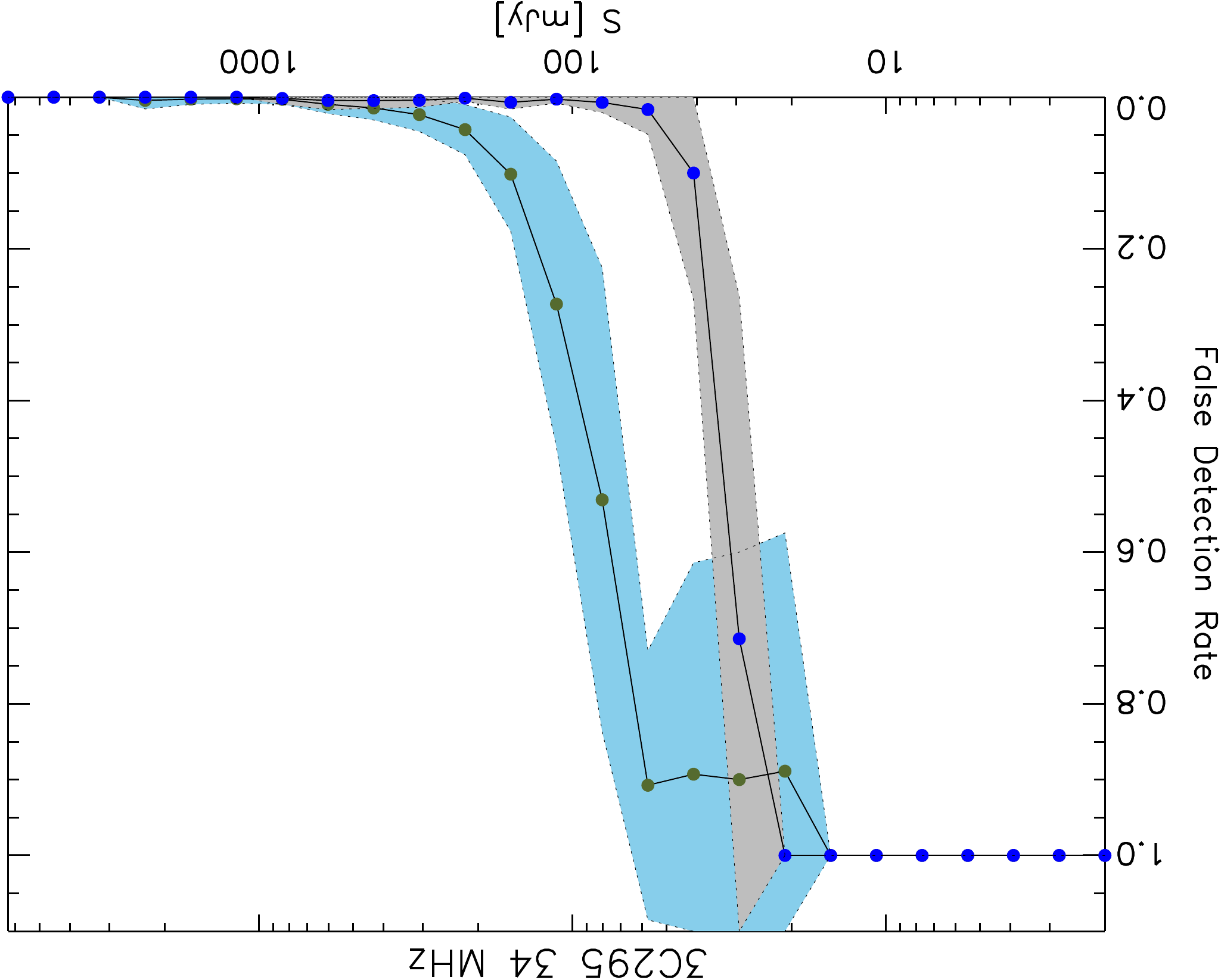}
\includegraphics[trim =0cm 0cm 0cm 0cm,angle=180, width=0.35\textwidth]{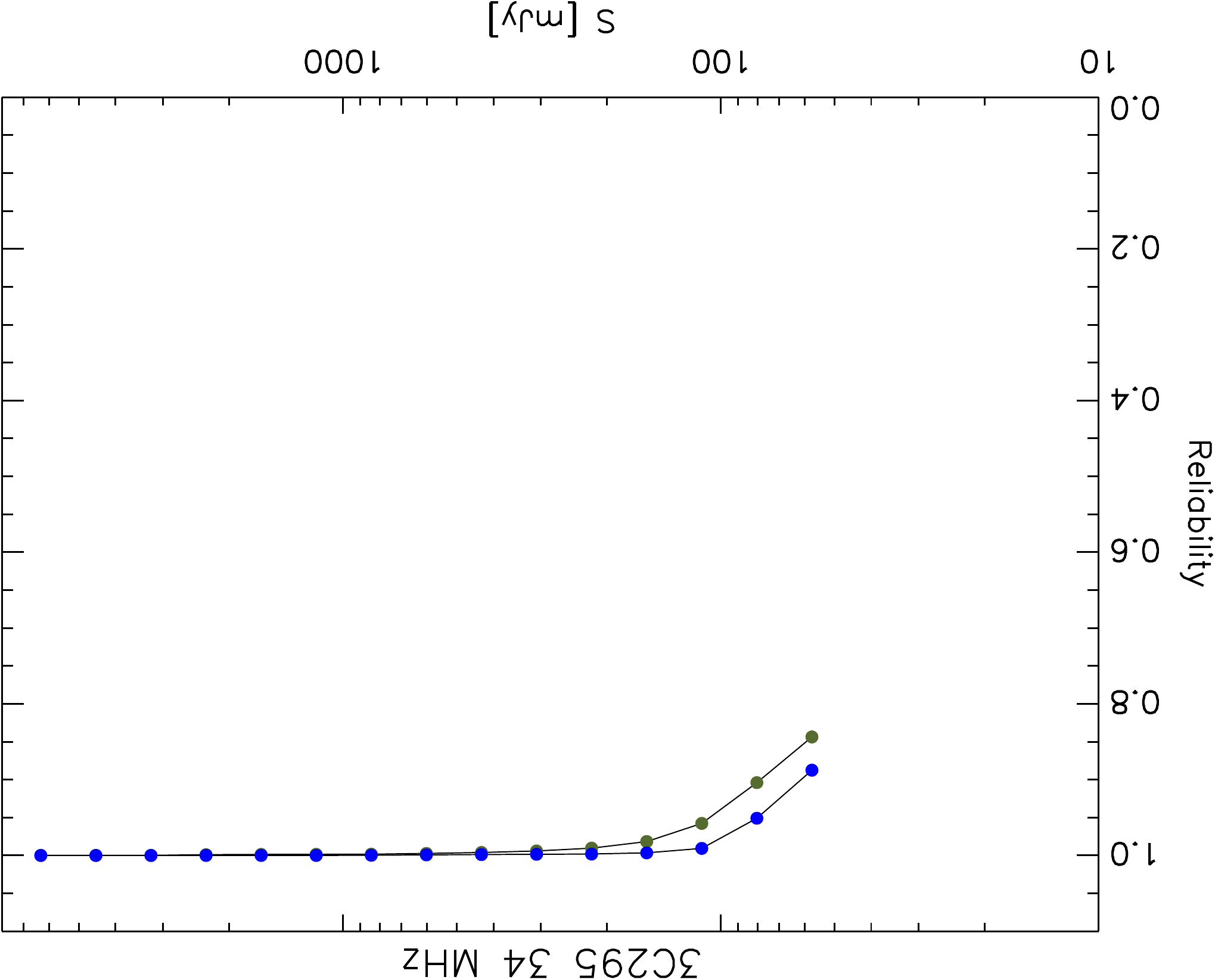}
\includegraphics[trim =0cm 0cm 0cm 0cm,angle=180, width=0.35\textwidth]{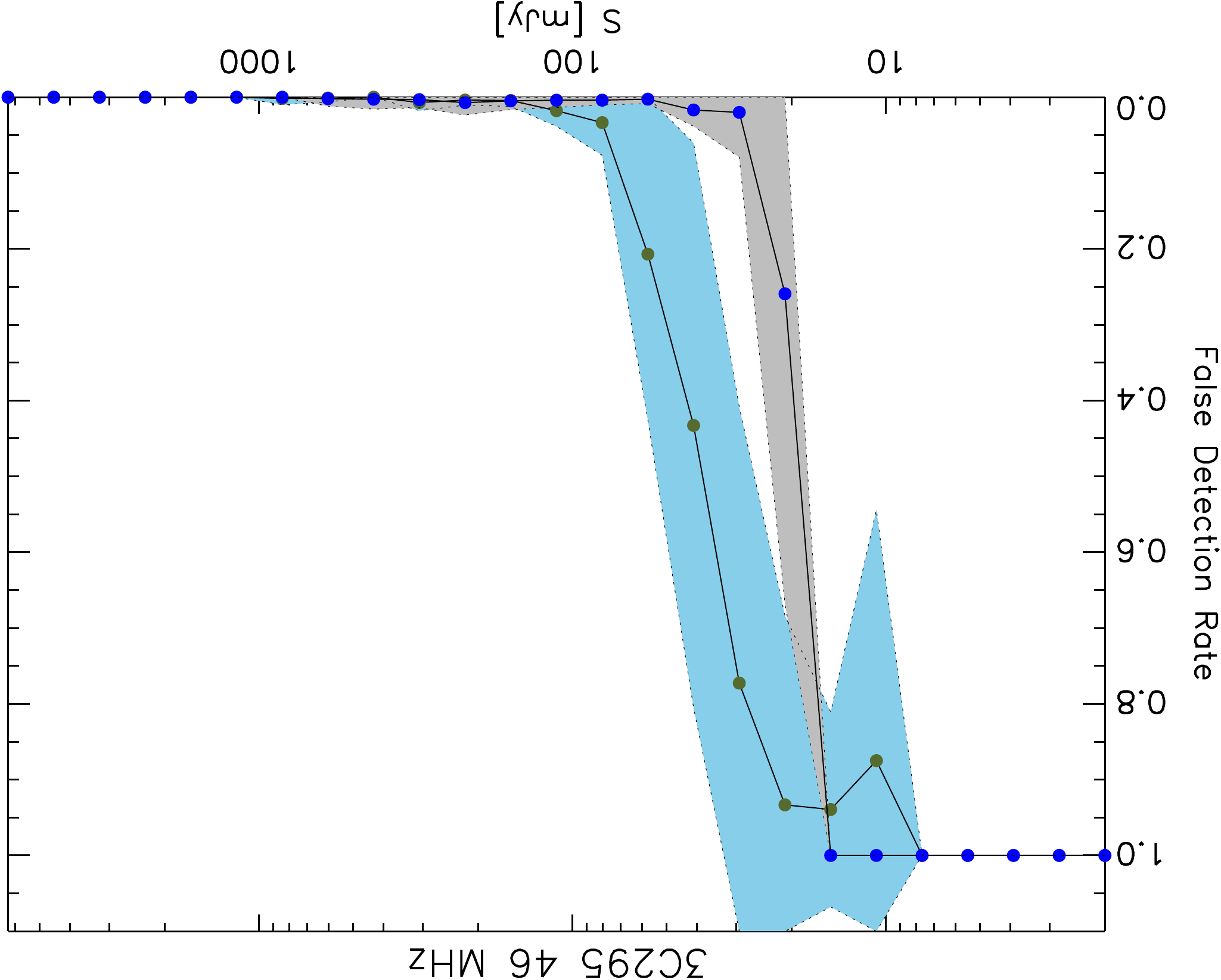}
\includegraphics[trim =0cm 0cm 0cm 0cm,angle=180, width=0.35\textwidth]{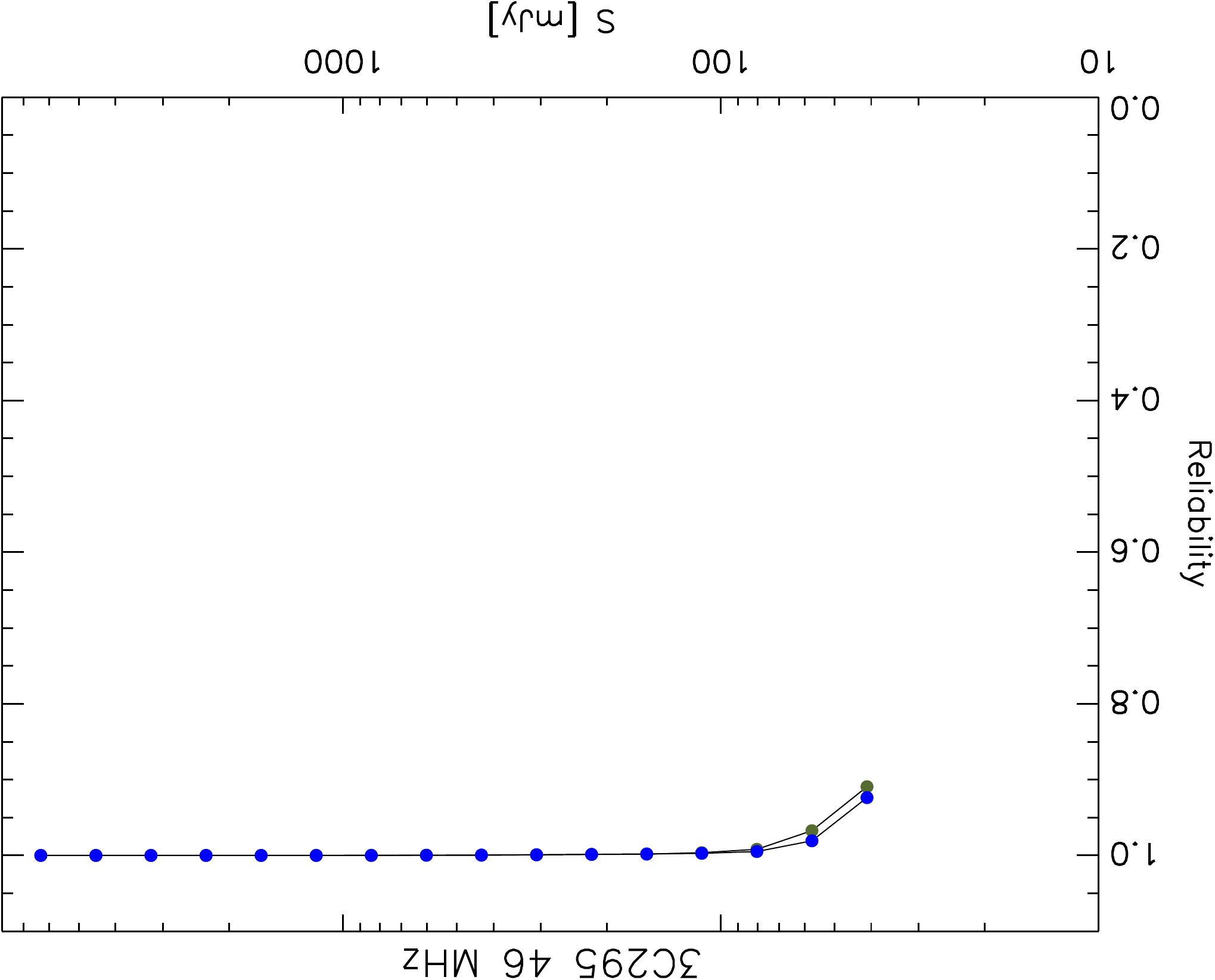}
\includegraphics[trim =0cm 0cm 0cm 0cm,angle=180, width=0.35\textwidth]{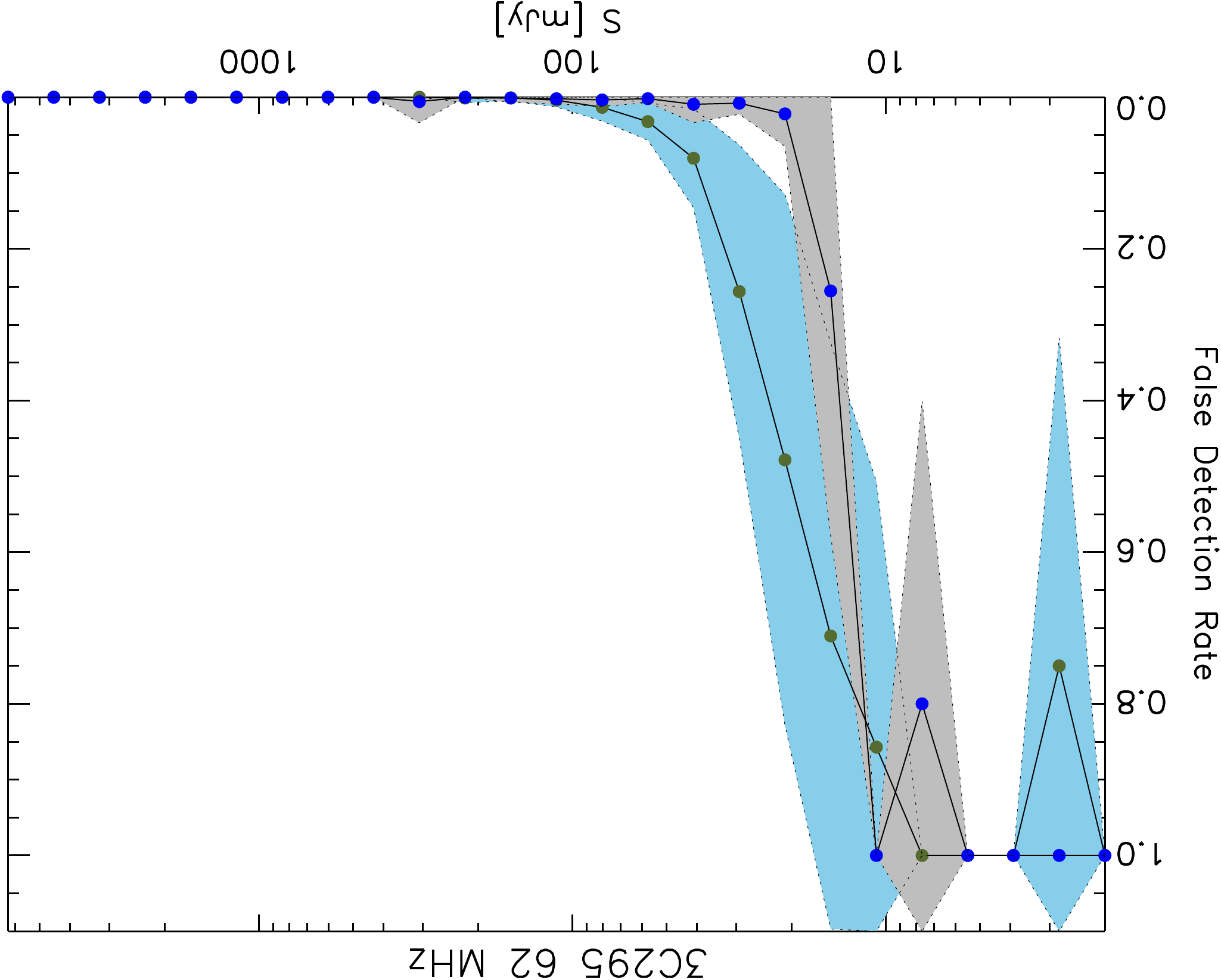}
\includegraphics[trim =0cm 0cm 0cm 0cm,angle=180, width=0.35\textwidth]{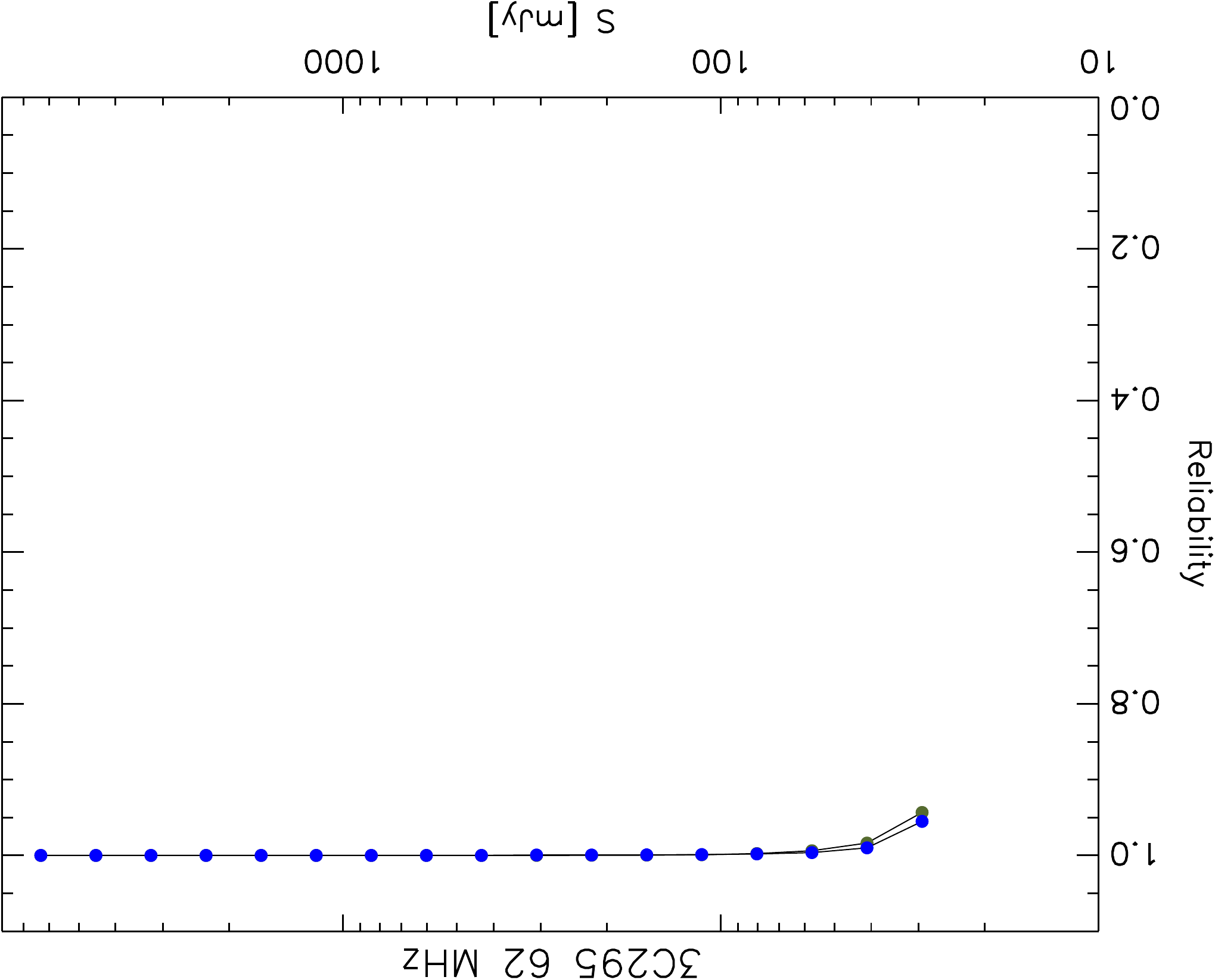}
\includegraphics[trim =0cm 0cm 0cm 0cm,angle=180, width=0.35\textwidth]{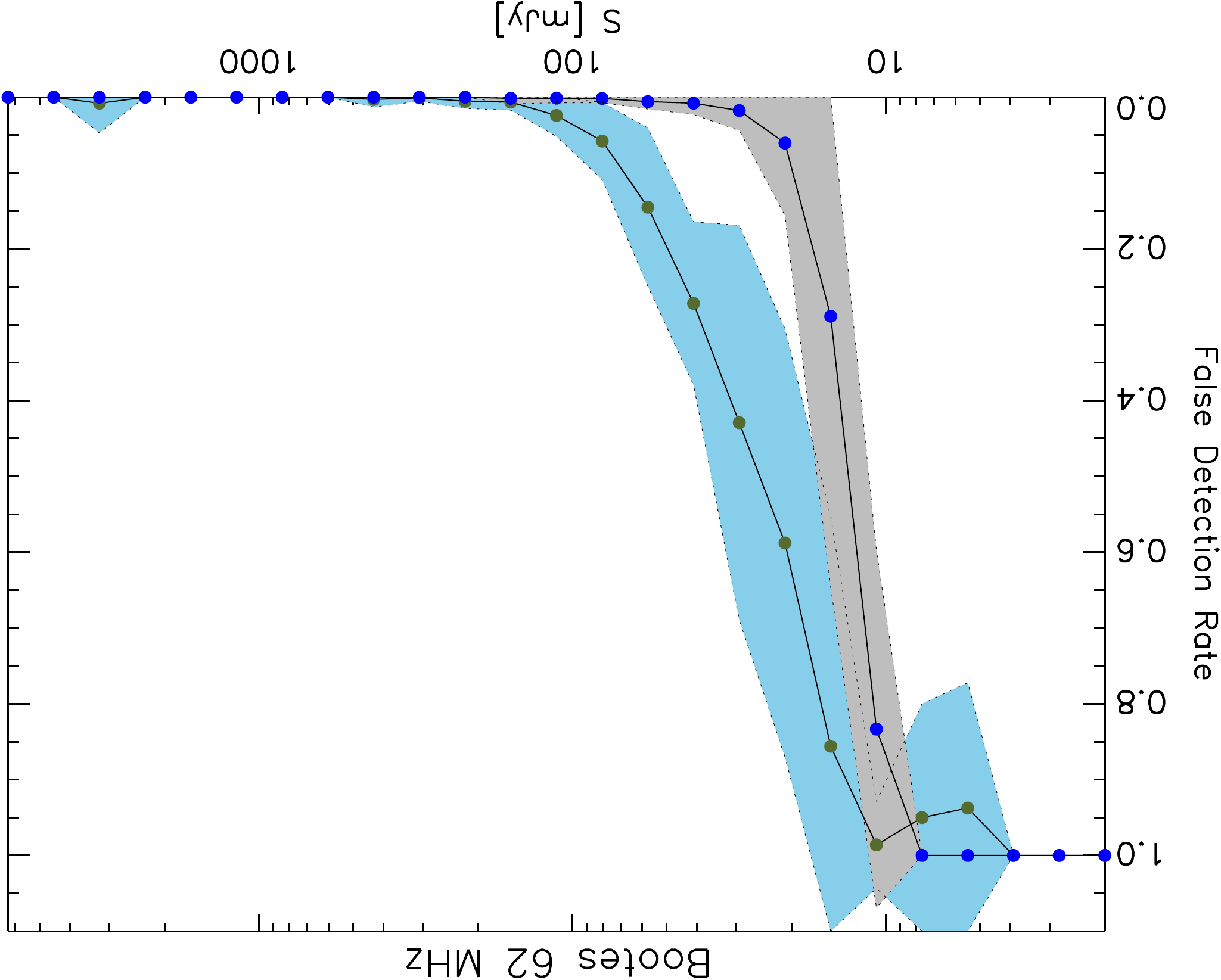}
\includegraphics[trim =0cm 0cm 0cm 0cm,angle=180, width=0.35\textwidth]{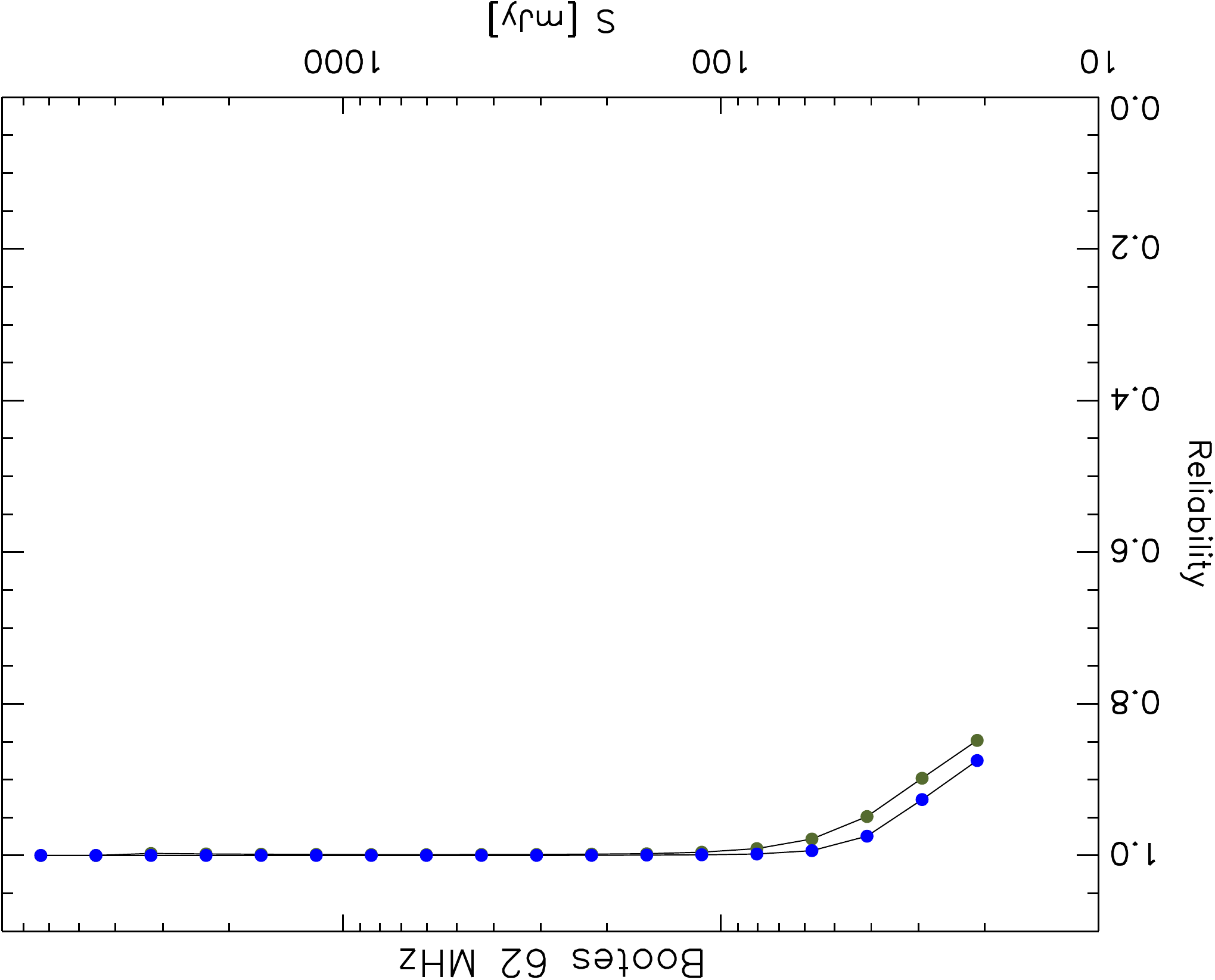}

\caption{False detection rate and reliability.  From \textit{top} to \textit{bottom:} 3C\,295 34, 46 and 62 MHz and Bo\"{o}tes 62 MHz. 
\textit{Left:} False detection rate as a function of peak flux density to local signal-to-noise ratio calculated from $25$ Monte-Carlo simulations. 
The solid line shows the mean of all $25$ randomly generated fields and the shaded areas show the $1\sigma$ uncertainty. The blue shaded areas and olive points include source smearing in the MC simulations (see main the text of Sect.~\ref{sec:completeness}). The grey shaded areas and blue points do not include source smearing. 
\textit{Right:} Estimated reliability of the catalogue as a function of integrated flux density limit accounting for the varying sensitivity across the field of view. The olive points include source smearing, the blue points do not.}
\label{fig:rely}
\end{figure*}

The reliability of the catalogue indicates how many sources above a given flux density are real. In {the \textit{left} panel of} Fig.~\ref{fig:rely},  the false detection rate $FDR$, i.e., the fraction of catalogued sources that do not have an input source, is plotted as a function of the integrated flux density. Integrating up from a  given limit and multiplying by the normalized source flux distribution, we can determine an estimate of the overall $FDR$ or reliability, $R = 1-FDR$, of the catalogue. {The reliability is plotted as a function of integrated flux density limit in the \textit{right} panel of Fig.~\ref{fig:rely}.} We thus estimate that the source list is $95$~per~cent reliable above a peak flux density of $42$~mJy (Bo\"otes $62$\,MHz) and $108$~mJy, $53$~mJy, and $32$~mJy (3C\,295 $34$, $46$ and $62$~MHz respectively). These estimates include  source smearing.

\section{Analysis}
\label{sec:analysis}
\subsection{Source counts}
\begin{figure}
\begin{center}
\includegraphics[ trim =0cm 0cm 0cm 0cm,angle=180, width=0.49\textwidth]{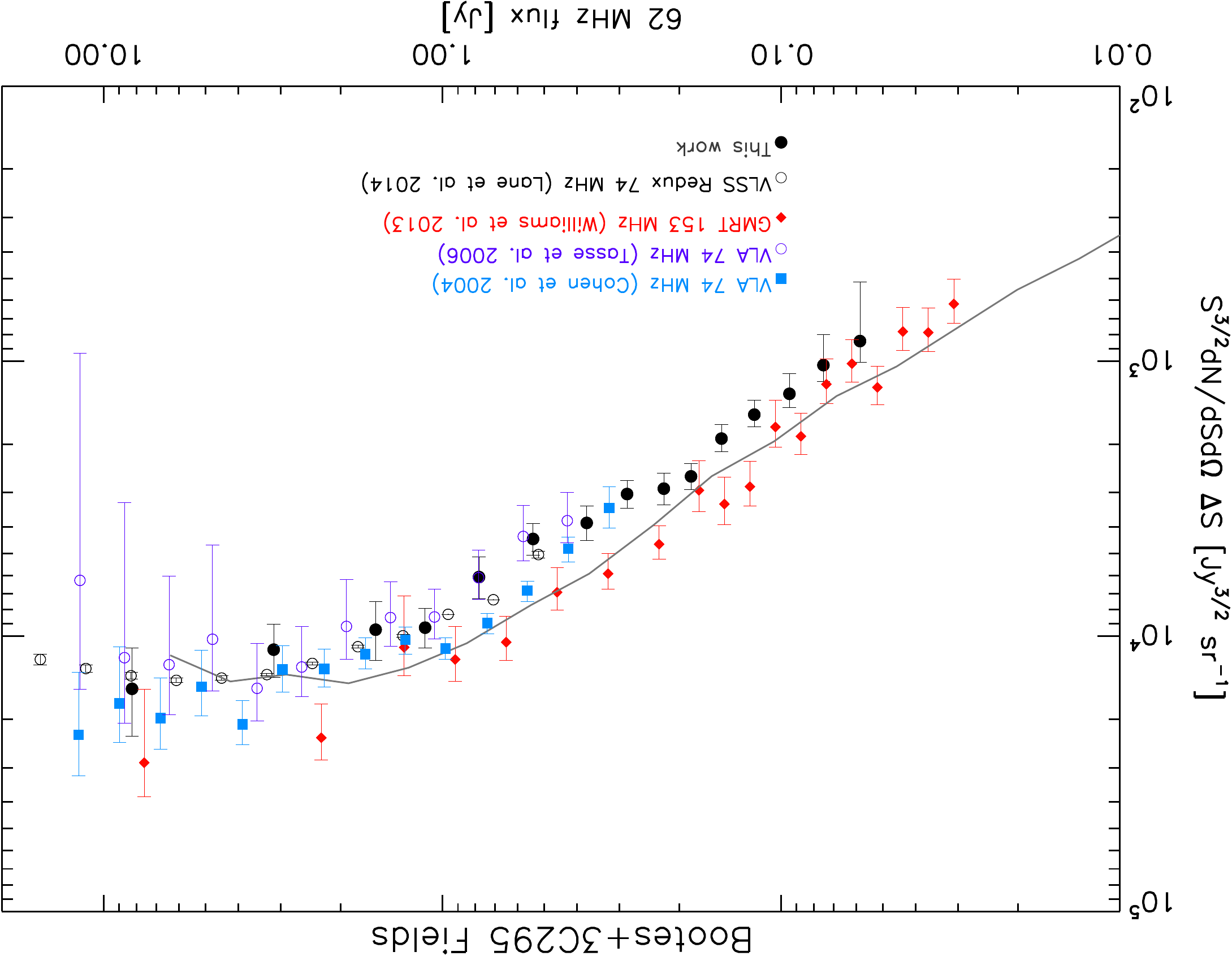}
\end{center}
\caption{Euclidean-normalized differential source counts at 62~MHz combining the Bo\"otes and 3C\,295 fields. The LOFAR points are indicated by the black symbols. Red diamonds are Bo\"otes field source counts at 153~MHz, scaled to 62~MHz using $\alpha=-0.7$. 
Black open circles, blue squares and purple open circles are 74~MHz differential source counts from \cite{2006A&A...456..791T,2004ApJS..150..417C,2014MNRAS.440..327L} and the solid grey line displays the counts from the 151~MHz SKADS S$^{3}$-SEX simulation \citep{2008MNRAS.388.1335W}. These are all scaled to 62~MHz assuming $\alpha=-0.7$.}
\label{fig:ctshigh}
\end{figure}

\begin{figure*}
\begin{center}
\includegraphics[trim =0cm 0cm 0cm 0cm,angle=180, width=0.49\textwidth]{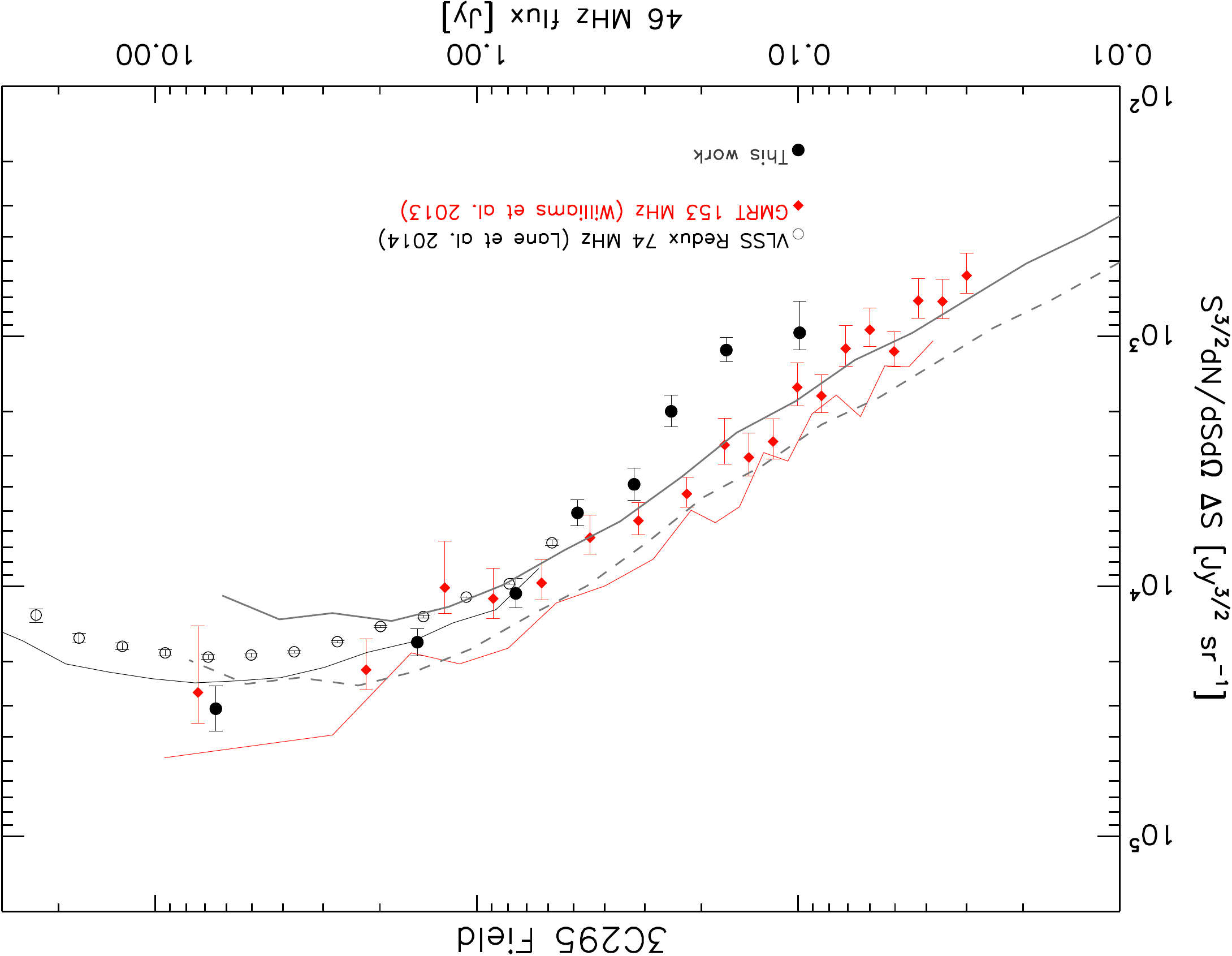}
\includegraphics[trim =0cm 0cm 0cm 0cm,angle=180, width=0.49\textwidth]{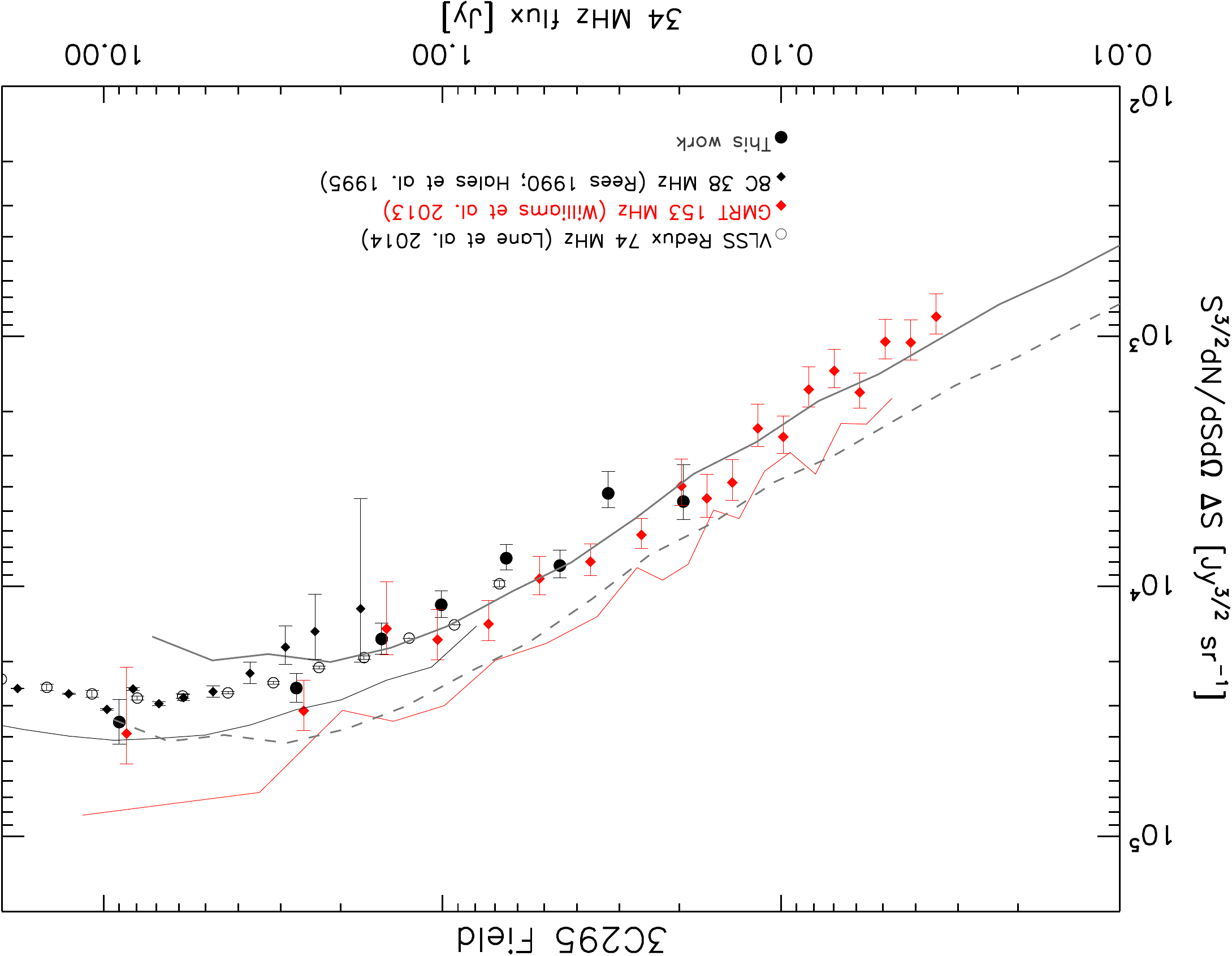}
\end{center}
\caption{Euclidean-normalized differential source counts at 46 (\textit{left}) and 34~MHz (\textit{right}) for the 3C\,295 field. The LOFAR points are indicated by the black circles. The red diamonds are Bo\"otes field source counts at 153~MHz and the black open circles show the VLSSr counts at 74~MHz \citep{2014MNRAS.440..327L}, both scaled with $\alpha=-0.5$. Red and black solid lines show the same source counts but scaled with $\alpha=-0.7$. The solid and dashed grey lines displays the counts from the 151~MHz SKADS S$^{3}$-SEX simulation \citep{2008MNRAS.388.1335W} scaled with $\alpha=-0.5$ and $\alpha=-0.7$, respectively. {For the 34~MHz panel we have also plotted the 8C source counts at 38~MHz with black diamonds. The 8C source counts are complete to a flux level of about 5~Jy. Below a flux density of 5~Jy we have corrected the source counts using the information provided in fig.~7 from \cite{1990MNRAS.244..233R}.}}
\label{fig:ctslow}
\end{figure*}

We use the source lists to compute the Euclidean-normalized source counts at 62, 46 and 34~MHz. For this, we have to take into account the large variations of the rms noise across the images \citep[e.g.,][]{1985ApJ...289..494W}. We choose the flux density bin sizes such that we have approximately 30--60 sources per bin, except for the first and last bins. We corrected these source counts using the Monte-Carlo simulations described in Sect.~\ref{sec:completeness} with the detected fraction of sources as a function of flux density. The propagated errors in the source counts are based on the Poissonian uncertainties and the uncertainties in the derived detection fraction, see Table~\ref{tab:counts}. The resulting Euclidean-normalized source counts are shown in Figs.~\ref{fig:ctshigh} and \ref{fig:ctslow}.

{Only a few source count studies exist below 100~MHz. \cite{2003ApJ...591..640C, 2004ApJS..150..417C} and \cite{2006A&A...456..791T} published source counts at 74~MHz with the VLA. We compare our source counts with those from \cite{2004ApJS..150..417C} and \cite{2006A&A...456..791T} because they go to fainter flux densities than those from \cite{2003ApJ...591..640C}. For that comparison, we rescale the \cite{2004ApJS..150..417C} and \cite{2006A&A...456..791T} flux densities to the newly adopted VLSSr flux-scale \citep{2012RaSc...47.....L}. The VLSSr counts are included as well \citep{2014MNRAS.440..327L}. We also compare with 153~MHz source counts from the GMRT for the Bo\"otes field \citep{2013A&A...549A..55W} because it overlaps to a large extent with our Bo\"otes field data.}

The combined source counts at 62~MHz from the Bo\"otes and 3C\,295  fields show very good agreement with the results at 74~MHz, which are scaled using $\alpha=-0.7$.  The 62~MHz counts reach significantly fainter (about a factor of 6), flux density levels than the 74~MHz counts. The 62~MHz counts fall slightly below the GMRT 153~MHz counts, if we scale these with a spectral index of $-0.7$. The simulated 151~MHz SKAD S$^{3}$-SEX counts \citep{2008MNRAS.388.1335W} closely follow the GMRT 153~MHz counts.

The 34~MHz source counts fall significantly below the extrapolated source counts from 153 and 74~MHz if we scale with $\alpha=-0.7$. This is also the case for the simulated 151~MHz SKAD S$^{3}$-SEX counts. The 46~MHz differential source counts show a similar situation, {although the difference is most pronounced below 0.5 Jy}. Scaling with $\alpha=-0.5$ gives a better agreement with the 34 and 46~MHz source counts, an indication that the average spectral index of the sources flattens towards lower frequencies, a result that has been reported before \citep[e.g.,][]{1992MNRAS.256..404L}. {However, part of the difference could also be caused by  field to field variations \citep{2013MNRAS.432.2625H}. To check this, we compared the separate source counts for the two fields at 62~MHz, instead of the combined counts that are shown in Fig.~\ref{fig:ctshigh}. We find that the 3C\,295 field source counts are generally about 20-30\% lower than for the Bo\"otes field (see Fig.~\ref{fig:ctshighsep}) so this could explain some of the difference. } 

Spectral flattening is expected {for some sources} because of absorption effects and low-frequency spectral indices are  flatter than high-frequency ones due to spectral ageing operating at higher frequencies. We note though that our flux reference 3C\,295 also incorporates a strong spectral turnover below $\sim60$~MHz and hence we have to be careful to conclude whether the flattening is intrinsic, or is caused by our uncertain calibrator flux-scale. Fortunately, the 8C 38~MHz counts allow for a more direct comparison at flux densities above $\sim 1$~Jy \citep{1990MNRAS.244..233R,1995MNRAS.274..447H}. We find good agreement between our 34~MHz sources counts and those from the 8C~survey. In addition, the 8C counts match up with the extrapolated counts from the VLSSr (at 74~MHz) and the GMRT  (at 153~MHz) using a spectral index scaling of $\alpha=-0.5$. {The 8C source counts at 38~MHz are not consistent with the VLSSr and GMRT counts if we scale with a spectral index of  $\alpha=-0.7$. This indeed shows that on average the spectral indices of sources flatten. It is important to note that the VLSSr and 8C counts are not affected by field to field variations given the large sky area they cover.}

\begin{table}
\begin{center}
\caption{Source Counts}
\tabcolsep=0.11cm
\begin{tabular}{lllll}
\hline
\hline
flux bin & raw counts & corrected counts &  normalized counts \\
Jy&  & &Jy$^{3/2}$~sr$^{-1}$ \\
\hline
\hline
62 MHz  \\
Bo\"otes+3C295   \\
\hline
0.051--0.066 & 32& 181& 844$^{+164}_{-329}$\\
0.066--0.084 & 47& 143&1,032$^{+150}_{-234}$\\
0.084--0.105 & 51& 119&1,312$^{+162}_{-204}$\\
0.105--0.135 & 62& 111&1,562$^{+167}_{-175}$\\
0.135--0.165 & 52& 78&1,909$^{+226}_{-210}$\\
0.165--0.204 & 65& 83&2,621$^{+300}_{-267}$\\
0.204--0.240 & 45& 53&2,904$^{+422}_{-354}$\\
0.240--0.330 & 65& 75&3,037$^{+376}_{-327}$\\
0.330--0.420 & 44& 49&3,869$^{+612}_{-505}$\\
0.420--0.660 & 55& 59&4,424$^{+634}_{-538}$\\
0.660--0.900& 31& 32 &6,087$^{+1,206}_{-945}$\\
0.900--1.35  & 36& 37 &9,295$^{+1,716}_{-1,367}$\\
1.35--1.80 & 16 & 16   &9,456$^{+2,792}_{-1,975}$\\
1.80--4.50 & 20 &  20 &11,175$^{+2,899}_{-2,132}$\\
4.5--12.0 & 7 &     7  &15,509$^{+7,595}_{-4,491}$\\
\hline
\hline
46 MHz  3C\,295  \\
\hline
0.072--0.126  &  27 & 155  &969$^{+165}_{-242}$ \\  
0.126--0.208  & 46  & 76 &1,136$^{+132}_{-127}$ \\ 
0.208--0.288  &  39 &  48 &1,998$^{+308}_{-278}$ \\ 
0.288--0.360  &  38 & 44 &3,913$^{+622}_{-542}$\\ 
0.360--0.612  &  65 & 72&5,087$^{+637}_{-577}$\\ 
0.612--0.900  &  55 &  54 &10,686$^{+1,535}_{-1,367}$\\ 
0.900--2.16  &  66 & 68 &16,747$^{+2,218}_{-1,994}$  \\ 
2.16--10.8 &  23 & 23  &30,898$^{+7,133}_{-5,812}$\\ 
\hline
\hline
34 MHz 3C\,295  \\
\hline
0.136--0.252 &  40 &505  &4,615$^{+843}_{-1323}$ \\  
0.252--0.396&  50 & 163  &4,257$^{+591}_{-773}$ \\ 
0.396--0.504 &  51 & 105 & 8,270$^{+984}_{-1090}$ \\ 
0.504--0.792 &  63 & 105&7,726$^{+883}_{-927}$\\ 
0.792--1.22&  61&  80 & 11,864$^{+1,467}_{-1,422}$\\ 
1.22--1.80&  46 &  53& 16,252$^{+2,409}_{-2,178}$\\ 
 1.80--3.60&  55 & 61 & 25,548$^{+3,574}_{-3,262}$  \\ 
3.60--14.4 &  23 & 25& 34,940$^{+7,840}_{-6,624}$\\ 
\hline
\hline
\end{tabular}
\label{tab:counts}
\end{center}
\end{table}

\subsection{Spectral indices}
\label{sec:spix}
{In the above section we found evidence for spectral flattening of sources towards lower frequencies. In this subsection we investigate the spectral properties of the detected LBA sources.}

\begin{figure*}
\begin{center}
\includegraphics[trim =0cm 0cm 0cm 0cm,angle=180, width=0.49\textwidth]{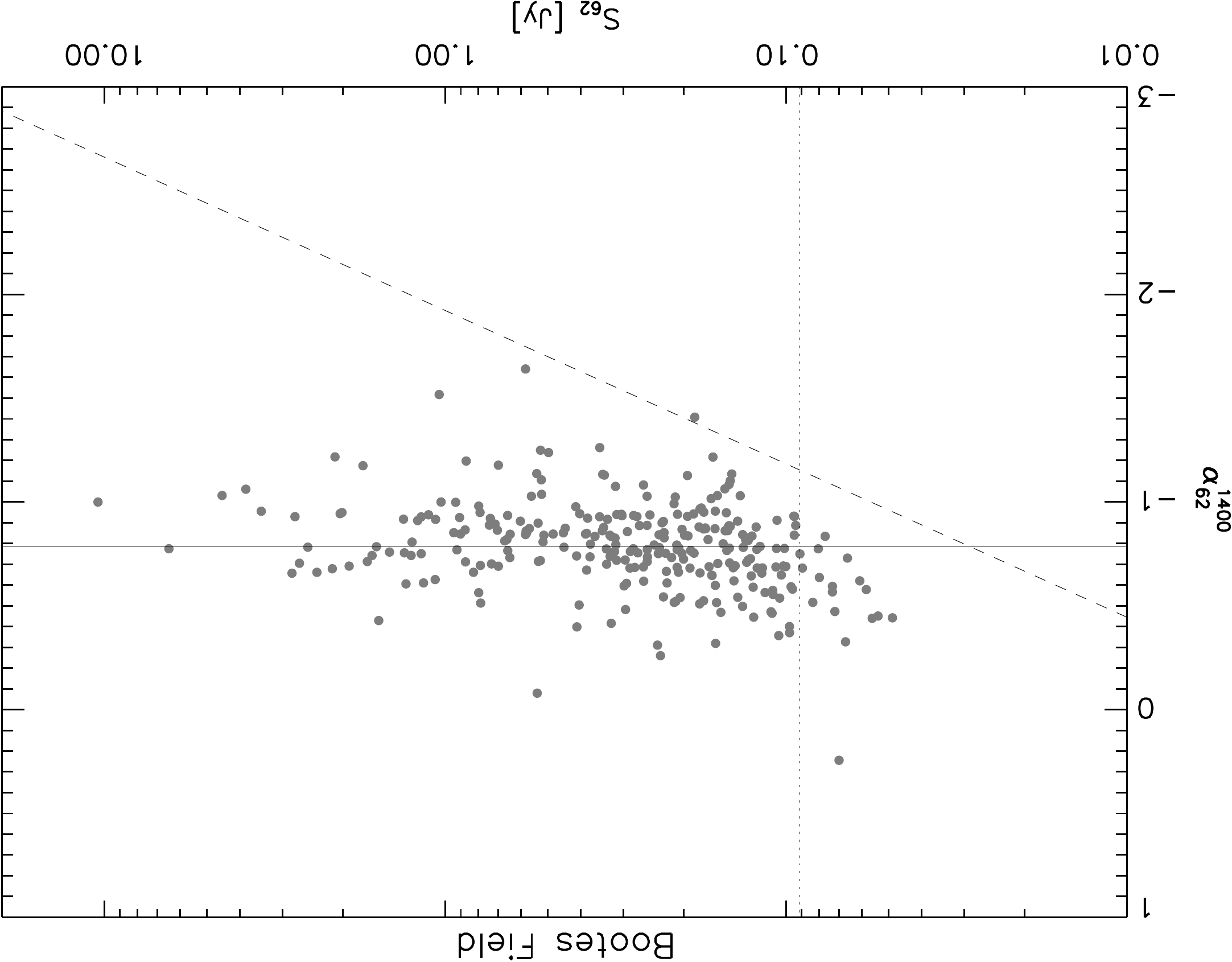}
\includegraphics[trim =0cm -0.7cm 0cm 0cm,angle=180, width=0.49\textwidth]{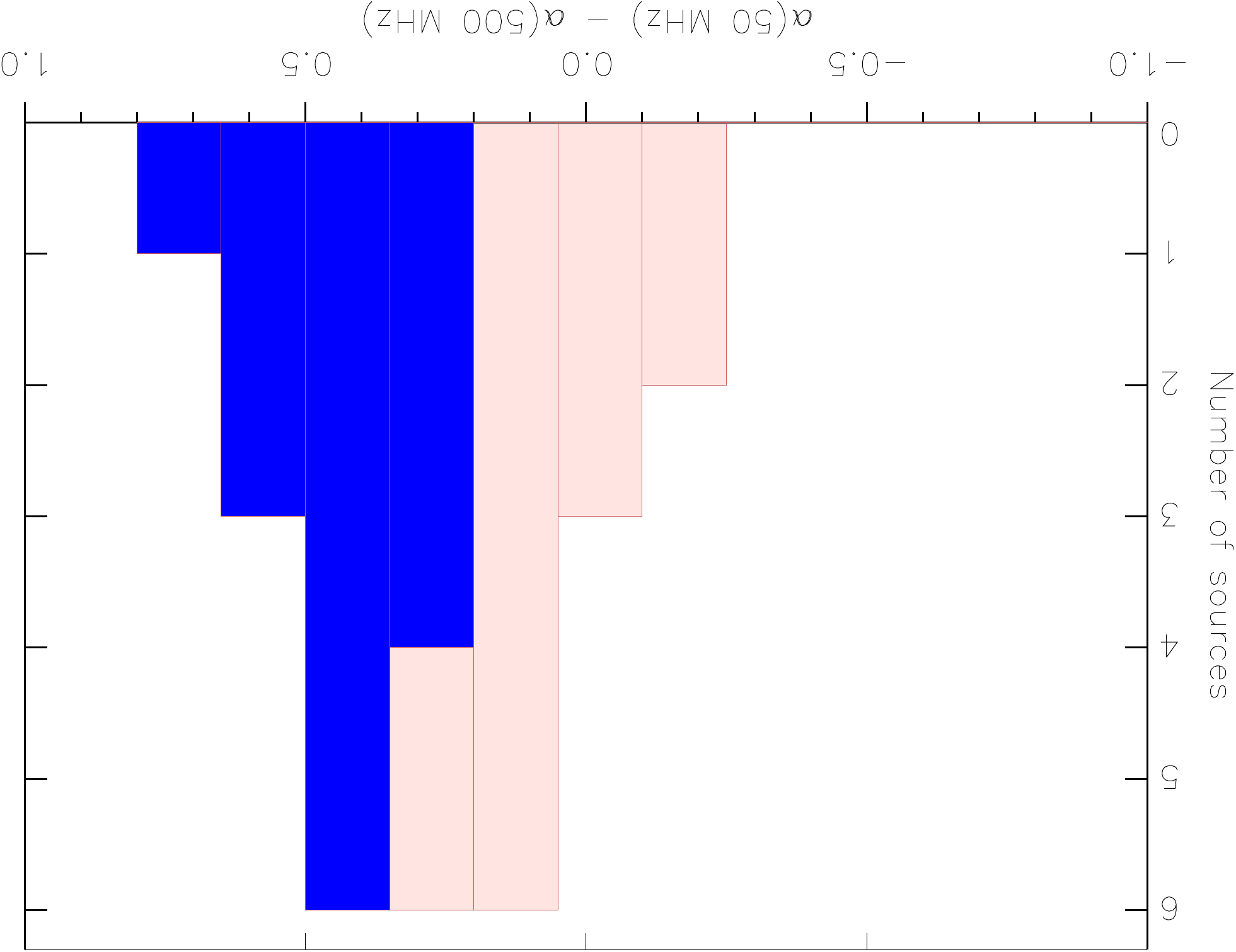}
\end{center}
\caption{{Left: Spectral index between 62 and 1400~MHz of sources in the  Bo\"otes field plotted against the integrated flux density. The solid line represents the average spectral index of the sources. The vertical dotted line is drawn at $10\sigma_{\rm{rms,avg}}$ and the dashed line indicates the completeness limit due to the NVSS sensitivity. Right: Histogram of spectral curvature between 500 and 50~MHz for bright 34~MHz sources. The spectral curvature was computed by fitting second order polynomials to the flux density measurements. The blue colors represent sources that cannot be properly fitted with power-law spectra and have $\lvert{a_2}\rvert > \sigma_{a_{2}}$, see Sect.~\ref{sec:spix}.} }
\label{fig:bootes_alphavsflux}
\end{figure*}

{For the  Bo\"otes field sources we search for counterparts in the NVSS and GMRT~153~MHz catalogs using a matching radius of 20\arcsec. If more than one counterpart to a 62~MHz source is found we add up the flux of all counterparts within the  20\arcsec~radius. In Fig.~\ref{fig:bootes_alphavsflux} we plot $\alpha^{1400}_{62}$ against the 62~MHz flux density. From this we find an average spectral index of $-0.79$. This average drops to $-0.74$ for $\alpha^{153}_{62}$ and increases to $-0.81$ for $\alpha^{1400}_{153}$. The average spectral index between 1400 and 153~MHz we find is within the range of previously reported values: $-0.87$, \citep{2013A&A...549A..55W},  $-0.79$ \citep{2011A&A...535A..38I}, $-0.78$ \citep{2010MNRAS.405..436I},  $-0.82$ \citep{2009MNRAS.392.1403S}, and $-0.85$ \citep{2007ASPC..380..237I}.} 

{We compute the same values for the 3C\,295 field, but starting with the 34~MHz source catalog. We find an average spectral index of $-0.81$ between 1400 and 34~MHz for the sources. This decreases to $-0.85$ between 1400 and 62~MHz and increases to $-0.64$ between 62 and 34~MHz, indicating that the average spectral index flattens towards lower frequencies. The value of $-0.64$ is somewhat steeper than the $-0.5$ suggested by the source count scalings. } 

{For the brighter, $S_{34} \gtrsim 1$~Jy, 34 MHz sources we also fitted the radio spectra with a second order polynomial ($\log_{10}{\left(S\right)} = a_0 + a_1\log_{10} {\left(\nu\right)}   + a_2\left(\log_{10}{\left(\nu\right)}\right)^2$), including the flux densities from the VLSSr, WENSS and NVSS surveys. In total we compute spectra for 27 sources, basically all 34~MHz sources that have a counterpart in the VLSSr survey (the VLSSr survey has a rms noise level of $\sim0.1$~Jy~beam$^{-1}$). From the polynomial fits we derive the spectral curvature between 500 and 50~MHz, i.e., the difference in the slope (spectral index) between 50 and 500~MHz. The resulting histogram is displayed in Fig.~\ref{fig:bootes_alphavsflux} (right panel). The histogram shows an excess of sources with curved spectra. We find that 14 sources have curved spectra (${a_2} < -\sigma_{a_{2}}$, where $\sigma_{a_{2}}$ is the uncertainty in $a_2$) , while 13 other sources have fits that are consistent with straight (power-law) spectra. None of these sources had an inverted spectrum, with $a_2  > \sigma_{a_{2}}$. The average spectral curvature of 0.3 is consistent with the  increase of the average spectral index from $\alpha^{1400}_{153} = -0.85$ to $\alpha^{62}_{34} = -0.64$ which we found earlier. This average was based on 133 sources so it shows that the spectral flattening is not only confined to the 27 brighter ($S_{34}\gtrsim 1$ Jy) 34~MHz sources.}

\subsection{Ultra-steep spectrum sources}
A large number of deep surveys at multiple wavelengths are available that cover the Bo\"otes field, in particular at radio wavelengths. We therefore carried out a search for sources which have ultra-steep radio spectra to select candidate HzRGs. USS sources that are detected at low frequencies could be missed by the higher frequency WENSS and/or NVSS survey due to their steep spectra. We therefore selected all sources detected at 62 and 153~MHz \citep[from][]{2013A&A...549A..55W}, but that are missed in either the WENSS or NVSS survey. In total we find 5 of these sources, see Table~\ref{tab:uss}. In addition, we selected sources from the 62~MHz source list that satisfied the criteria $\alpha^{153}_{62} < -1.1$  and $\alpha^{1400}_{153} < -1.1$, since a large part of our field overlaps with the deep 1.4~GHz WSRT survey of the Bo\"otes field from \cite{2002AJ....123.1784D} which can be used to compute the spectral indices.  Three additional sources were found in this way (Table~\ref{tab:uss}). 

{\cite{2008AJ....135.1793C} also searched for HzRGs in the Bo\"otes field. They selected candidate HzRGs with $S_{1400} > 1$~mJy in  a matched 325~MHz/1.4~GHz sample. The five sources with the steepest radio spectra and without optical counterparts were followed up with deep K-band imaging. None of the sources listed in Table~\ref{tab:uss} are reported by \cite{2008AJ....135.1793C}, as all but one of the sources from  \citeauthor{2008AJ....135.1793C} have $\alpha^{1400}_{325} > -1.0$. The source with the steepest spectral index from \citeauthor{2008AJ....135.1793C} (J142631+341557) is not detected in our LOFAR 62~MHz image. The source is detected at 153~MHz with an integrated flux of $20\pm5$~mJy, giving $\alpha^{1400}_{153} = -1.3$.}

\begin{table*}
\begin{center}
\caption{USS sources}
\begin{tabular}{llllllll}
\hline
\hline
source & RA, DEC$^{a}$ &$S_{62}$ & $S_{153}$ & $S_{1400}$$^{b}$ & $\alpha^{153}_{62}$ & $\alpha^{1400}_{153}$   \\ 
               & (J2000) &mJy        &  mJy    &   mJy     & &  \\
\hline
J143117.9+351549$^{*,c}$ & 14$^{\rm{h}}$31$^{\rm{m}}$18$^{\rm{s}}$.1	+35\degr15\arcmin50\arcsec&$252\pm41$  & $53\pm11$ & $1.74 \pm 0.08$ & $-1.72\pm0.29$ & $-1.54\pm0.10$ &   \\
J143127.4+343506$^{*, d}$ &14$^{\rm{h}}$31$^{\rm{m}}$27$^{\rm{s}}$.3	+34\degr35\arcmin07\arcsec  &$140\pm23$  & $40\pm8 $  &   $3.00\pm0.12$  & $-1.39\pm0.31$ & $-1.17\pm0.09$ & \\
J143236.1+333251$^{*}$ & 14$^{\rm{h}}$32$^{\rm{m}}$36$^{\rm{s}}$.3 +33\degr32\arcmin54\arcsec  &$65\pm14$  & $23\pm5$ & $2.00\pm0.09$  &   $-1.15\pm0.34$  & $-1.10 \pm0.10$ & \\ 
J143345.9+353856$^{*}$ & 14$^{\rm{h}}$33$^{\rm{m}}$46$^{\rm{s}}$.0	+35\degr38\arcmin55\arcsec  &$153\pm26$    &$57\pm12$&   $2.86\pm0.14^{c}$   & $-1.09\pm0.30$ & $-1.35 \pm 0.10$ &  \\
J143501.0+342531$^{*}$ & 14$^{\rm{h}}$35$^{\rm{m}}$01$^{\rm{s}}$.0 +34\degr25\arcmin31\arcsec&$173\pm28$ &$55\pm11$& $2.18\pm0.09$&  $-1.27\pm0.28$ & $-1.46\pm 0.09$ & \\
J143426.1+342809$^{c, e}$ &  14$^{\rm{h}}$34$^{\rm{m}}$25$^{\rm{s}}$.6 +34\degr28\arcmin19\arcsec     &  $341\pm54$ & $115\pm24$  & $10.1\pm0.5$& $-1.20\pm0.29$  &$-1.10\pm 0.10 $ &\\ 
J143506.8+350058 &  14$^{\rm{h}}$35$^{\rm{m}}$06$^{\rm{s}}$.9 +35\degr00\arcmin59\arcsec  &  $581\pm91$   & $141\pm29$ & $4.02\pm0.16$     &   $-1.57  \pm  0.29$     &  $-1.61 \pm0.09$ & \\
J143520.5+345949 &  14$^{\rm{h}}$35$^{\rm{m}}$20$^{\rm{s}}$.5 +34\degr59\arcmin50\arcsec&  $185\pm31$  &  $57\pm12$    & $1.55\pm0.07$        &  $-1.30  \pm    0.30$    &   $-1.63 \pm0.10 $ & \\
\hline
\hline
\end{tabular}
\label{tab:uss}
\end{center}
$^{*}$ source not detected in the NVSS and/or WENSS survey \\
$^{a}$ 1.4~GHz position from \cite{2002AJ....123.1784D}\\
$^{b}$ 1.4~GHz flux density from \cite{2002AJ....123.1784D}\\
$^{c}$ Source detected in the HerMES survey \citep{2010MNRAS.409...48R,2012MNRAS.424.1614O,2012MNRAS.419..377S}\\  
$^{d}$ Position and flux density from the 1.4~GHz FIRST survey \citep{1997ApJ...475..479W}\\ 
$^{e}$ Position from the 1.4~GHz FIRST survey and flux density from the NVSS survey\\ 

\vspace{10mm}
\end{table*}

{For the USS sources we identify candidate counterparts in the NOAO Deep Wide-Field Survey (NDWFS) I-band images. For the optical identification we use the likelihood ratio technique \citep{1992MNRAS.259..413S,2008A&A...490..879T}. In this way we obtain a probability $P(i)$ that candidate $i$ is the true optical counterpart to a given radio source.  For the radio position we take the GMRT 153~MHz position, or when available the 1.4~GHz FIRST position. We then obtain flux measurements for all candidate counterparts (with $P(i) > 5\%$) from the NOAO Deep Wide Field Survey \citep[NDWFS, $B_W$, $R$, $I$, $K$;][]{1999ASPC..191..111J}, the Flamingos Extragalactic Survey \citep[FLAMEX, $J$, $K_s$;][]{2006ApJ...639..816E}, the zBootes survey \citep[$z'$;][]{2007ApJS..169...21C}, the Spitzer Deep Wide Field Survey \citep[SDWFS, $\lbracket3.6\rbracket$, $\lbracket4.5\rbracket$, $\lbracket5.8\rbracket$, $\lbracket8.0\rbracket$;][]{2009ApJ...701..428A}, GALEX GR5 \citep[NUV, FUV;][]{2007ApJS..173..682M} and the MIPS AGN and Galaxy Evolution Survey \citep[MAGES $\lbracket24\rbracket$;][]{2010AAS...21547001J} to obtain photometric redshifts ($z_{\rm{phot}}$).
For the spectral energy distribution (SED) and $z_{\rm{phot}}$-fitting we require measurements in at least 5 bands. The fitting is performed using both the {\tt LRT} code from \cite{2008ApJ...676..286A} and  {\tt EAZY}\footnote{{\tt EAZY} does not use the 24 micron band for the fitting} \citep{2008ApJ...686.1503B} for comparison. A much more extensive description of the radio counterpart identification and SED fitting will be given in Williams et al. (in prep). The results of the fitting are summarized in Table~\ref{tab:photoz}. Figures showing the SEDs for each source and I band, IRAC 4.5~micron, IRAC 8.0~micron and MIPS 24~micron postage stamps, with GMRT (and FIRST where there is a source) contours, are shown in Appendix~\ref{sec:eazy}.}

We find that the photometric redshifts of the sources are mostly in the $0.7 \lesssim z_{\rm{phot}} \lesssim 2.5$ range. Given the correlation between optical brightness and redshift, counterparts without photo-z's are likely located at a higher redshift. For J143127.4+343506 and J143345.9+353856 the differences between the {\tt EAZY} and {\tt LRT} codes are substantial. We note that the {\tt LRT} code is supposed to do a better job of fitting AGN and {\tt LRT} also takes into account the upper limits. Larger USS samples are needed to detect more distant objects as they are more rare. However, steep-spectrum selection also misses a significant fraction of HzRGs \citep[e.g.,][]{2009MNRAS.398L..83J} and not all USS sources are associated with HzRGs \citep[e.g.,][]{2001MNRAS.326.1563J,2007MNRAS.375.1349C,2009A&A...508...75V}. Therefore a combination of deep radio and optical/NIR survey data will be a more powerful way of identifying HzRGs by searching for optically/NIR faint counterparts to the radio sources \citep[e.g.,][]{2006MNRAS.366.1265B,2012MNRAS.420.2644K}.

\begin{table*}[htp]
\caption{USS Sources SED fit results}
 \begin{center}
\begin{tabular}{lllllllll}
\hline
\hline
source & gmrt id & $P_{match}$ & $N_{\rm{bands}}$ &\multicolumn{3}{c}{$z_{eazy}$}        & \multicolumn{2}{c}{$z_{lrt}$} \\
       &         & \%         & & $z_a$ & $z_m$ &  $\chi^2/{\textrm{n.d.f.}}$ & $z$  & $\chi^2/{\textrm{n.d.f.}} $  \\
\hline 
J143520.5+345949 & 428 & \textbf{99.6} &11 & 0.746 & $0.743^{+0.056}_{-0.056}$ &  1.793 &  0.70 & 9.37  \\ [+1.5pt] 
J143506.8+350058 & 440 & \textbf{86.7} & 8   & 1.988 & $2.016^{+0.285}_{-0.279}$ &  1.054 &  2.46 & 1.32  \\ [+1.5pt] 
J143506.8+350058 & 440 & 13.0 & 2   & - & - &  - &  - & - \\ [+1.5pt] 
J143501.0+342531 & 445 & \textbf{99.4} & 11 & 1.400 & $1.380^{+0.171}_{-0.175}$ & 0.749 &  1.34 & 1.91  \\ [+1.5pt] 
J143426.1+342809 & 485 & \textbf{55.3} &4&- & - &  - &  - & - \\ [+1.5pt] 
J143426.1+342809 & 485 & 41.6 &3&- & - &  - &  - & - \\ [+1.5pt] 
J143345.9+353856 & 517 & \textbf{99.8} & 6 &1.871 & $1.935^{+0.125}_{-0.137}$ & 1.534 &  2.45 & 1.93  \\ [+1.5pt] 
J143236.1+333251 & 591 &\textbf{99.1} &11&0.967 & $1.033^{+0.096}_{-0.089}$ & 1.677 &  1.06 & 20.59  \\ [+1.5pt] 
J143127.4+343506 & 667 & \textbf{76.6} & 8 &1.815 & $1.851^{+0.208}_{-0.182}$ &  1.549 &  0.32 & 0.28  \\ [+1.5pt] 
J143127.4+343506 & 667 & 23.2 &1 & - & -  & - &  - & - \\ [+1.5pt] 
J143117.9+351549 & 679 & \textbf{92.4} &7&1.548 & $2.253^{+0.750}_{-0.695}$  & 0.107 &  1.56 & 0.09  \\ [+1.5pt]  \hline
\hline
 \end{tabular}\label{tab:photoz}
 \end{center}
The radio source name and GMRT radio source ID are given in Cols. 1 and 2; Col. 3 gives the probability that a given
source is the true optical counterpart to the radio source. The highest probability match is marked in boldface; The number of bands available for SED fitting ($N_{\rm{bands}}$) is given in Col. 4.; Cols. 5 and 6 give the redshift obtained via {\tt EAZY}, with $z_a$ the redshift at the minimum $\chi^2$, and $z_m$ the redshift marginalized over the $p(z)$ distribution, with the $68\%$ confidence intervals; The reduced $\chi^2$ of the fit is listed in Col. 7; The fitted redshift from the {\tt LRT} code and corresponding reduced $\chi^2$ are given in Cols. 8 and 9. \\
$^a$
No SED/$z_{\rm{phot}}$ fitting could be performed since there are less than 5 flux measurements were available.
 \end{table*}

\section{Conclusions}
\label{sec:conclusions}

We have presented the results of LOFAR LBA observations of the Bo\"otes and 3C\,295 fields. In our 62~MHz Bo\"otes field image, with a central noise level of 4.8~mJy~beam$^{-1}$, we detect a total of 329 sources over a 19.4~deg$^2$ area. Our images of the 3C\,295 field cover an area from 17 to 52.3~deg$^2$ from 62 to 34~MHz, respectively. We reach central noise levels of 5.3, 8.2 and 12~mJy~beam$^{-1}$ at 62, 46 and 34~MHz for the 3C\,295 field. In total we detect 329, 367, 392 sources at  62, 46 and 34~MHz. 

From our source lists, we derive the deepest differential source counts at 62, 46 and 34~MHz to date. At 62~MHz the source counts are in good agreement with 74~MHz counts from VLA observations and scaling with a spectral index of $-0.7$. At 34~MHz the measured source counts fall significantly below extrapolated source counts from 74 and 153~MHz, using a spectral index scaling of $-0.7$. {Instead, we find that a spectral index scaling of $-0.5$ provides a  better match to the observed 34~MHz source counts.  Our 34~MHz source counts are also consistent with those obtained from the 38~MHz 8C survey. {In addition, evidence for spectral flattening is found from the increase of the average radio spectral index from high to low frequencies. From  polynomial fits to the individual flux densities  of bright ($\gtrsim 1$ Jy) 34~MHz sources, we conclude that about half of these sources have curved spectra. The curved spectra of these sources could be caused by absorption effects, as well as by spectral ageing.}
 
We also selected sources with steep radio spectra  ($\alpha < -1.1$) in the Bo\"otes field to find candidate high-z radio galaxies. We identified optical counterparts to these sources and fitted the SEDs to obtain photometric redshifts. We conclude that most of these USS sources seem to be located in the $ 0.7 \lesssim  z  \lesssim 2.5$ range.

\acknowledgments
{\it Acknowledgments:}
We thank the anonymous referee for useful comments. LOFAR, the Low Frequency Array designed and constructed by ASTRON, has facilities in several countries, that are owned by various parties (each with their own funding sources), and that are collectively operated by the International LOFAR Telescope (ILT) foundation under a joint scientific policy.
This work made use of images and/or data products provided by the NOAO Deep Wide-Field Survey \citep{1999ASPC..191..111J}, which is supported by the National Optical Astronomy Observatory (NOAO). NOAO is operated by AURA, Inc., under a cooperative agreement with the National Science Foundation.

Support for this work was provided by NASA through Einstein Postdoctoral
grant number PF2-130104 awarded by the Chandra X-ray Center, which is
operated by the Smithsonian Astrophysical Observatory for NASA under
contract NAS8-03060. Chiara Ferrari acknowledges financial support by the {\it ``Agence 
Nationale de la Recherche''} through grant ANR-09-JCJC-0001-01.

\appendix

\section{A. Extended sources at 62~MH\lowercase{z}}
\label{sec:extim}
Figures \ref{fig:cutouts_bootes} and \ref{fig:cutouts_3c295} show the 62 MHz LOFAR images of extended sources in the Bo\"otes and 3C\,295 fields.

\begin{figure*}
\begin{center}
\includegraphics[trim =0cm 0cm 0cm 0cm,angle=180, width=0.24\textwidth]{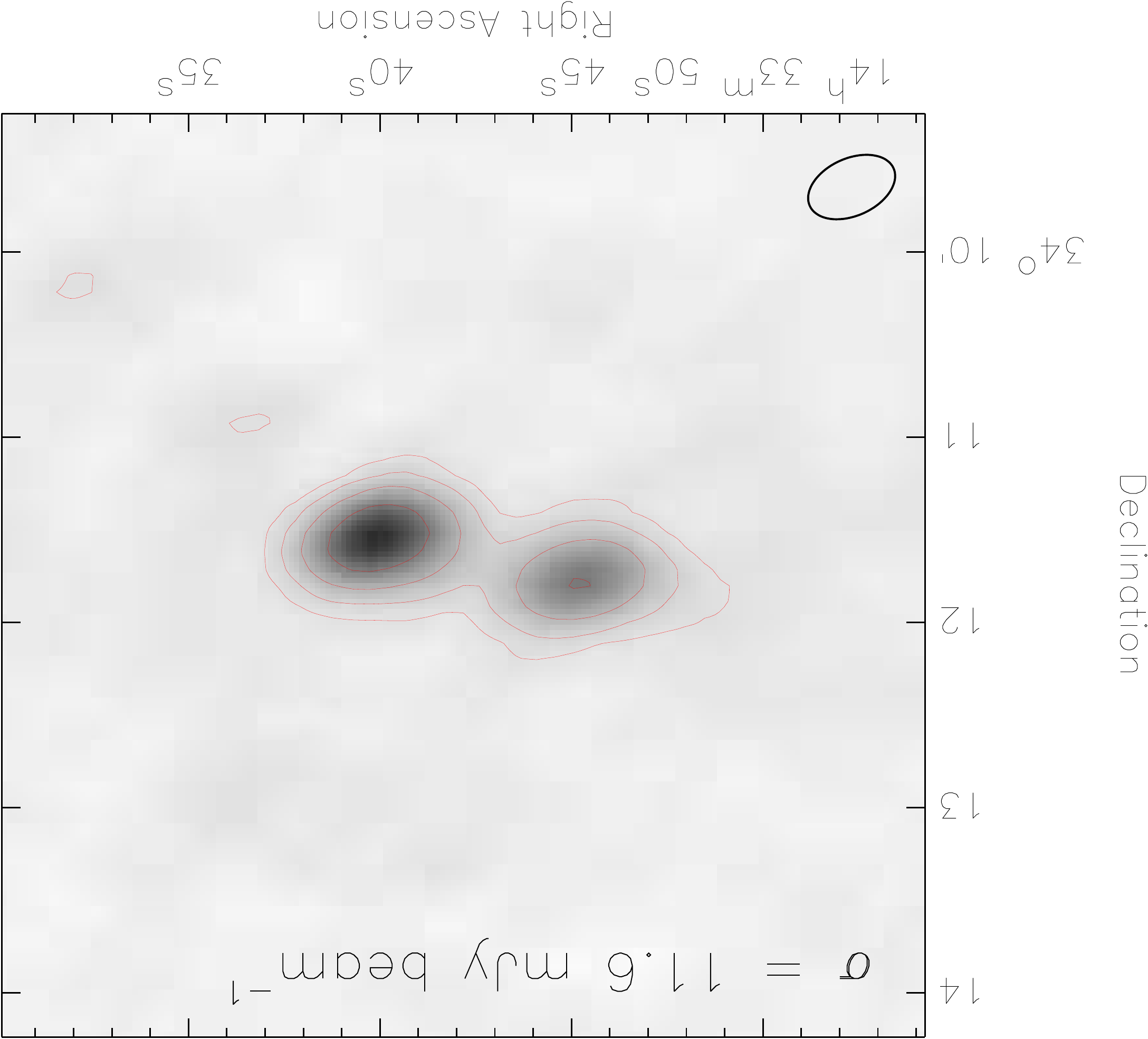}
\includegraphics[trim =0cm 0cm 0cm 0cm,angle=180, width=0.24\textwidth]{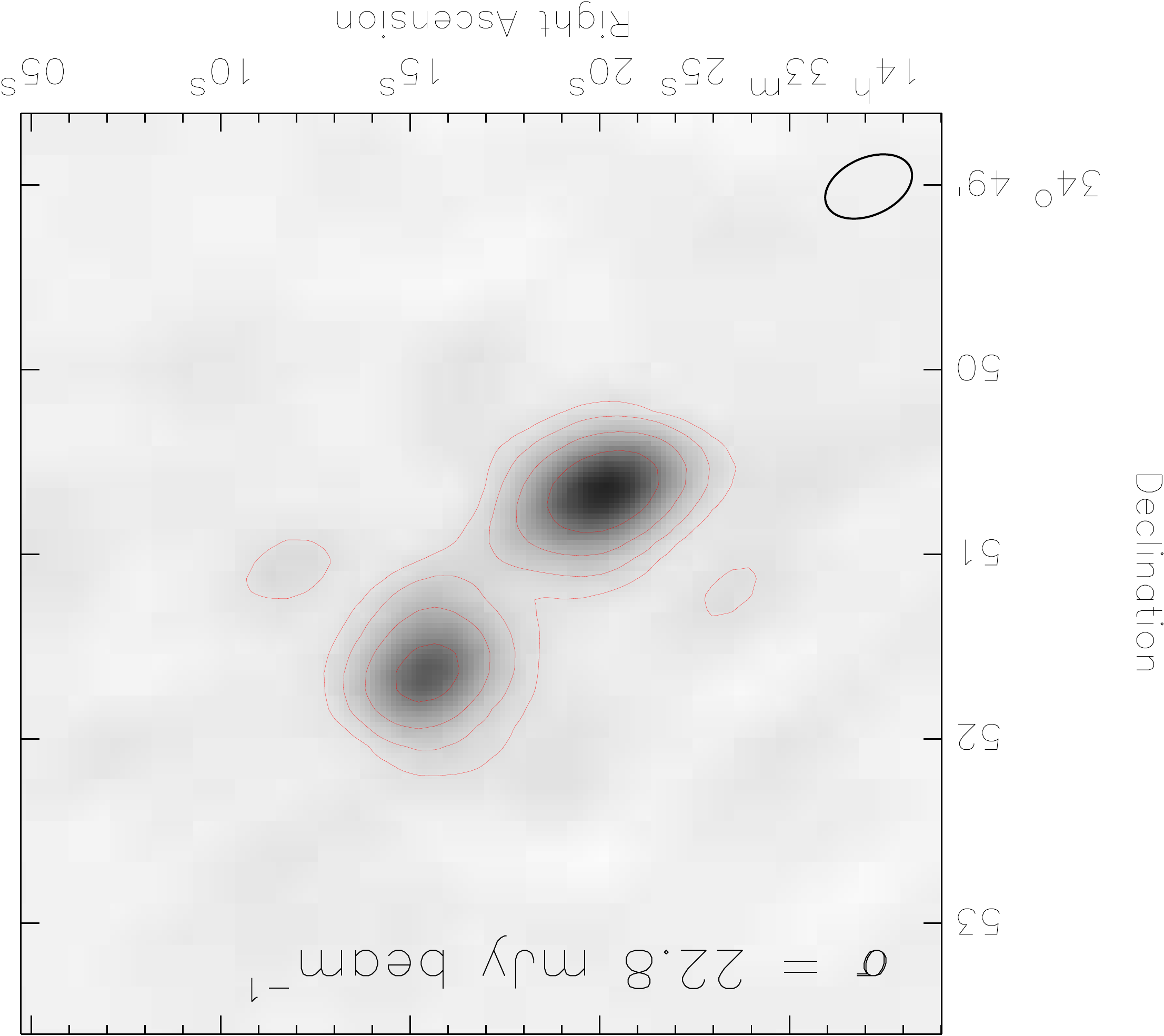}
\includegraphics[trim =0cm 0cm 0cm 0cm,angle=180, width=0.24\textwidth]{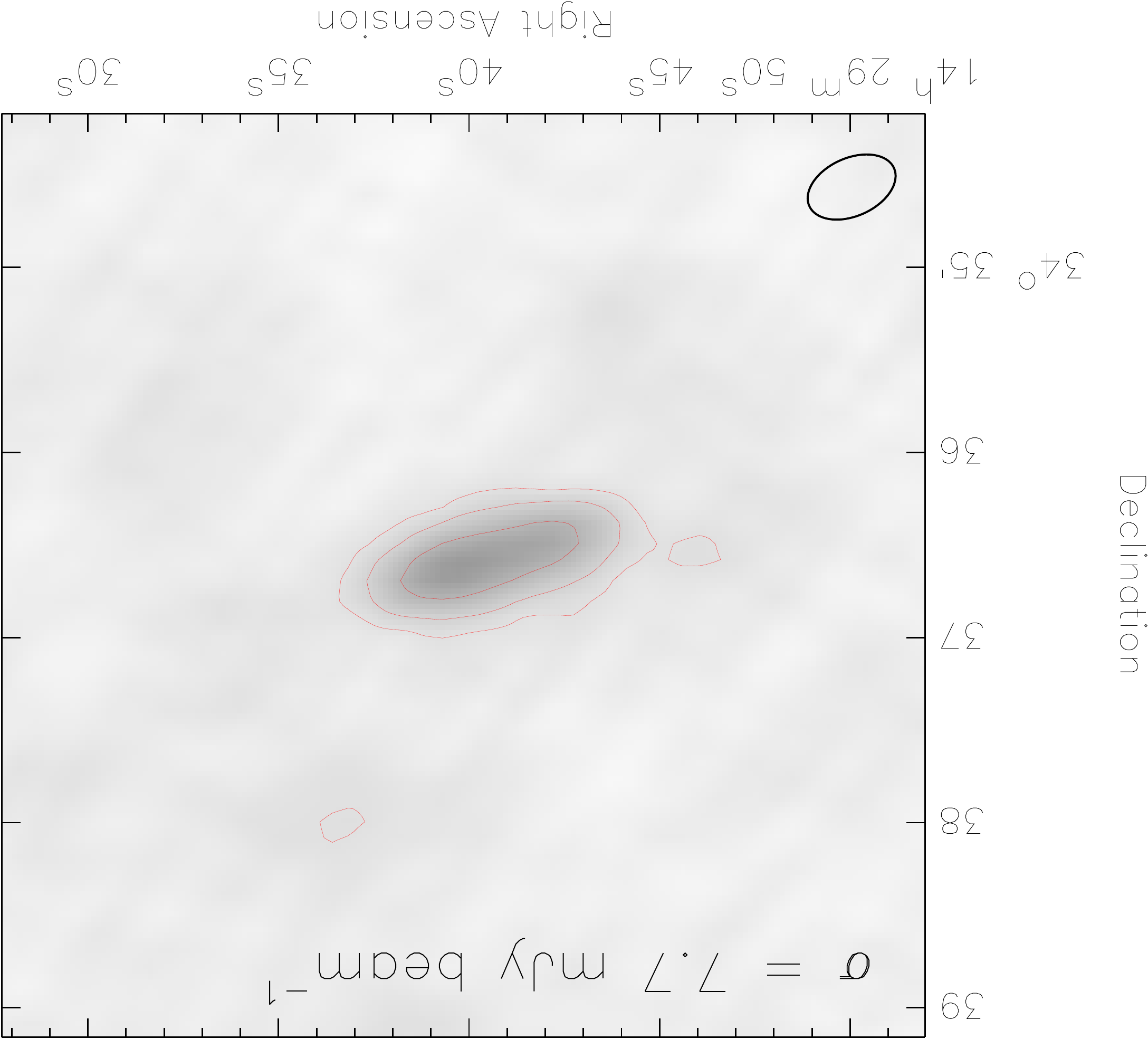}
\includegraphics[trim =0cm 0cm 0cm 0cm,angle=180, width=0.24\textwidth]{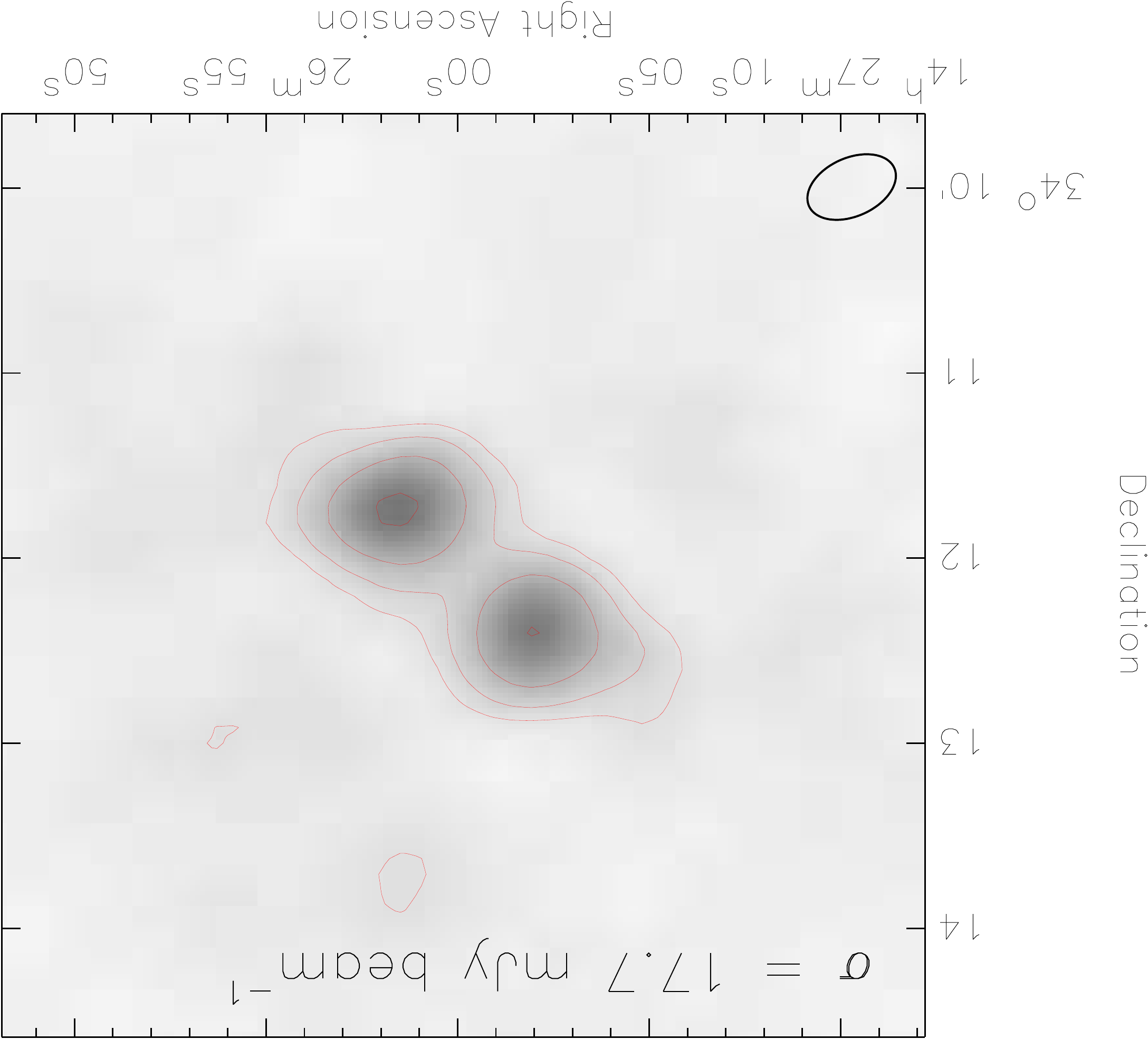}
\includegraphics[trim =0cm 0cm 0cm 0cm,angle=180, width=0.24\textwidth]{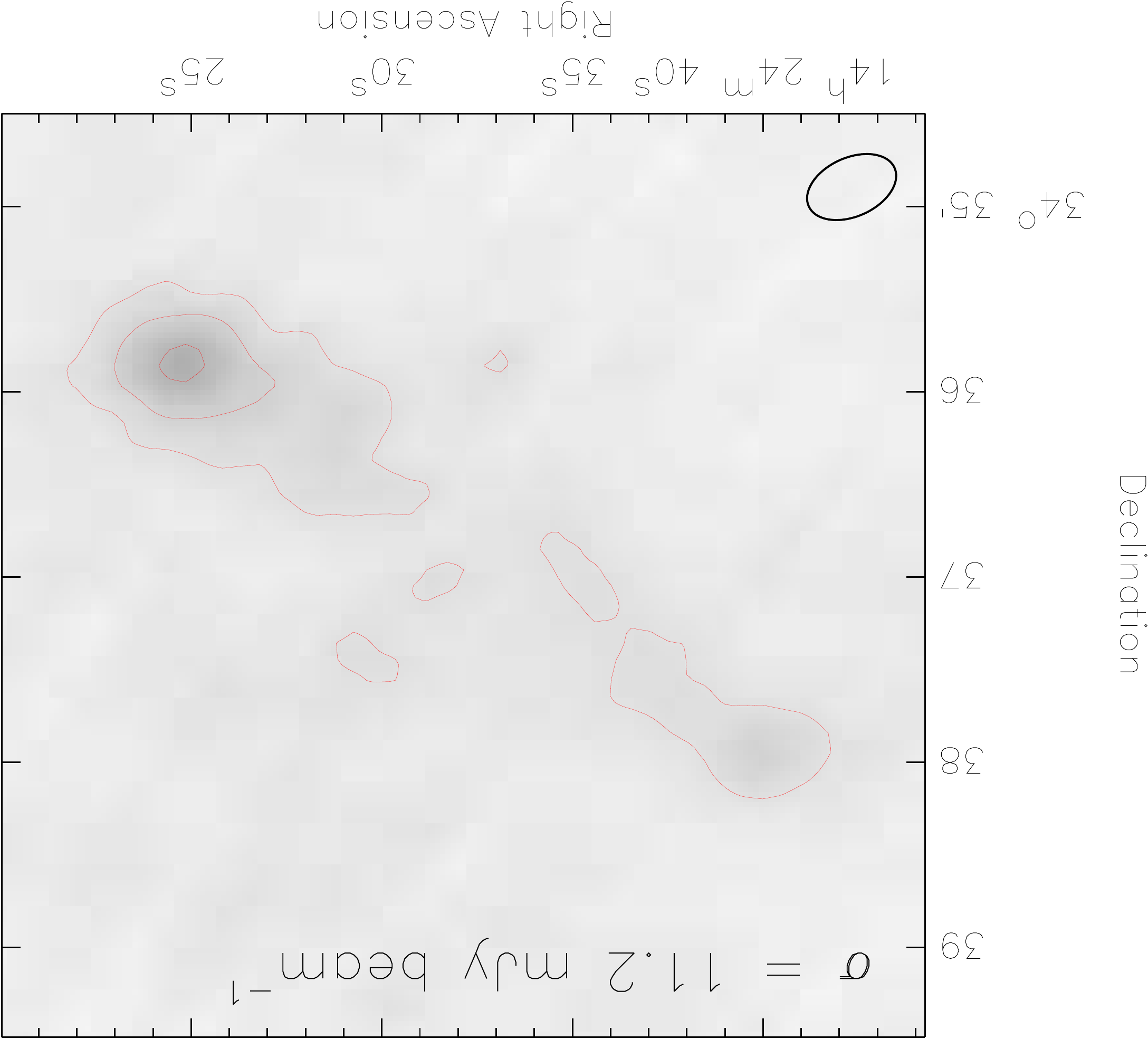}
\includegraphics[trim =0cm 0cm 0cm 0cm,angle=180, width=0.24\textwidth]{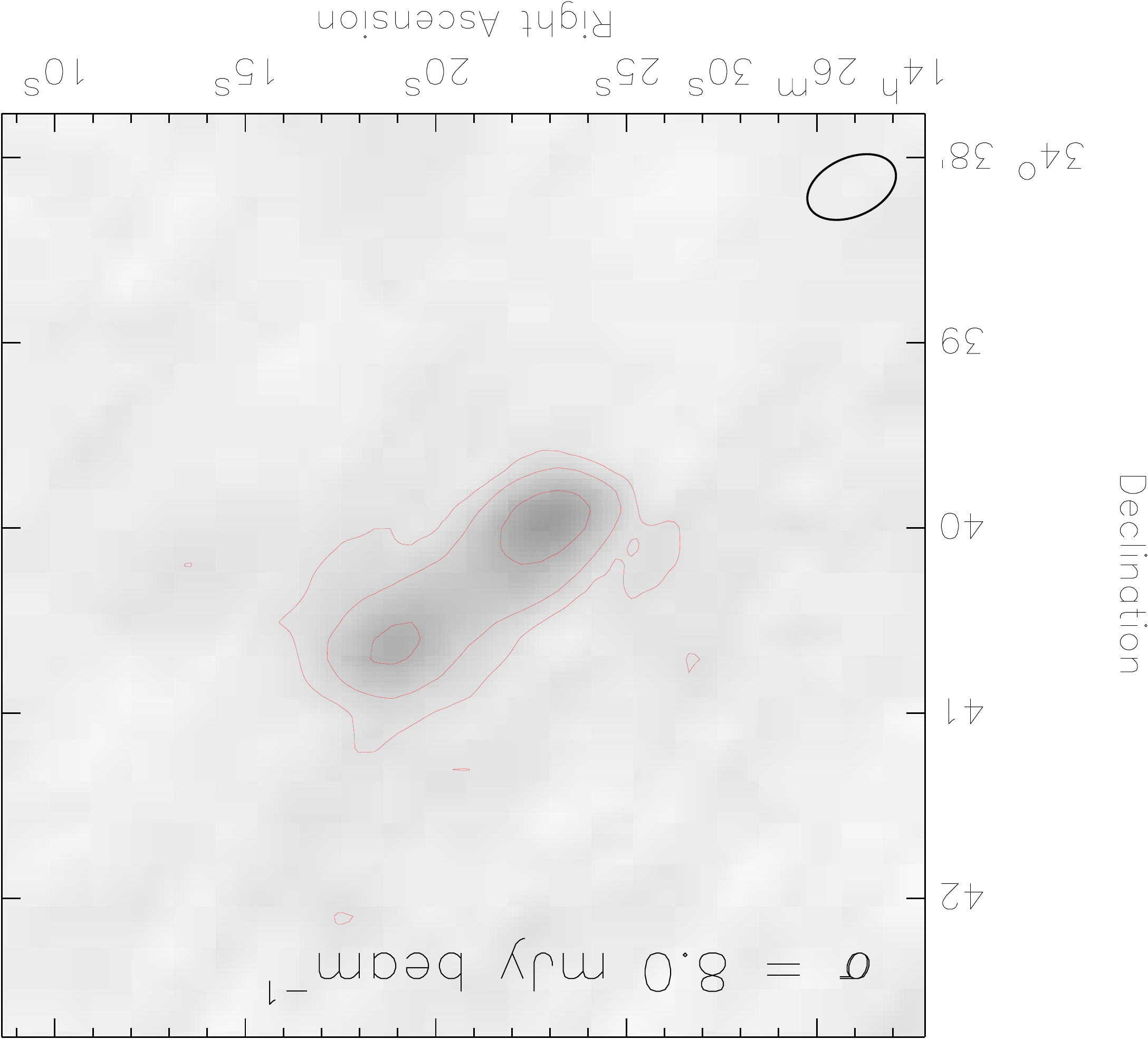}
\includegraphics[trim =0cm 0cm 0cm 0cm,angle=180, width=0.24\textwidth]{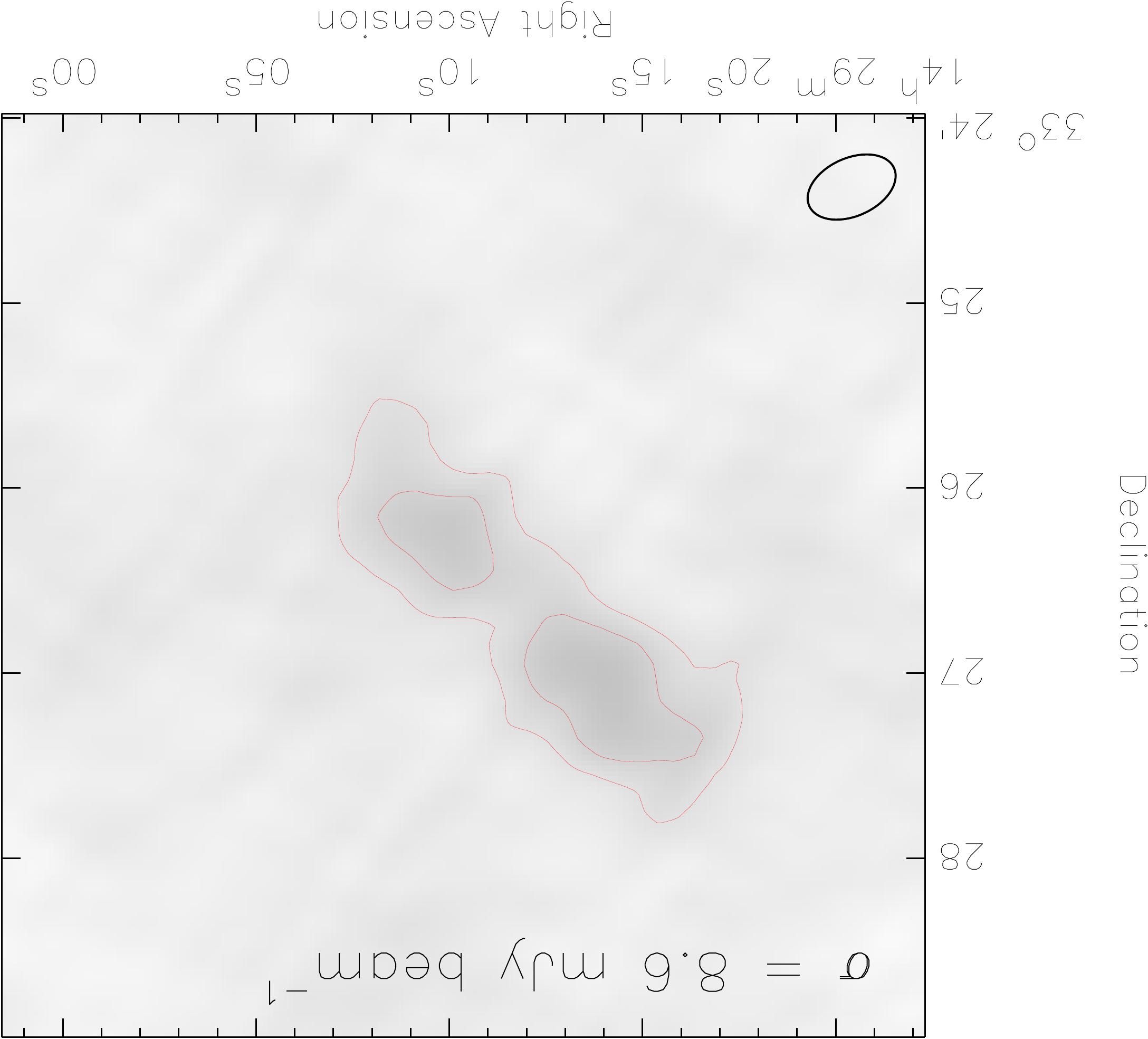}
\includegraphics[trim =0cm 0cm 0cm 0cm,angle=180, width=0.24\textwidth]{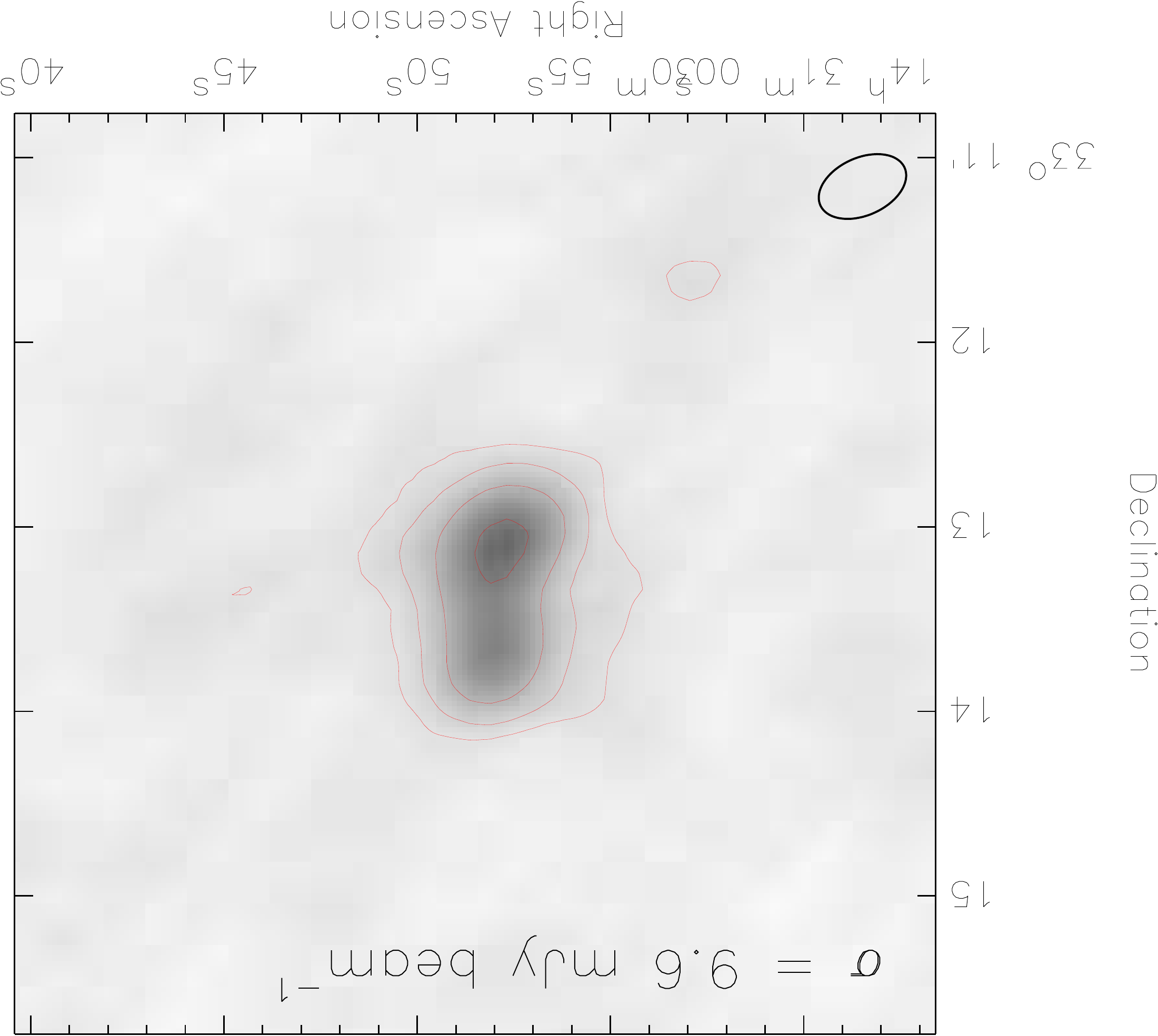}
\includegraphics[trim =0cm 0cm 0cm 0cm,angle=180, width=0.24\textwidth]{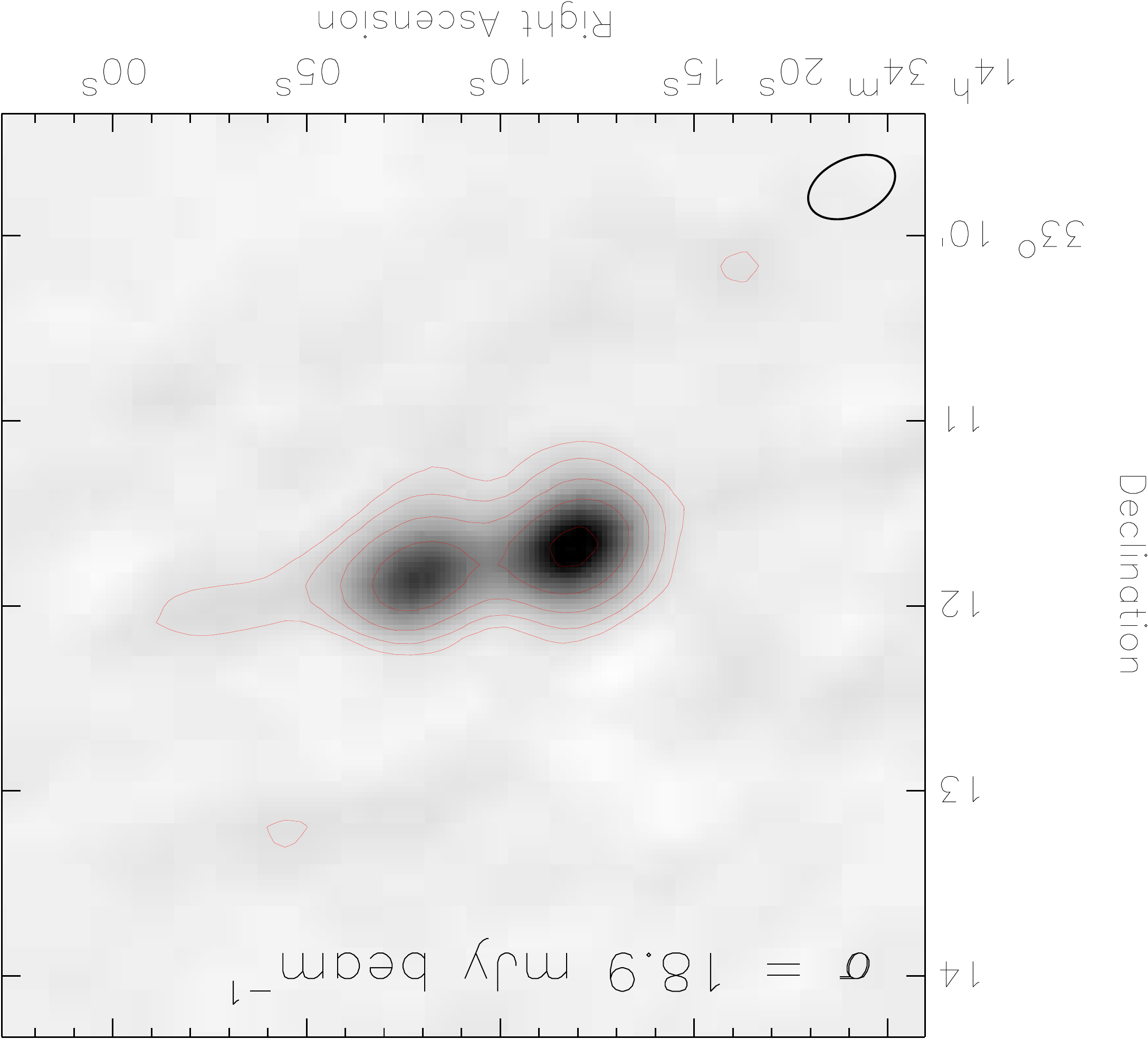}
\includegraphics[trim =0cm 0cm 0cm 0cm,angle=180, width=0.24\textwidth]{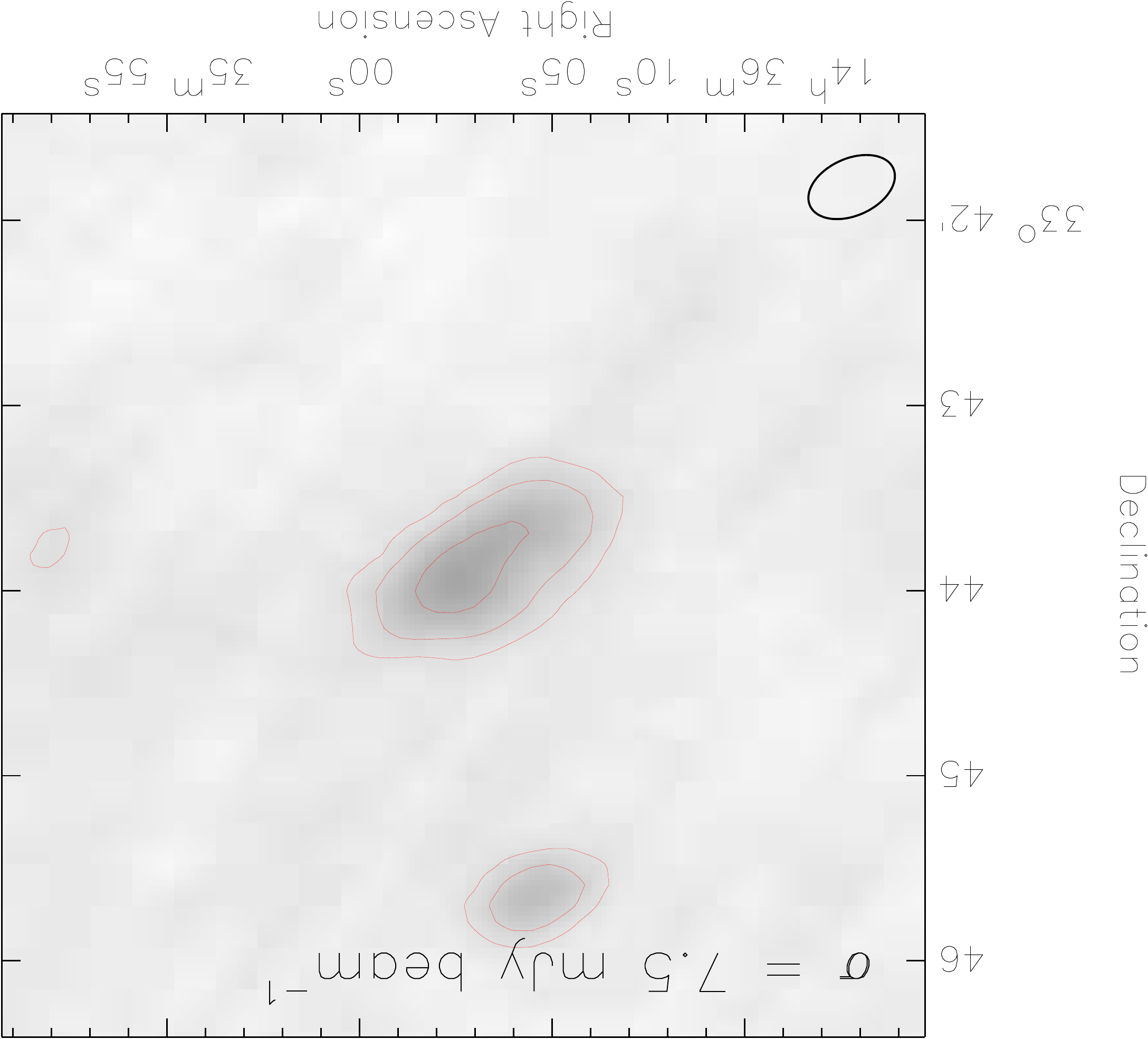}
\includegraphics[trim =0cm 0cm 0cm 0cm,angle=180, width=0.24\textwidth]{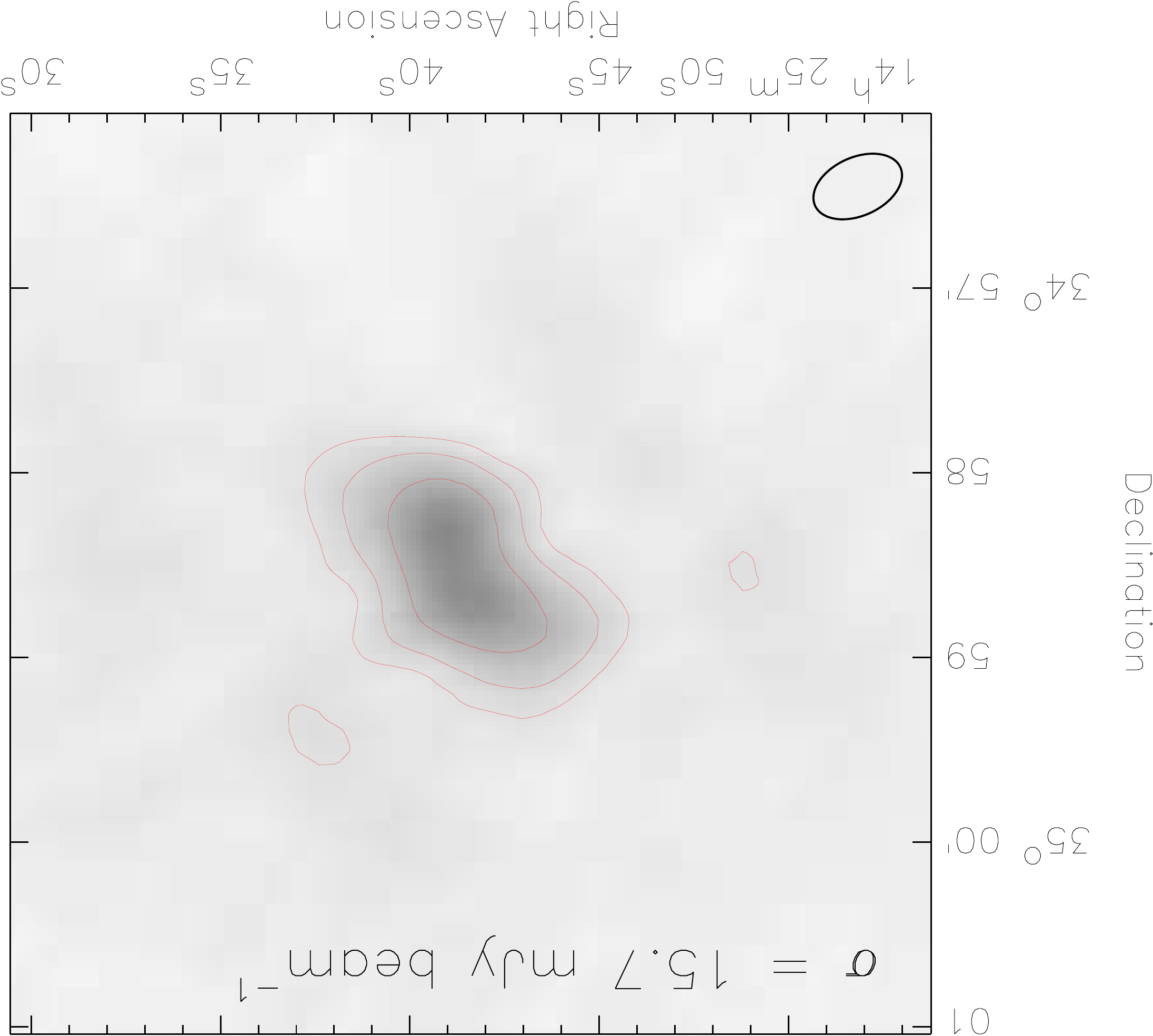}
\includegraphics[trim =0cm 0cm 0cm 0cm,angle=180, width=0.24\textwidth]{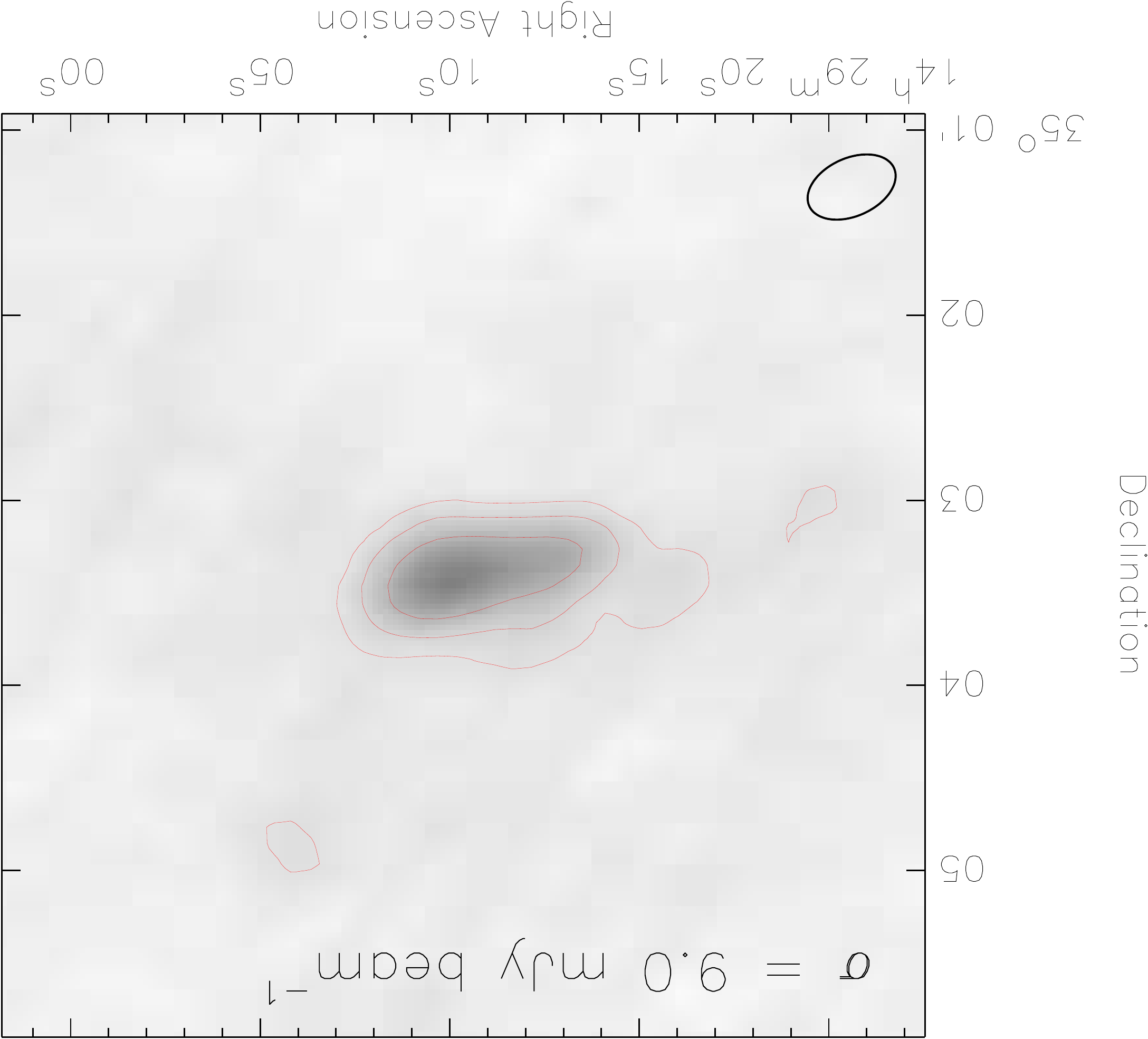}
\includegraphics[trim =0cm 0cm 0cm 0cm,angle=180, width=0.24\textwidth]{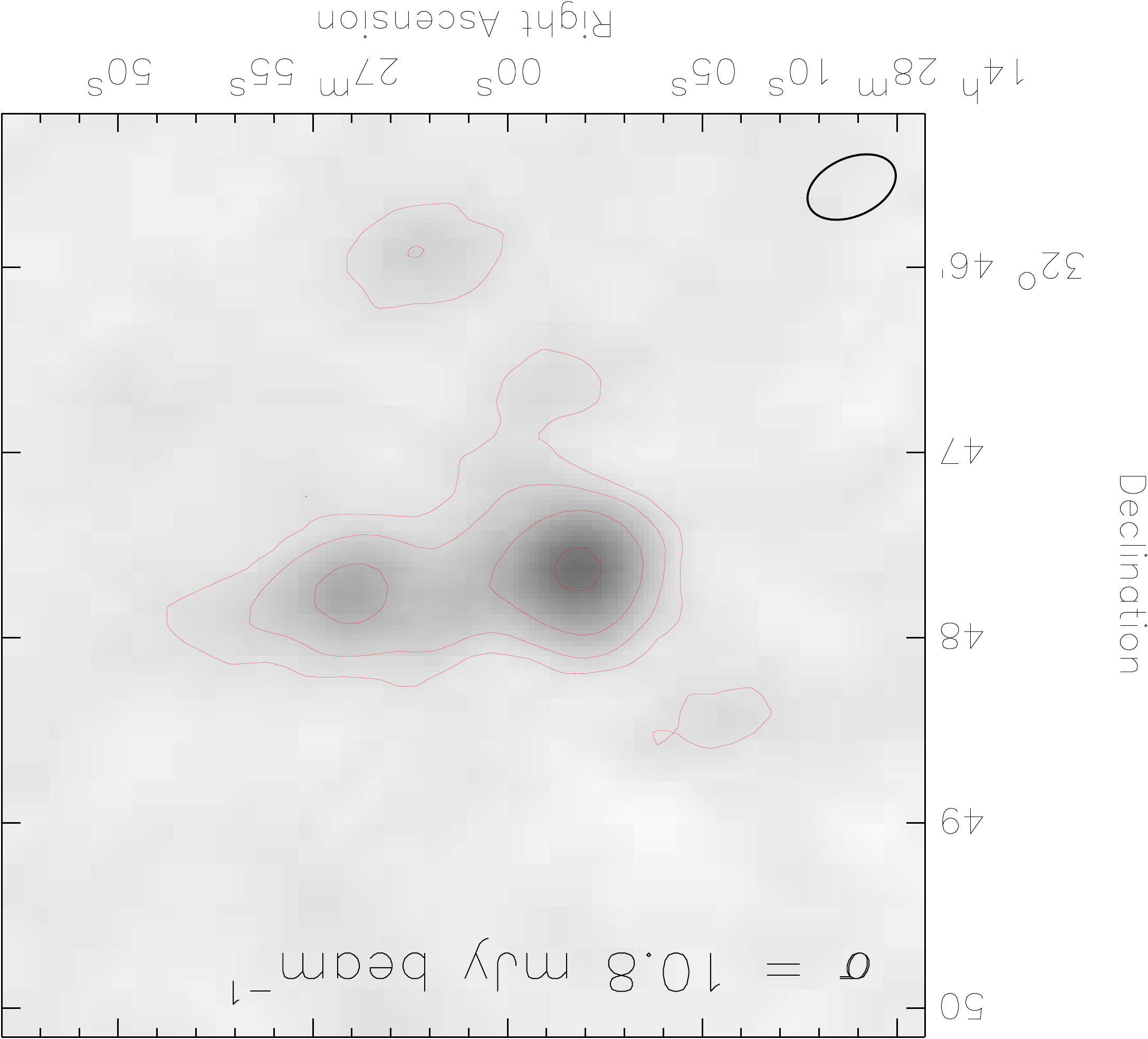}
\includegraphics[trim =0cm 0cm 0cm 0cm,angle=180, width=0.24\textwidth]{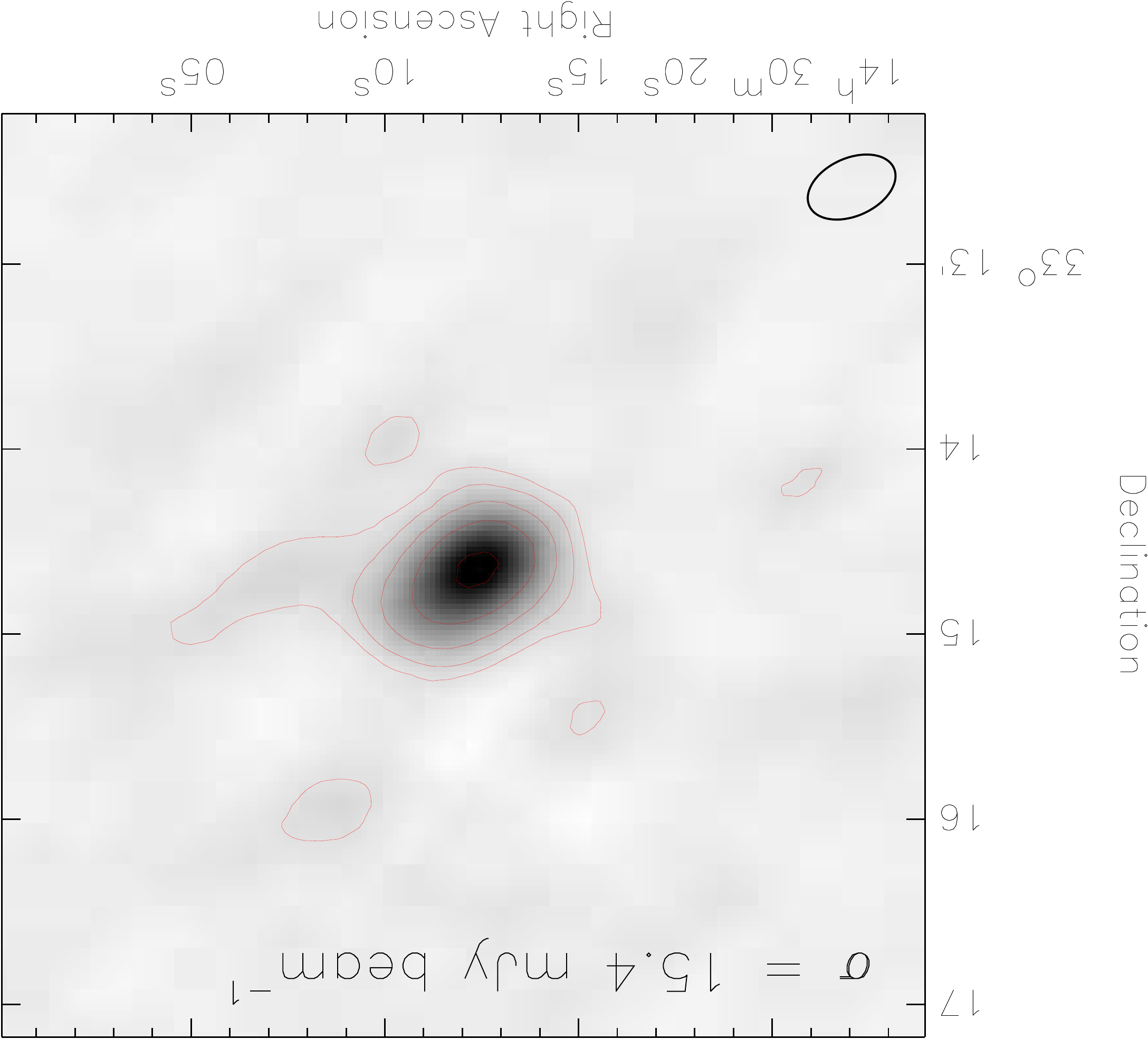}
\includegraphics[trim =0cm 0cm 0cm 0cm,angle=180, width=0.24\textwidth]{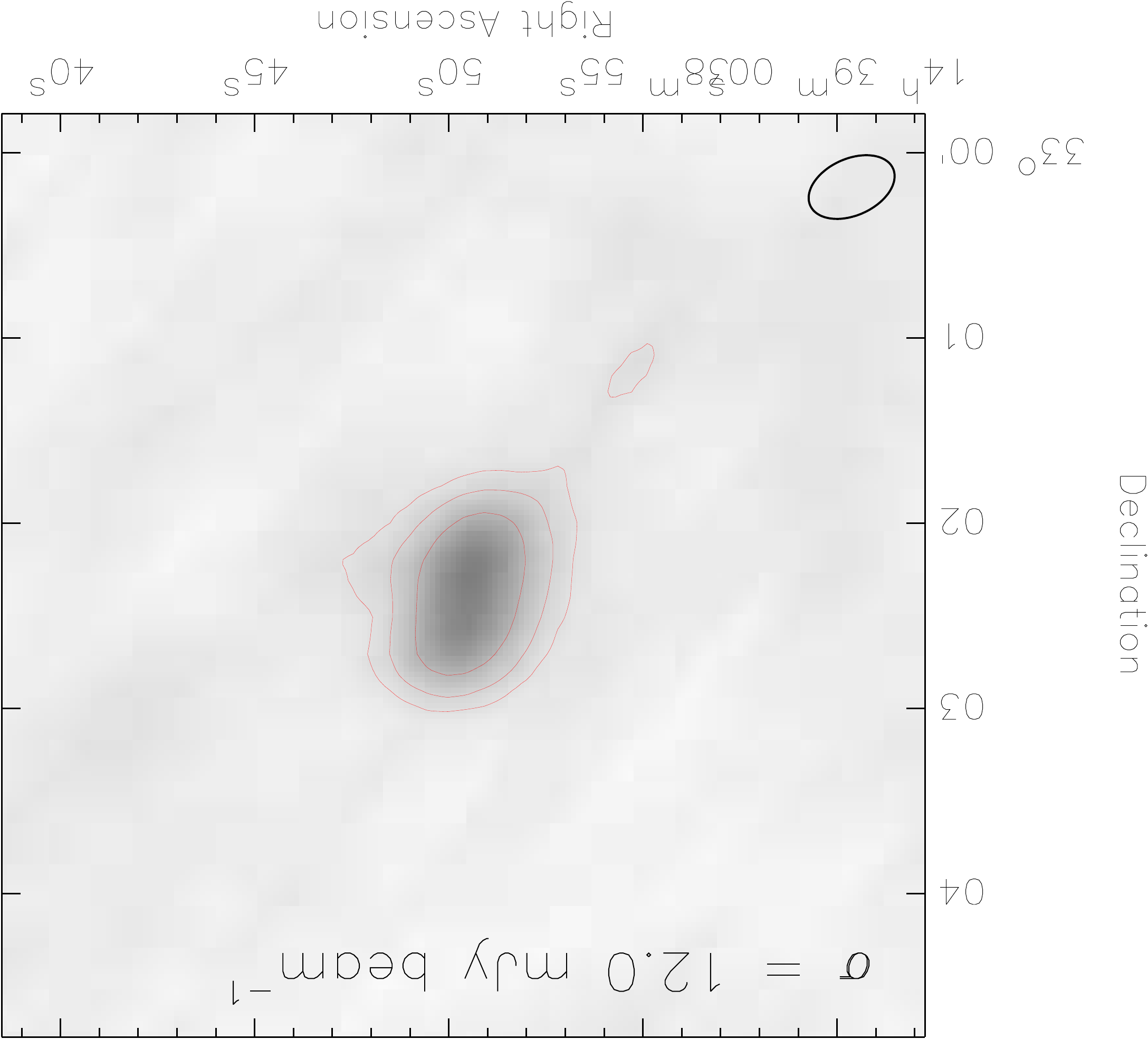}

\end{center}
\caption{Images of resolved sources in the Bo\"otes field at 62~MHz. Contour levels are drawn at $[1,2,4,8,\ldots] \times 3\sigma_{\rm{local\mbox{ }rms}}$, with $\sigma_{\rm{local\mbox{ }rms}}$ reported in each image. The beam size is shown in the bottom left corner of the images.}
\label{fig:cutouts_bootes}
\end{figure*}

\begin{figure*}
\begin{center}
\includegraphics[ trim =0cm 0cm 0cm 0cm,angle=180, width=0.24\textwidth]{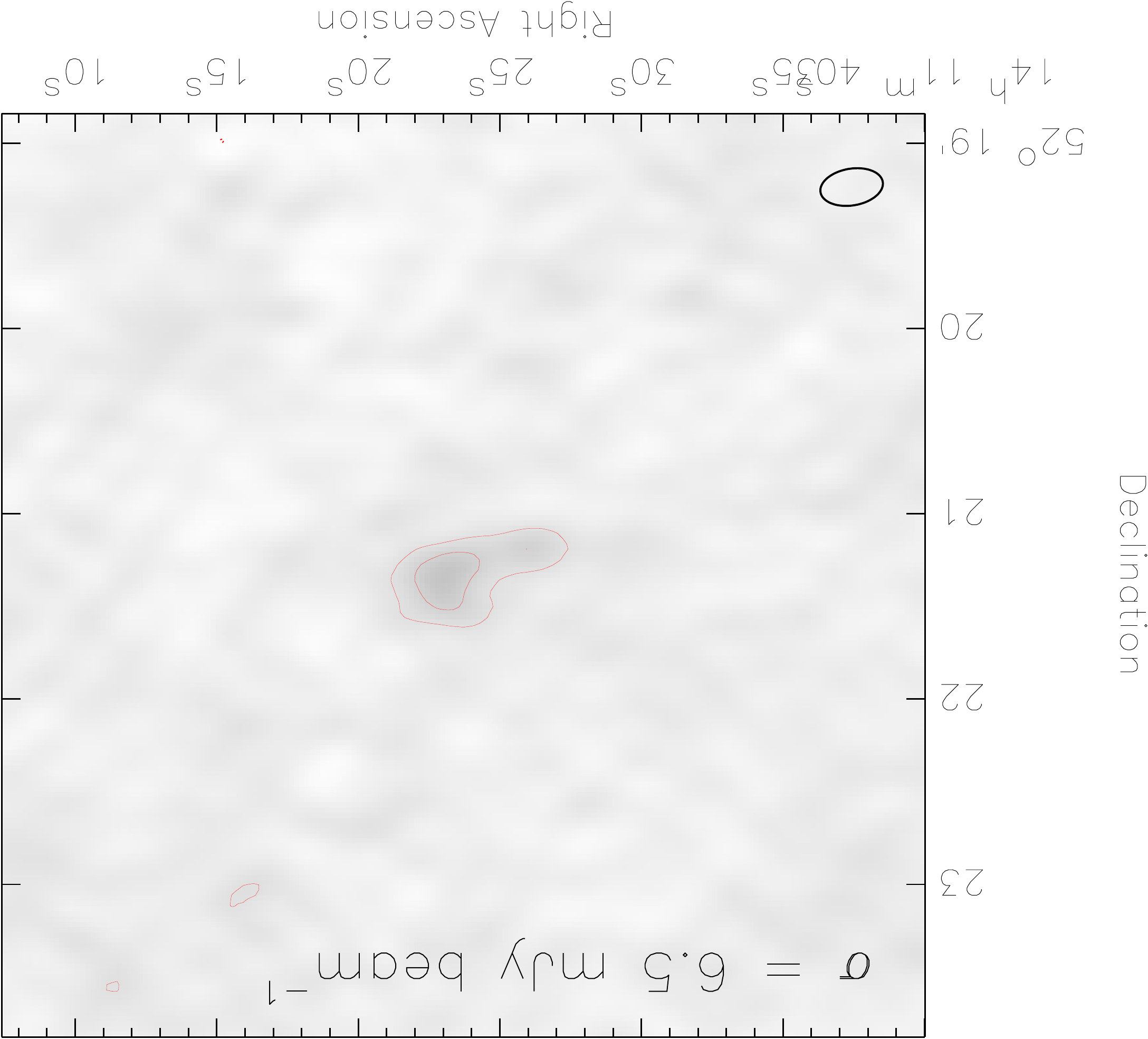}
\includegraphics[ trim =0cm 0cm 0cm 0cm,angle=180, width=0.24\textwidth]{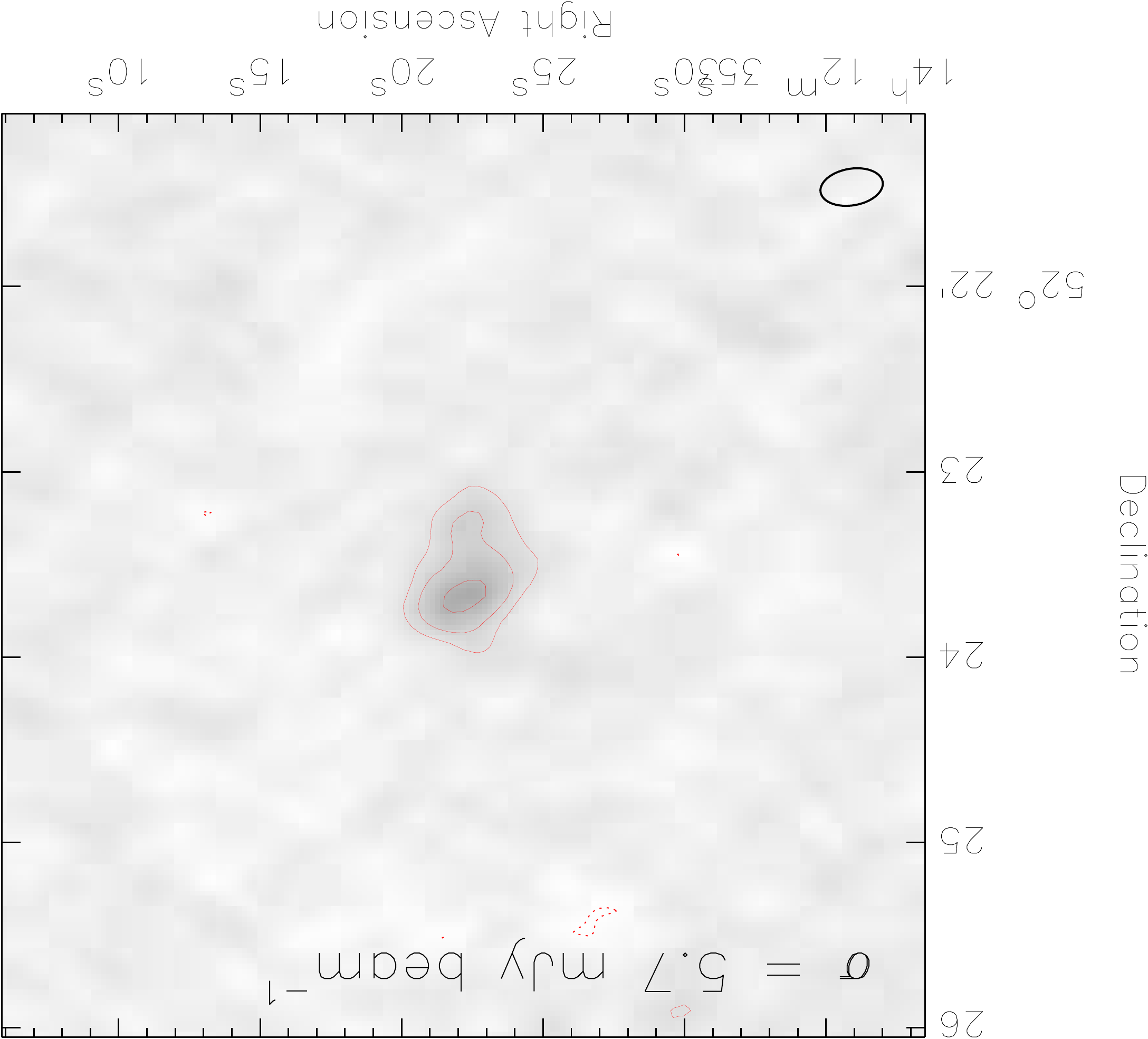}
\includegraphics[ trim =0cm 0cm 0cm 0cm,angle=180, width=0.24\textwidth]{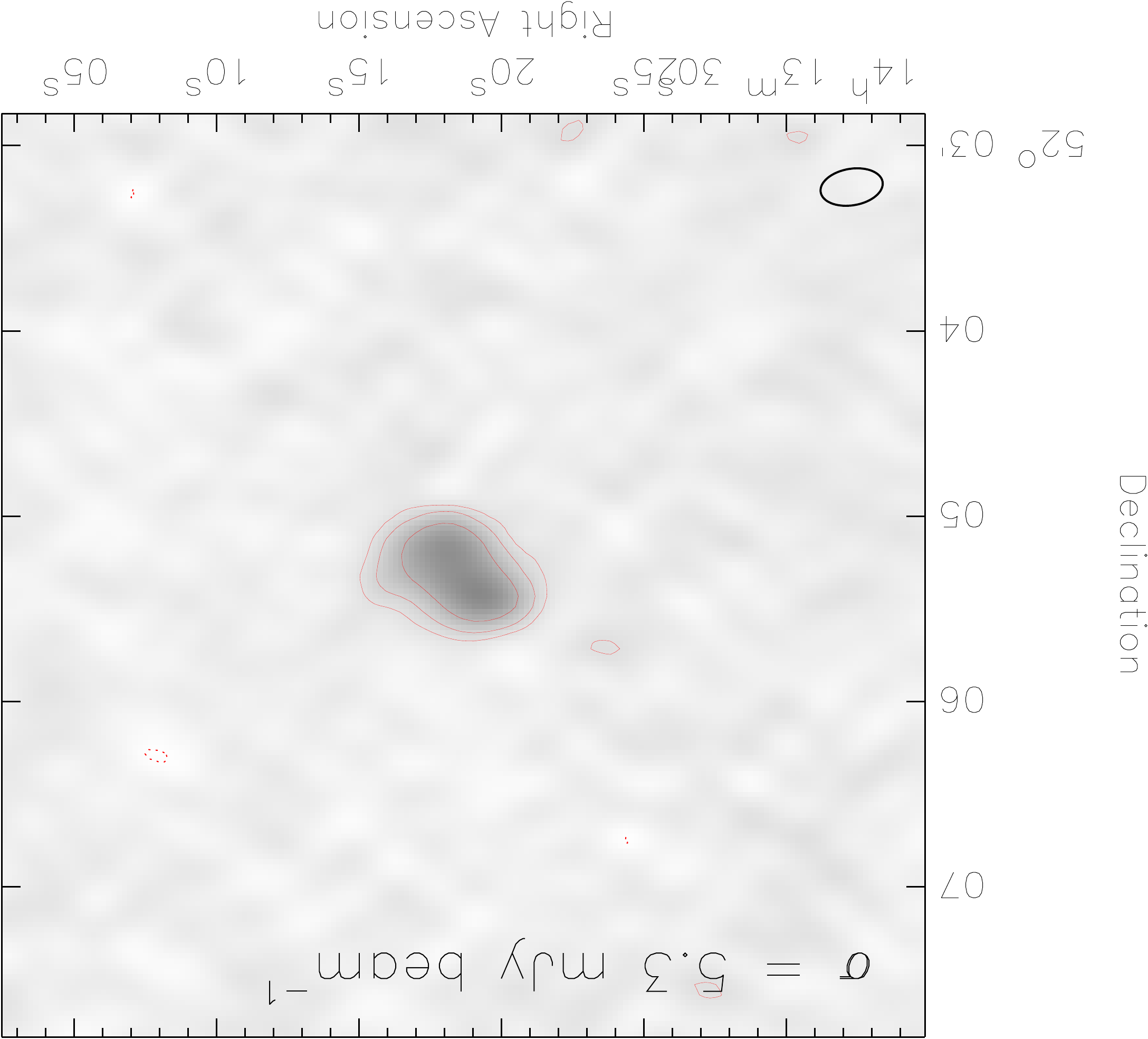}
\includegraphics[ trim =0cm 0cm 0cm 0cm,angle=180, width=0.24\textwidth]{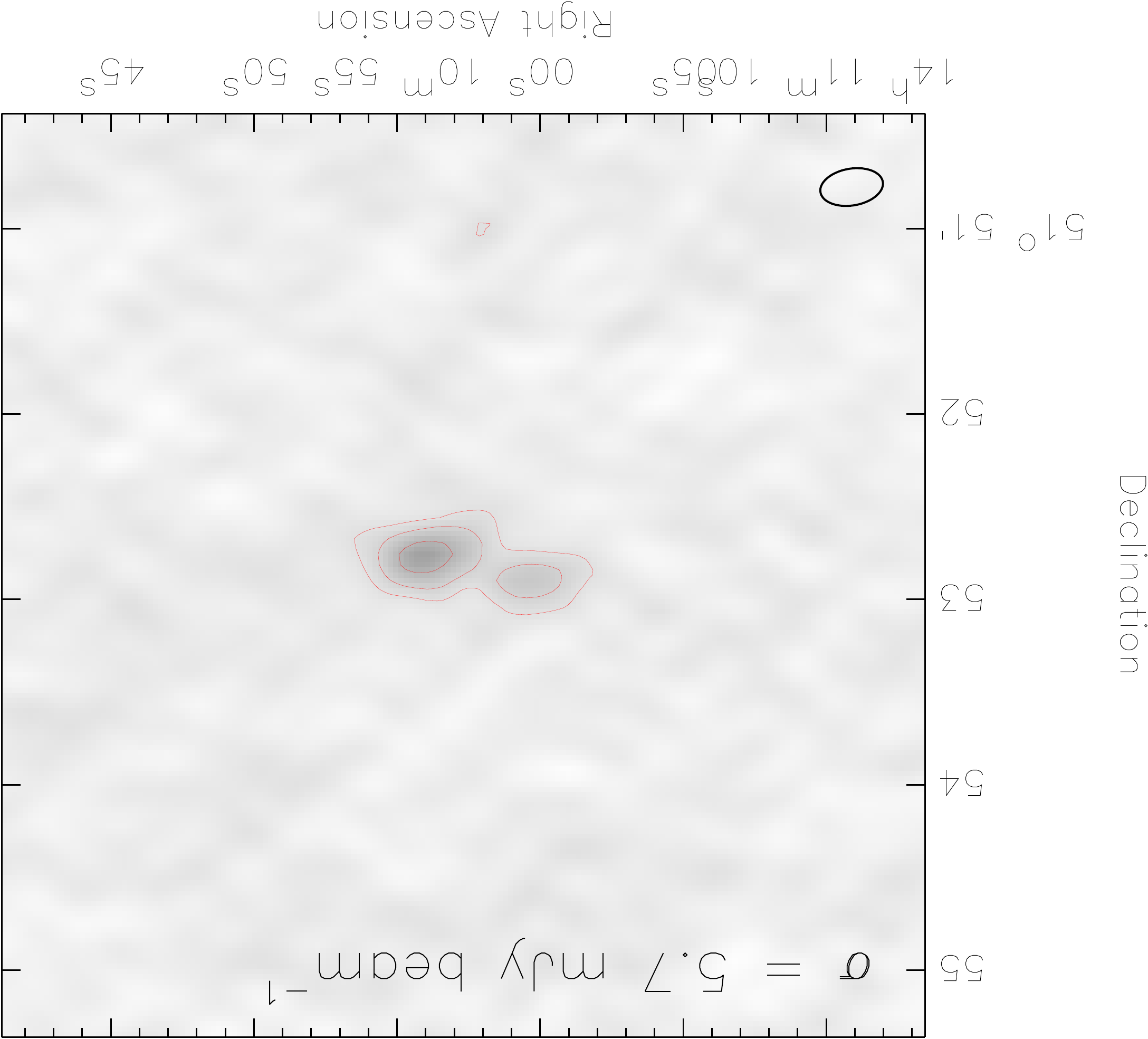}
\includegraphics[ trim =0cm 0cm 0cm 0cm,angle=180, width=0.24\textwidth]{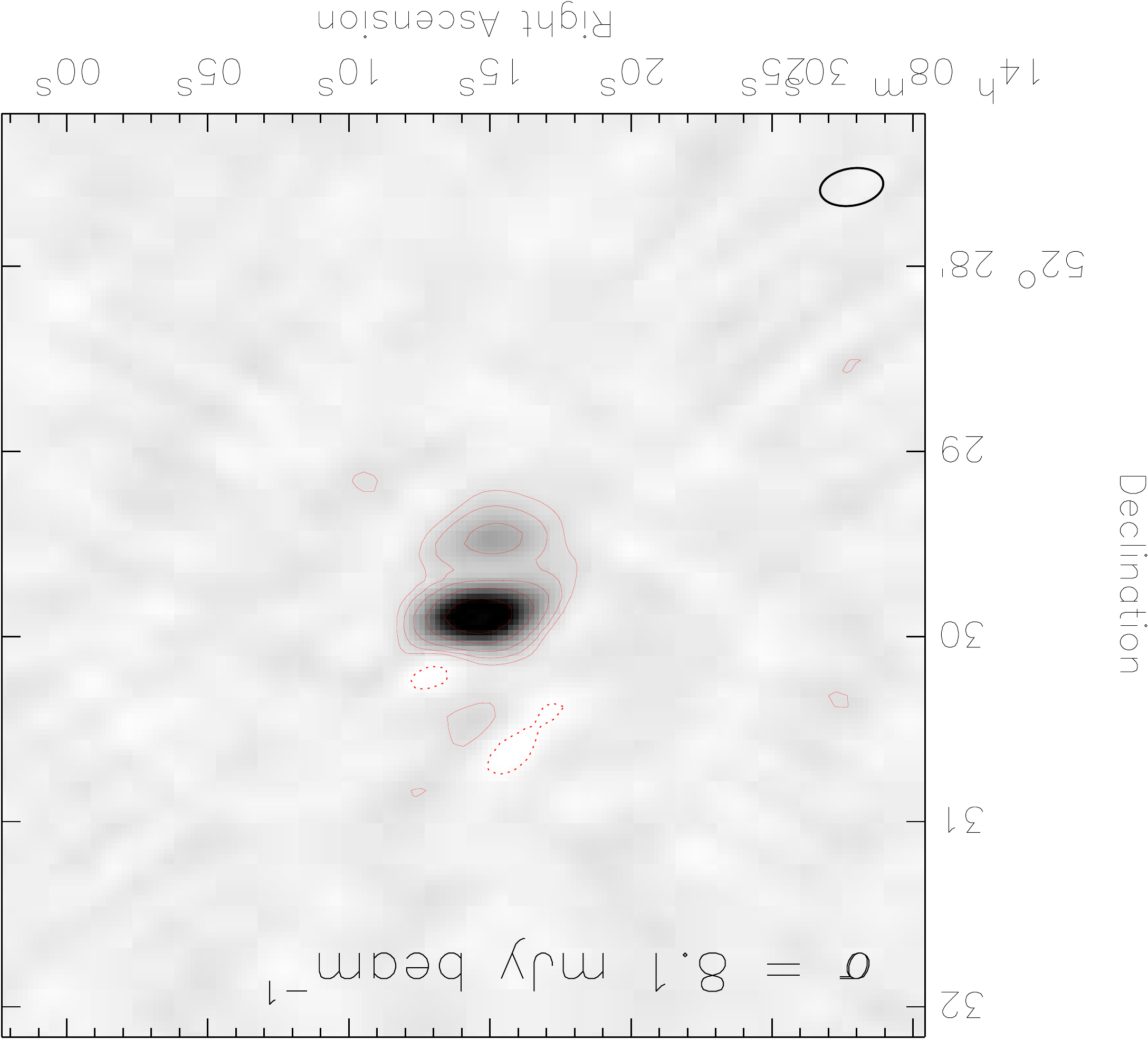}
\includegraphics[ trim =0cm 0cm 0cm 0cm,angle=180, width=0.24\textwidth]{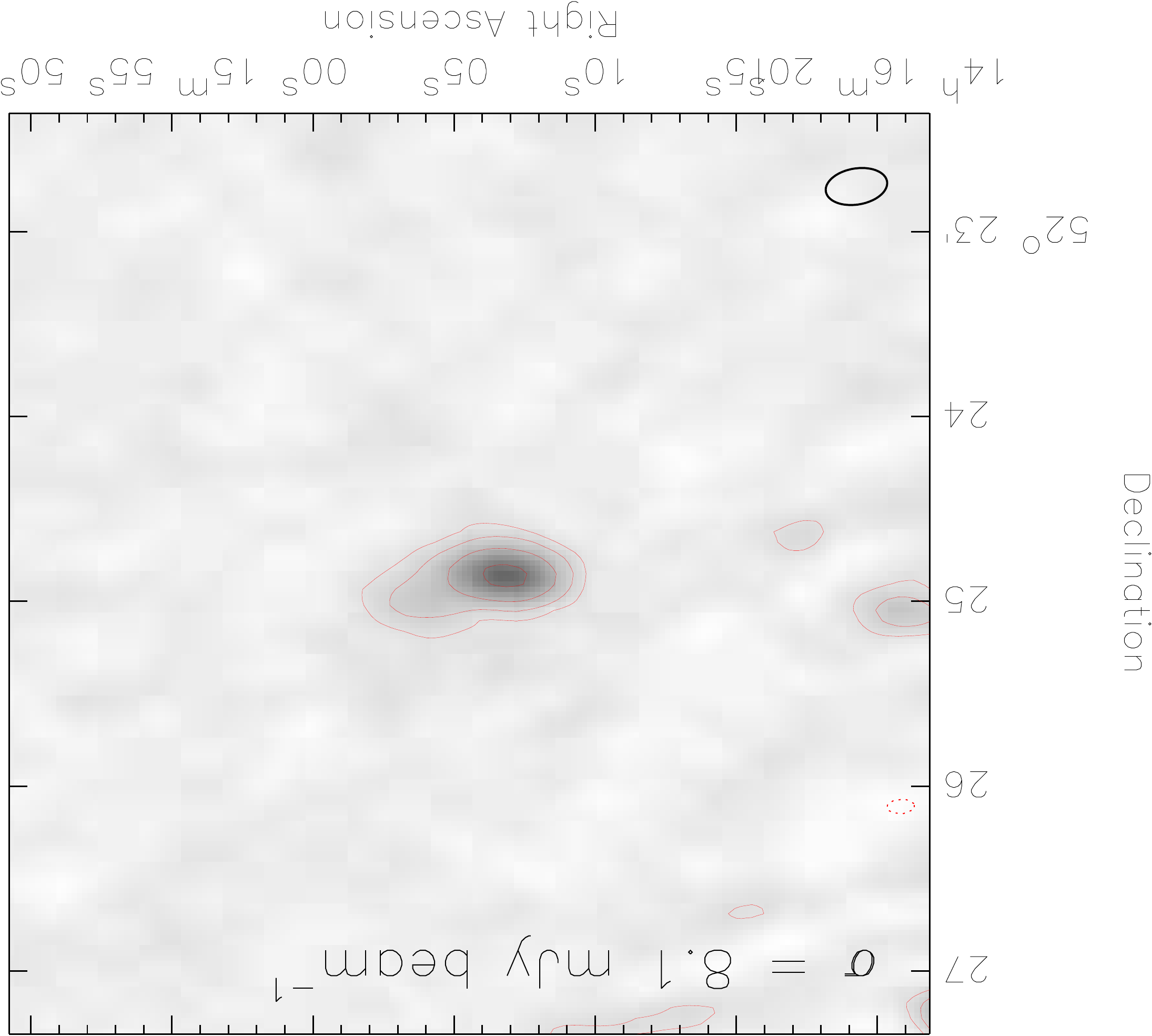}
\includegraphics[ trim =0cm 0cm 0cm 0cm,angle=180, width=0.24\textwidth]{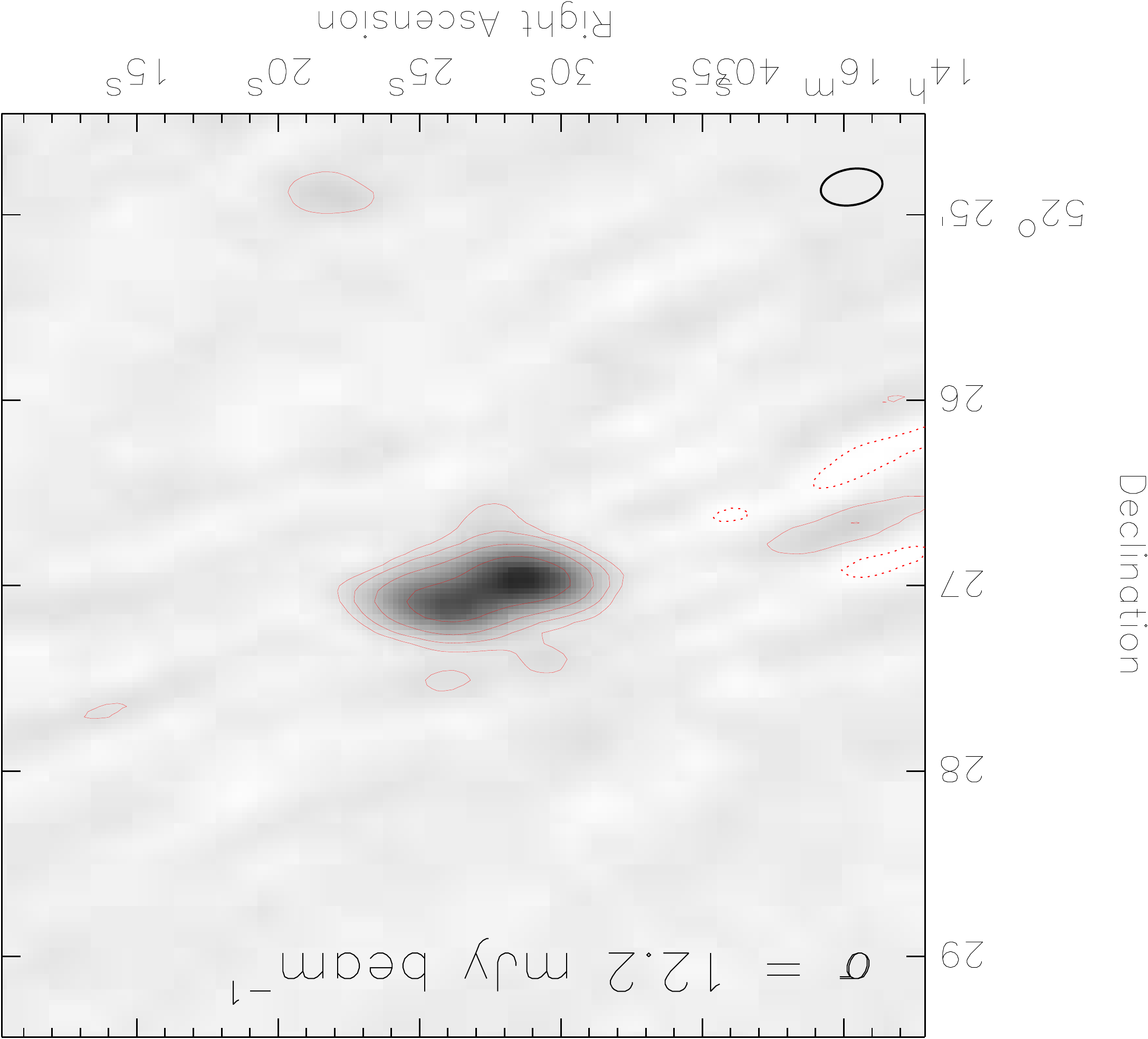}
\includegraphics[ trim =0cm 0cm 0cm 0cm,angle=180, width=0.24\textwidth]{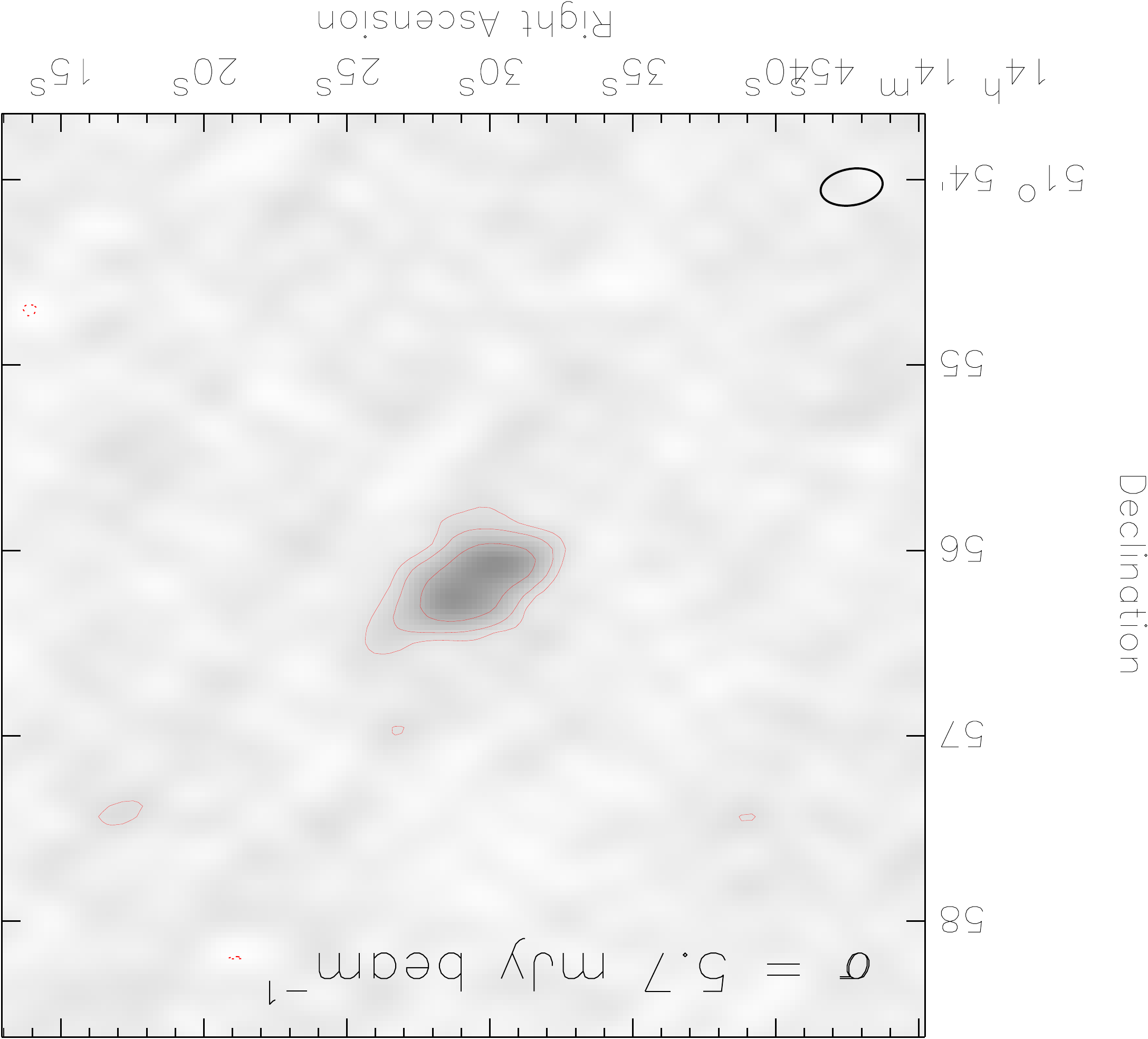}
\includegraphics[ trim =0cm 0cm 0cm 0cm,angle=180, width=0.24\textwidth]{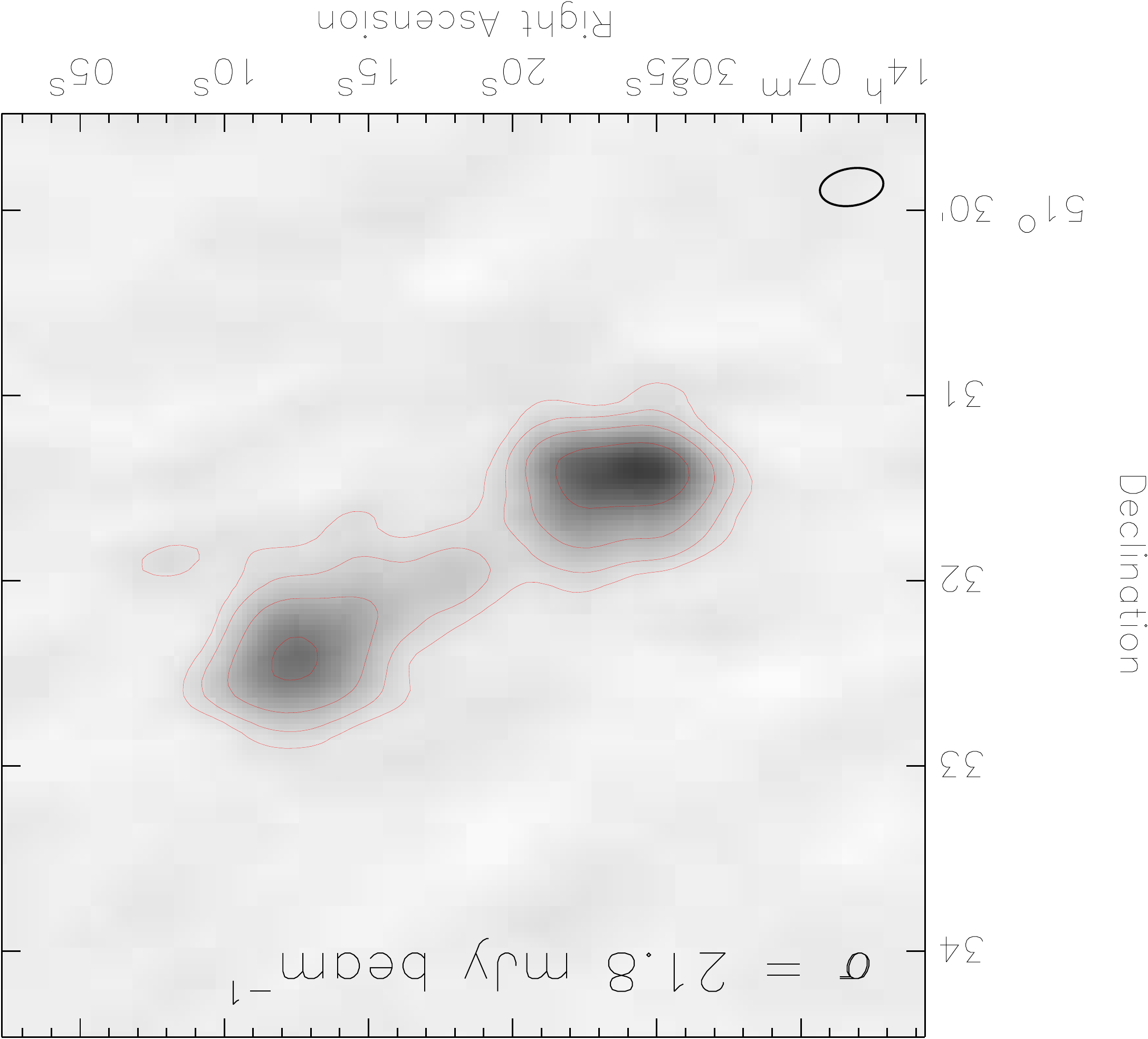}
\includegraphics[ trim =0cm 0cm 0cm 0cm,angle=180, width=0.24\textwidth]{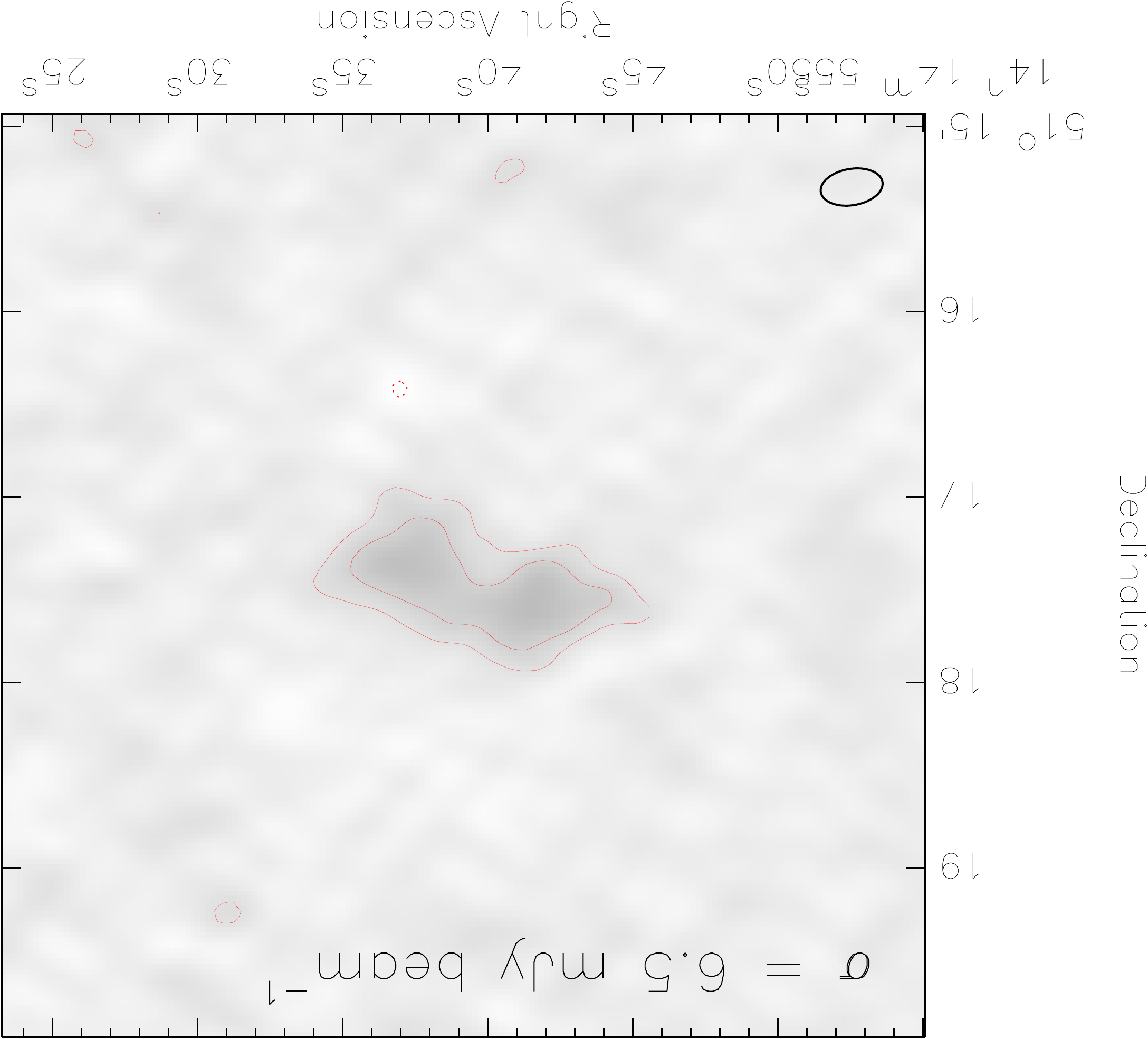}
\includegraphics[ trim =0cm 0cm 0cm 0cm,angle=180, width=0.24\textwidth]{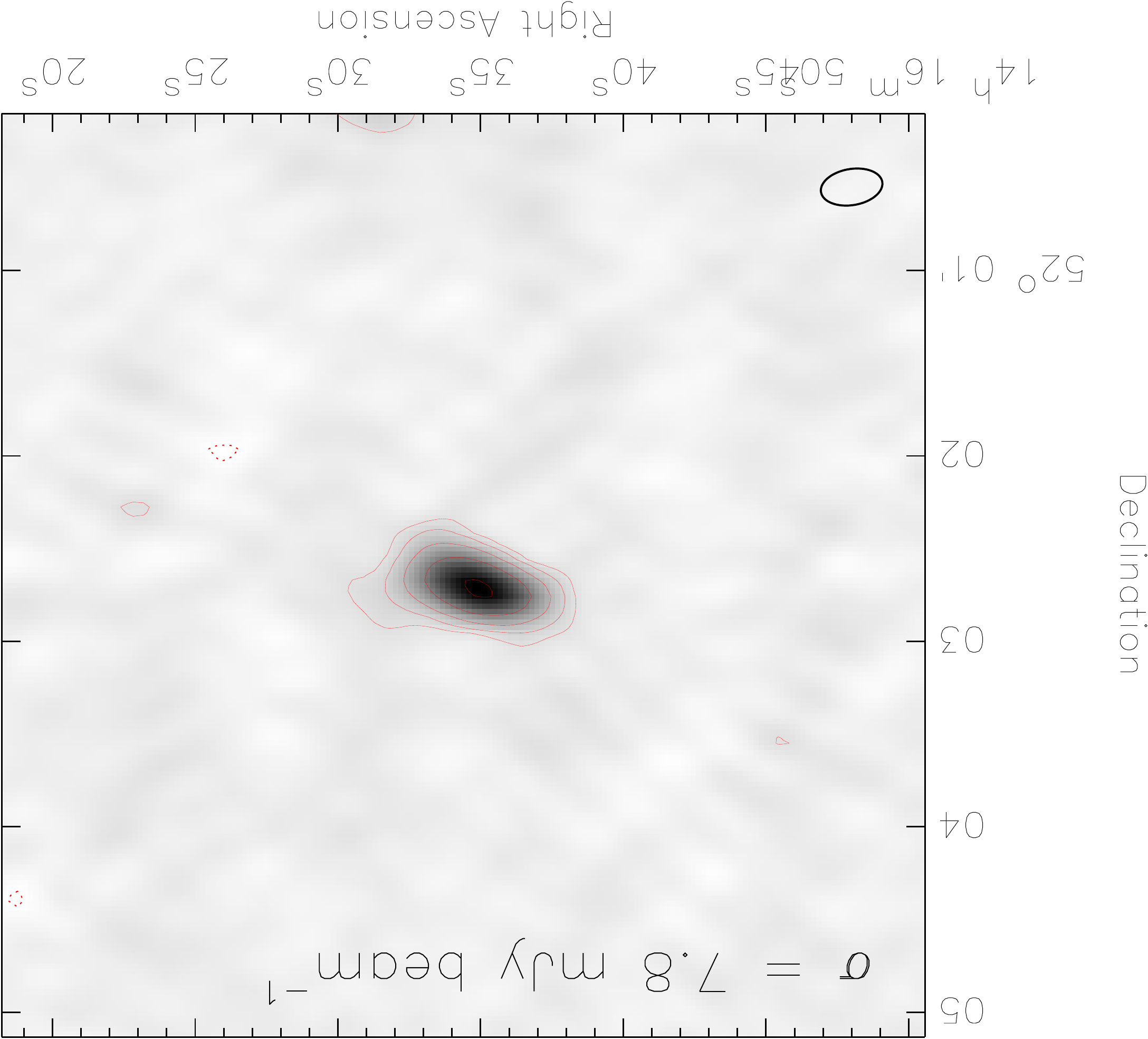}
\includegraphics[ trim =0cm 0cm 0cm 0cm,angle=180, width=0.24\textwidth]{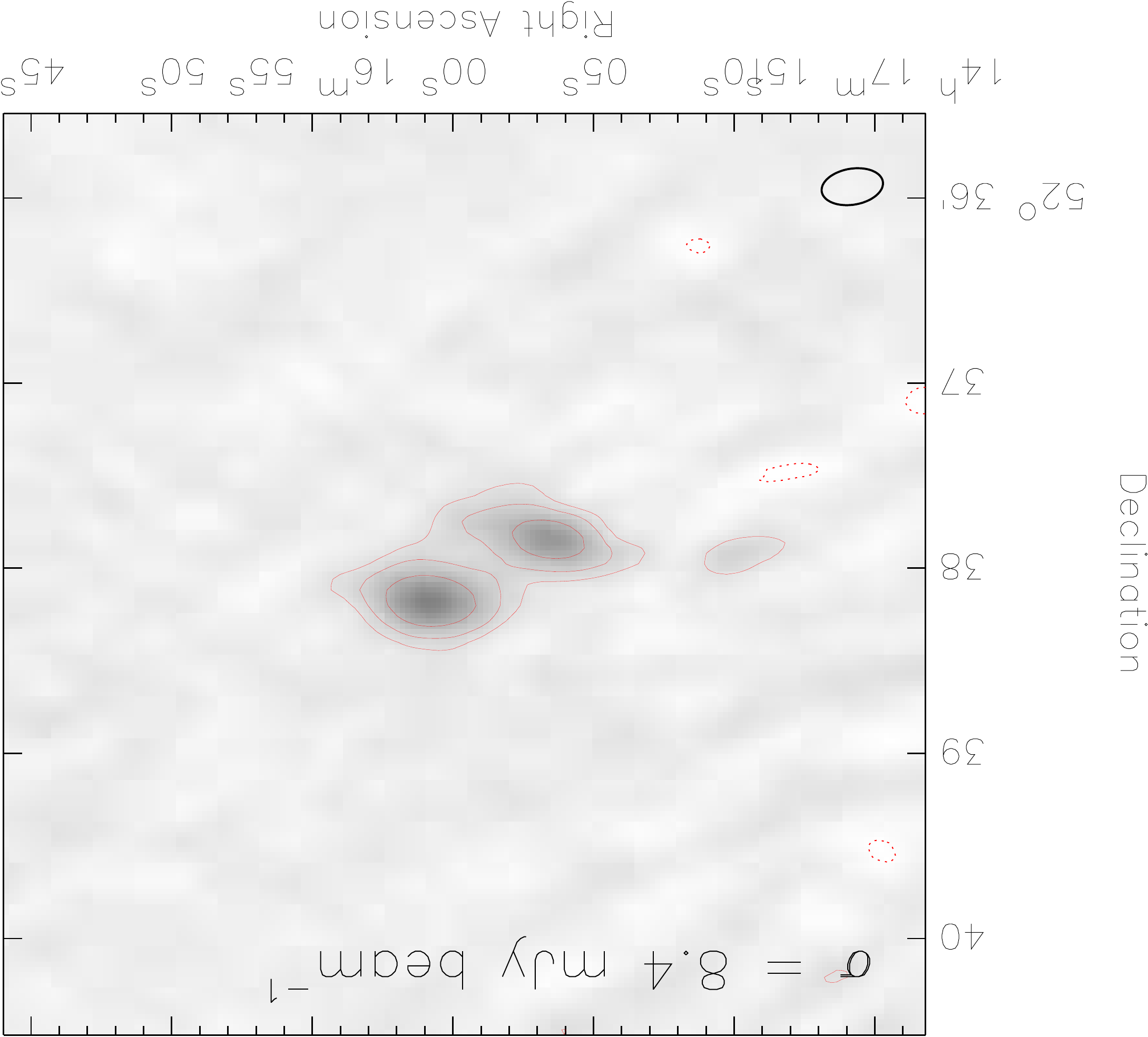}
\includegraphics[ trim =0cm 0cm 0cm 0cm,angle=180, width=0.24\textwidth]{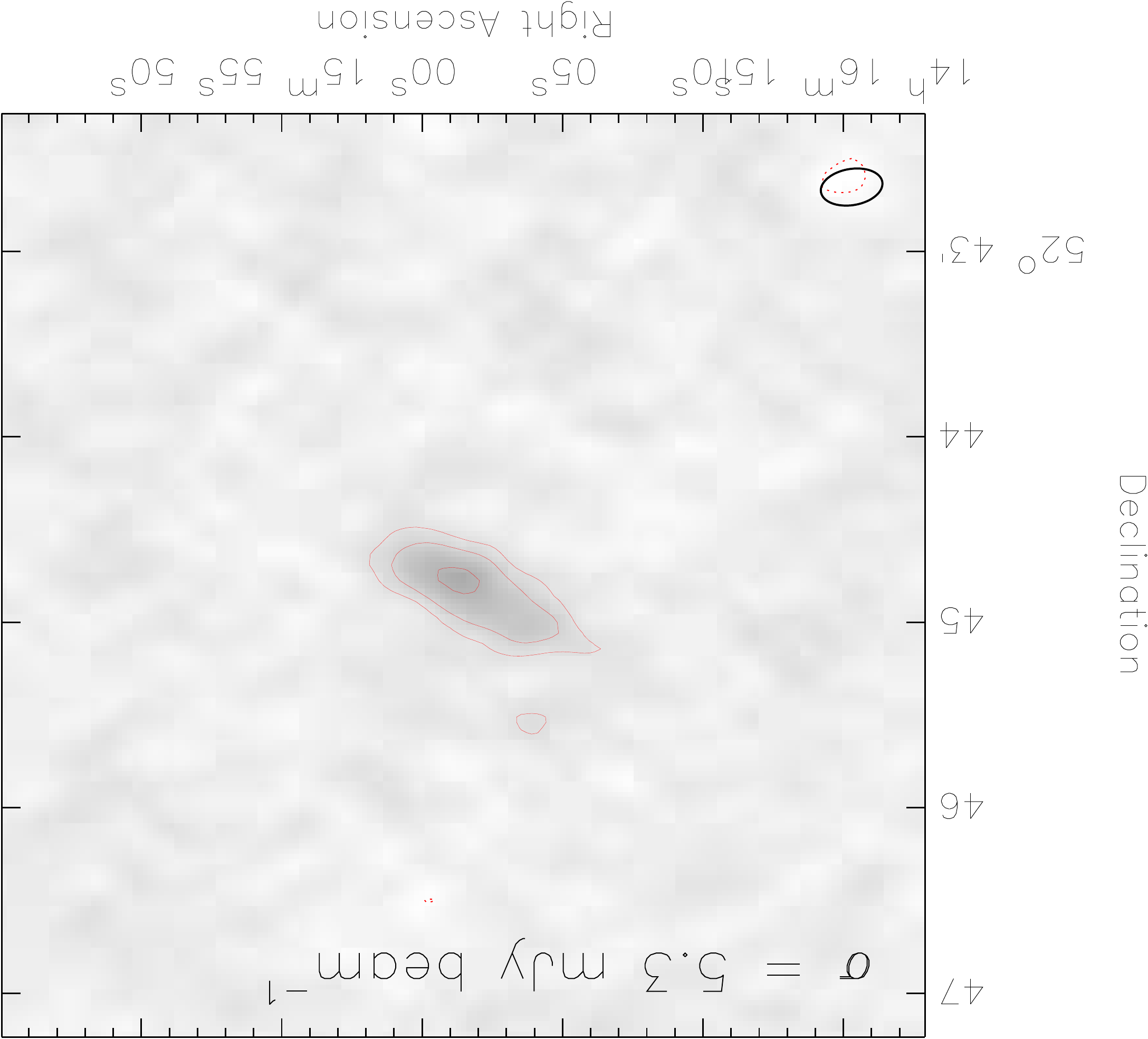}
\includegraphics[ trim =0cm 0cm 0cm 0cm,angle=180, width=0.24\textwidth]{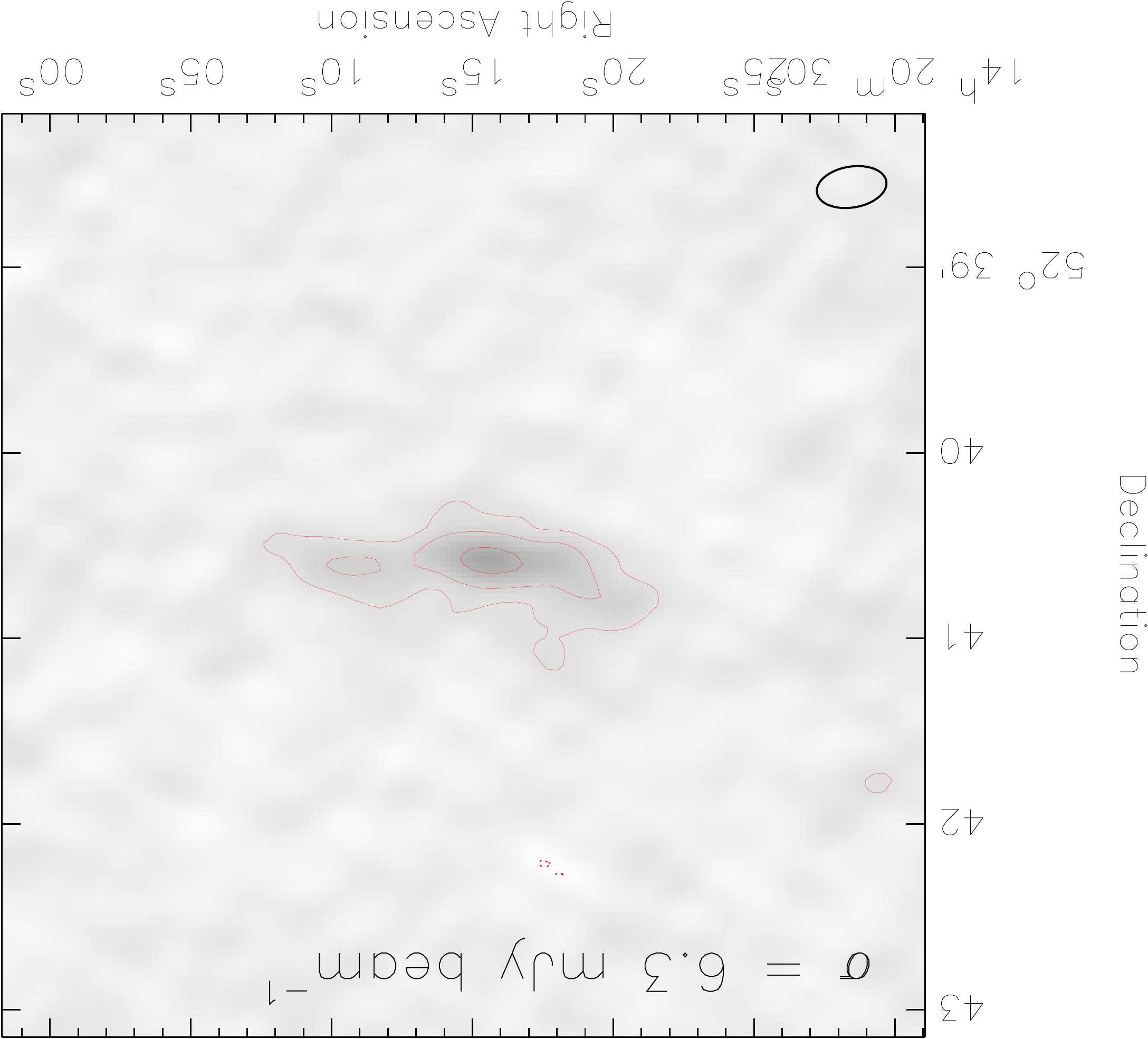}
\includegraphics[ trim =0cm 0cm 0cm 0cm,angle=180, width=0.24\textwidth]{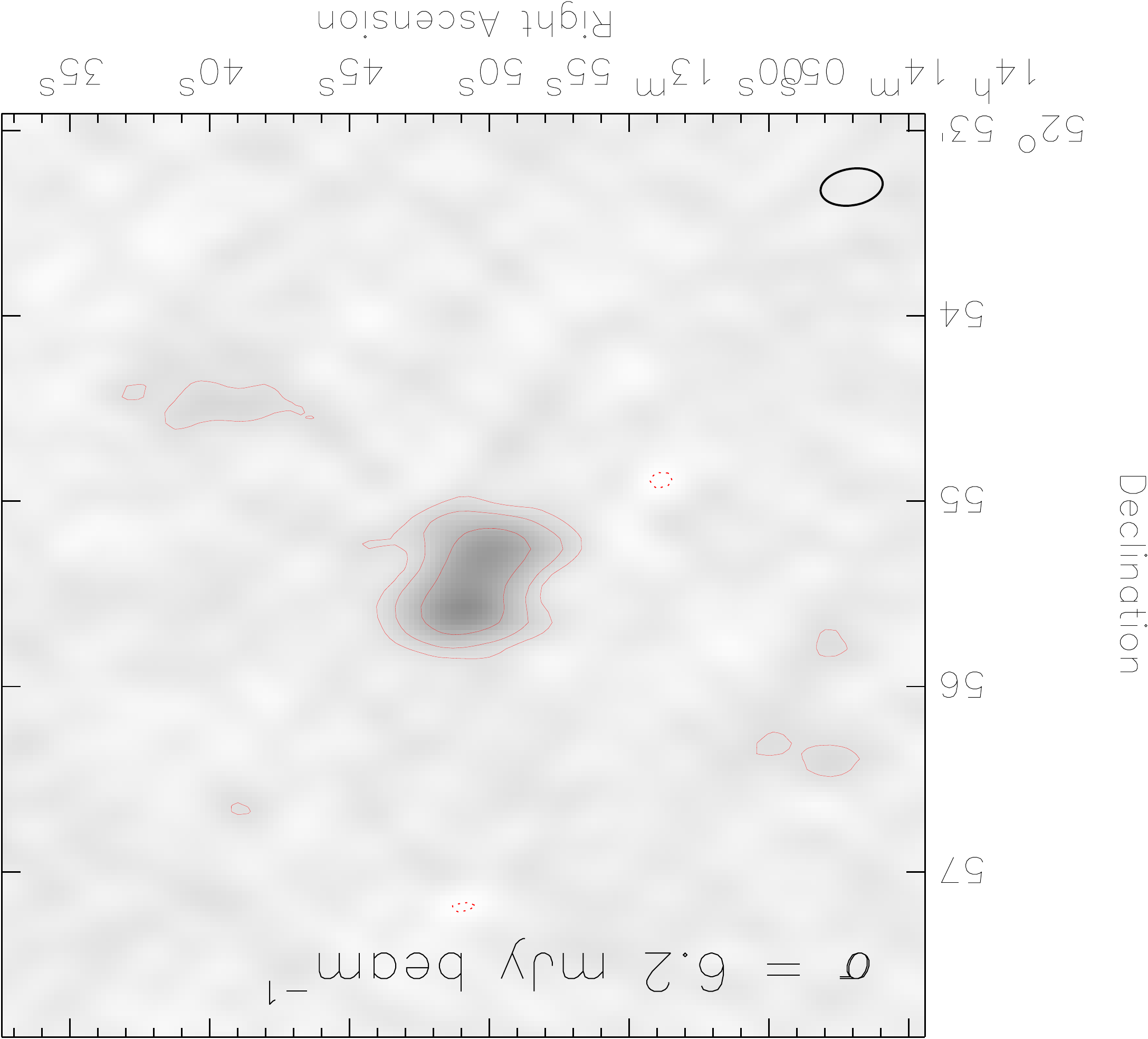}
\includegraphics[ trim =0cm 0cm 0cm 0cm,angle=180, width=0.24\textwidth]{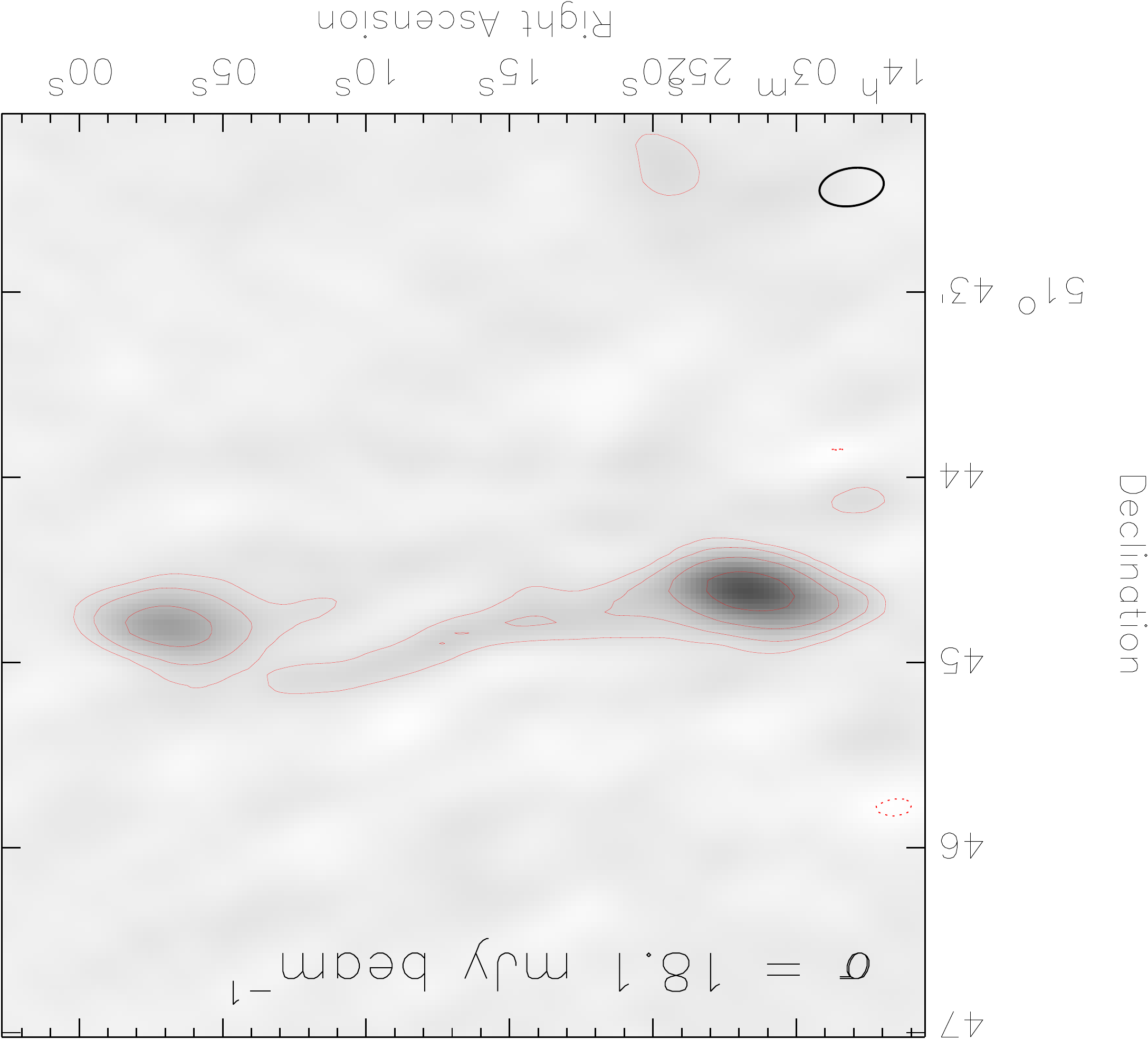}
\includegraphics[ trim =0cm 0cm 0cm 0cm,angle=180, width=0.24\textwidth]{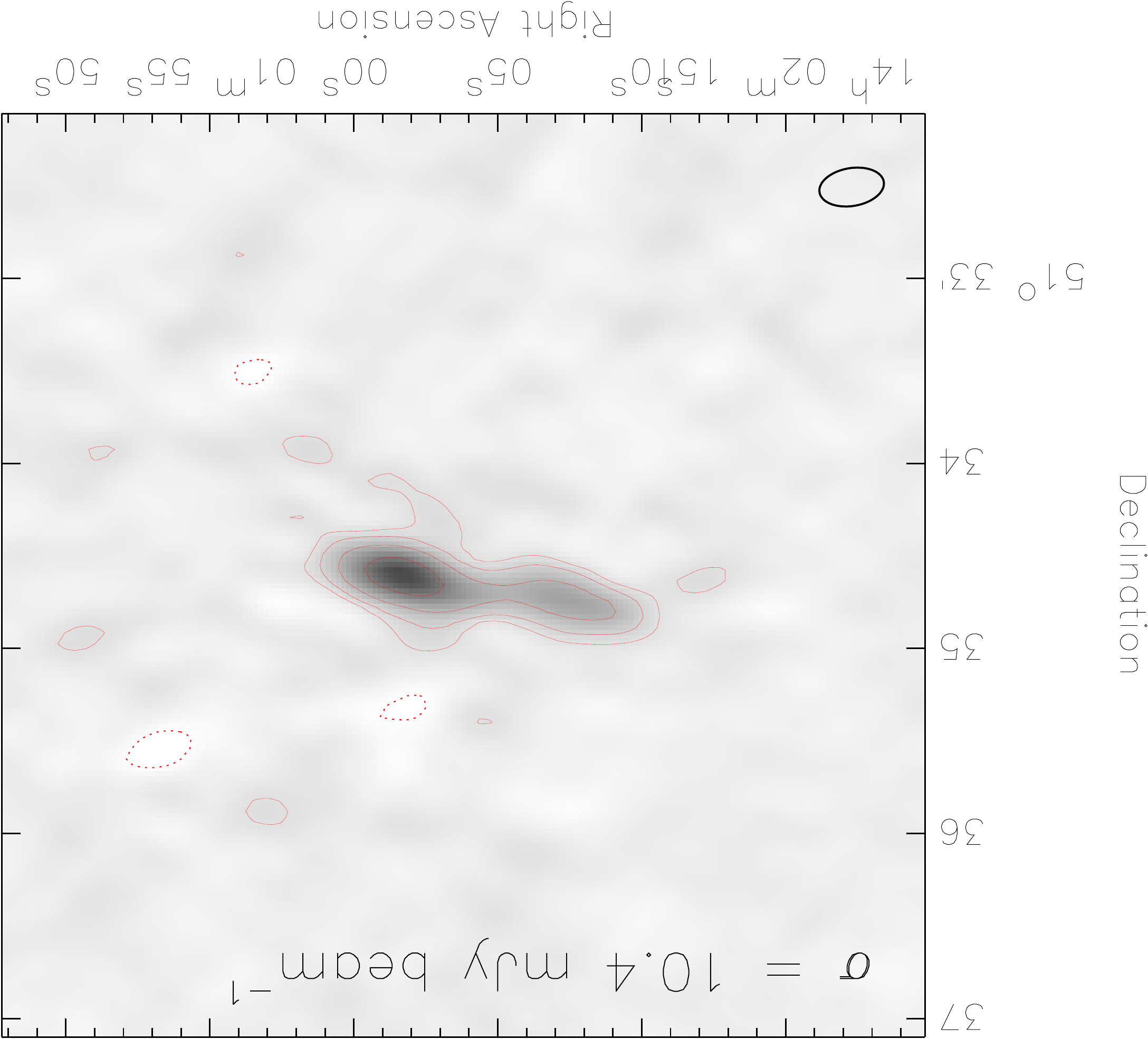}
\includegraphics[ trim =0cm 0cm 0cm 0cm,angle=180, width=0.24\textwidth]{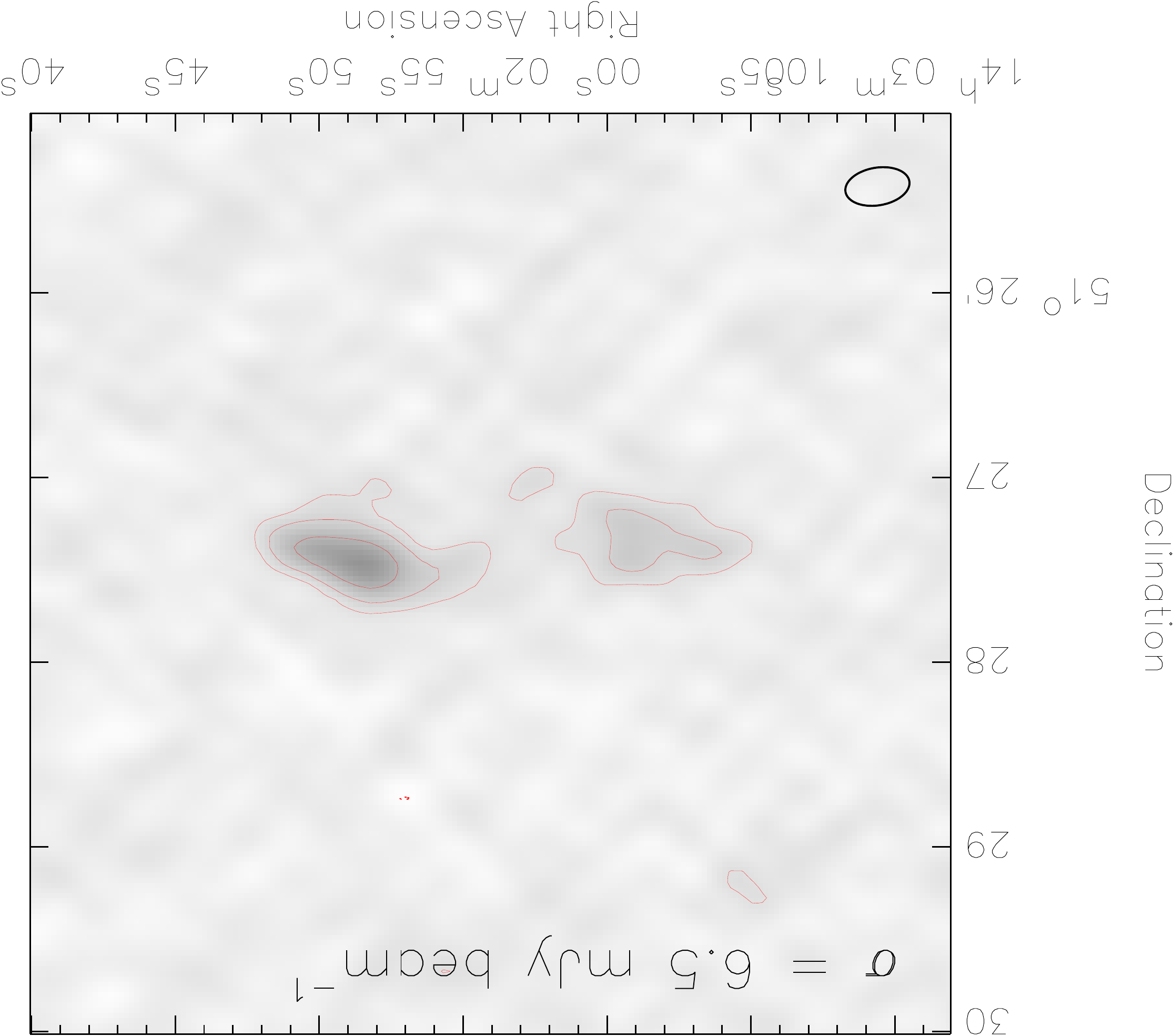}
\end{center}
\caption{Images  of resolved sources in the 3C\,295 field at 62~MHz. Contour levels are drawn at $[1,2,4,8,\ldots] \times 3\sigma_{\rm{local\mbox{ }rms}}$, with $\sigma_{\rm{local\mbox{ }rms}}$ reported in each image. The beam size is shown in the bottom left corner of the images.}
\label{fig:cutouts_3c295}
\end{figure*}

\clearpage
\section{B. SED \& photo-z fitting results}
\label{sec:eazy}
{Figure2~\ref{fig:sedss} and \ref{fig:sedss2} shows the SEDs for each counterpart to a USS source and I band, IRAC 4.5~micron, IRAC 8.0~micron and MIPS 24~micron postage stamps, with GMRT (and FIRST where there is a source) contours overlaid.}

\begin{figure*}
\centering
\includegraphics[width=0.49\textwidth]{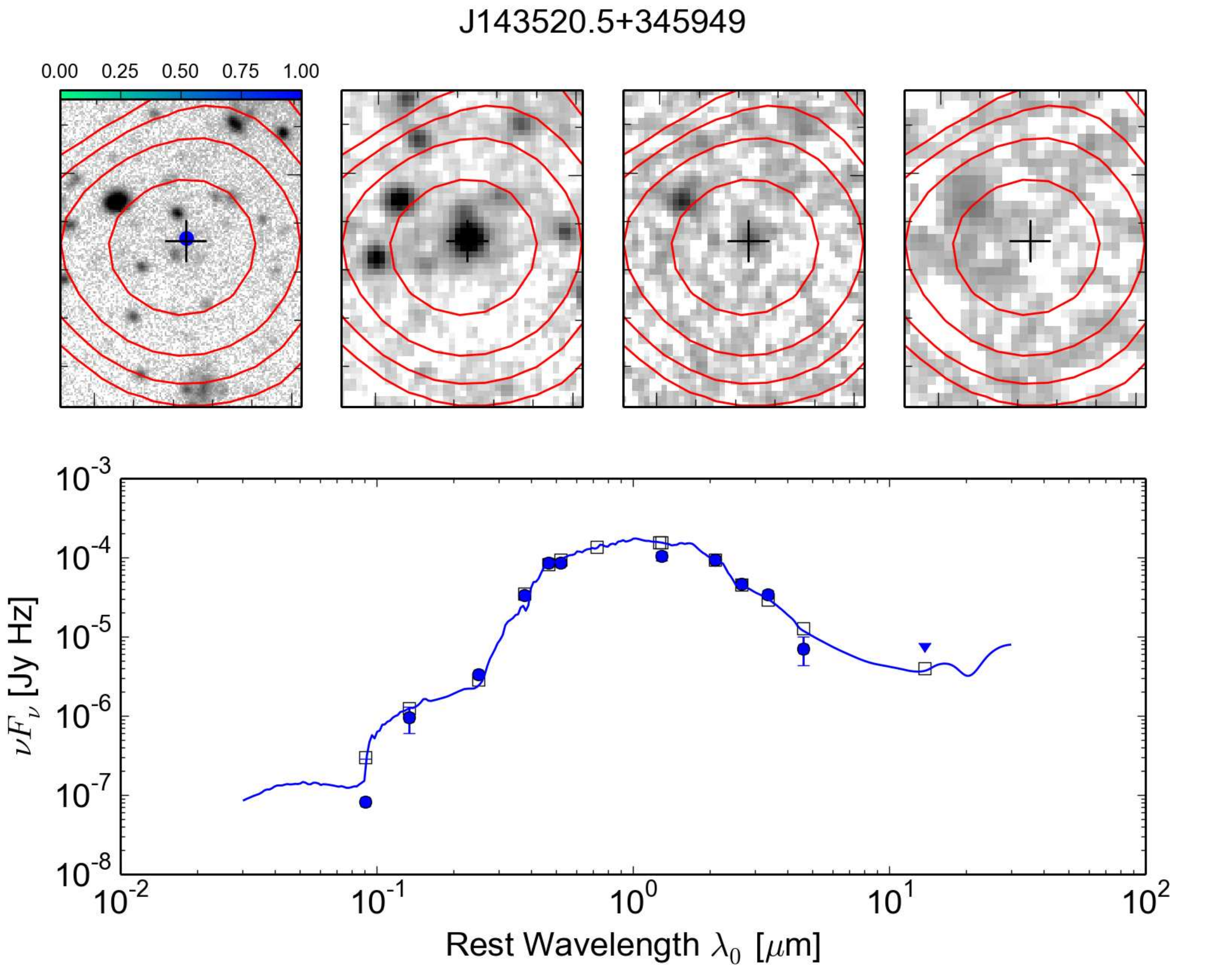}
\includegraphics[width=0.49\textwidth]{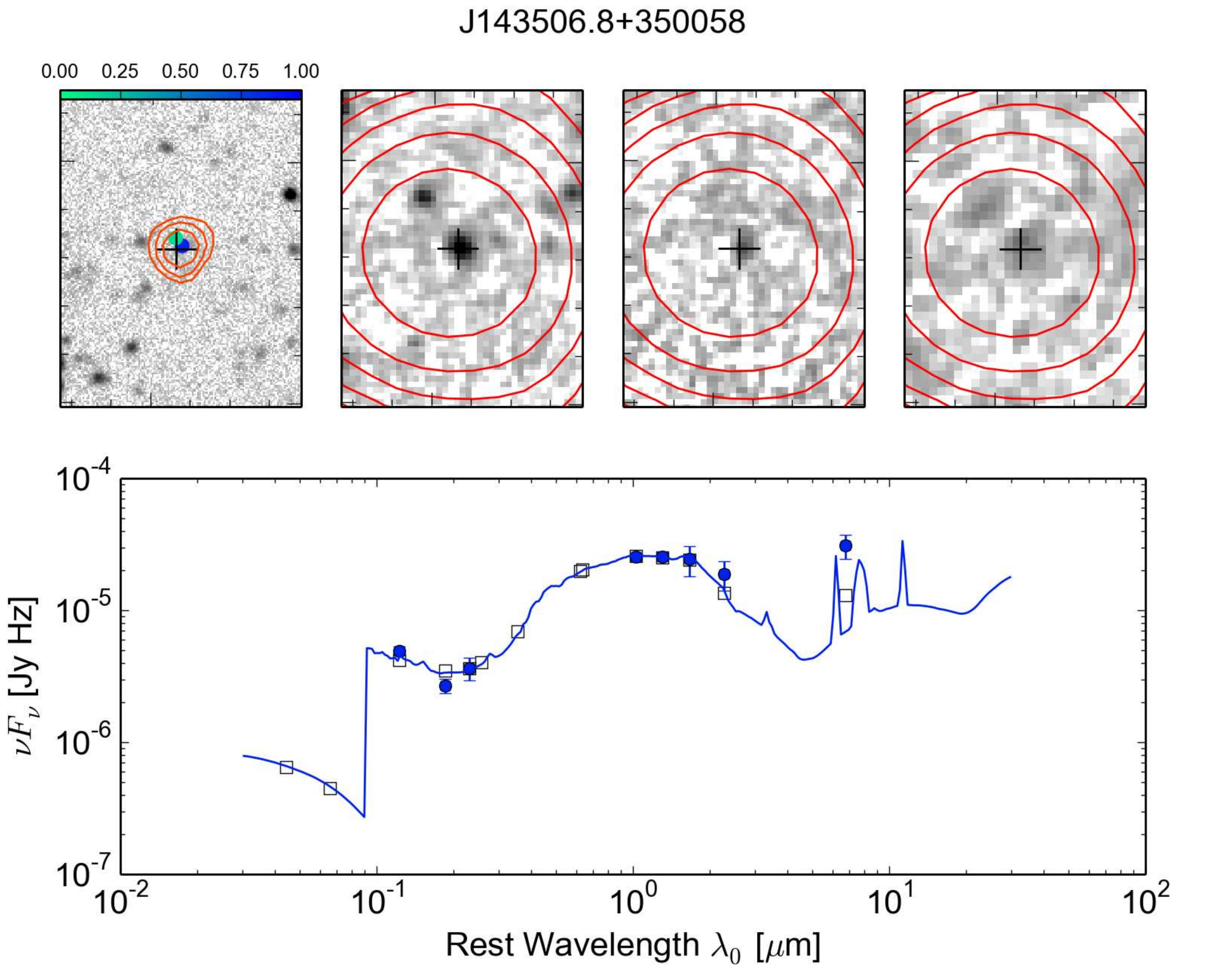}\\
\caption{Top panels: postage stamps showing NDWFS I-band, IRAC 4.5~micron, IRAC 8.0~micron and MIPS 24~micron images. GMRT 153~MHz (red) and FIRST 1.4~GHz (orange, when available) contours are overlaid. Radio contour levels are drawn at $[1, 2, 4, \ldots] \times 3\sigma_{\rm{rms}}$. A black cross indicates the GMRT radio position and the color scale at the top of the I-band image shows the probability that the I-band source, marked with a colored point, is the true optical counterpart. Bottom panels: Spectral energy distribution and best fitted {\tt LRT} model for the optical counterpart(s). The flux measurements were taken from \cite{1999ASPC..191..111J,2006ApJ...639..816E,2007ApJS..169...21C,2009ApJ...701..428A,2007ApJS..173..682M,2010AAS...21547001J}. }
\label{fig:sedss}
\end{figure*}

\begin{figure*}
\centering
\includegraphics[width=0.49\textwidth]{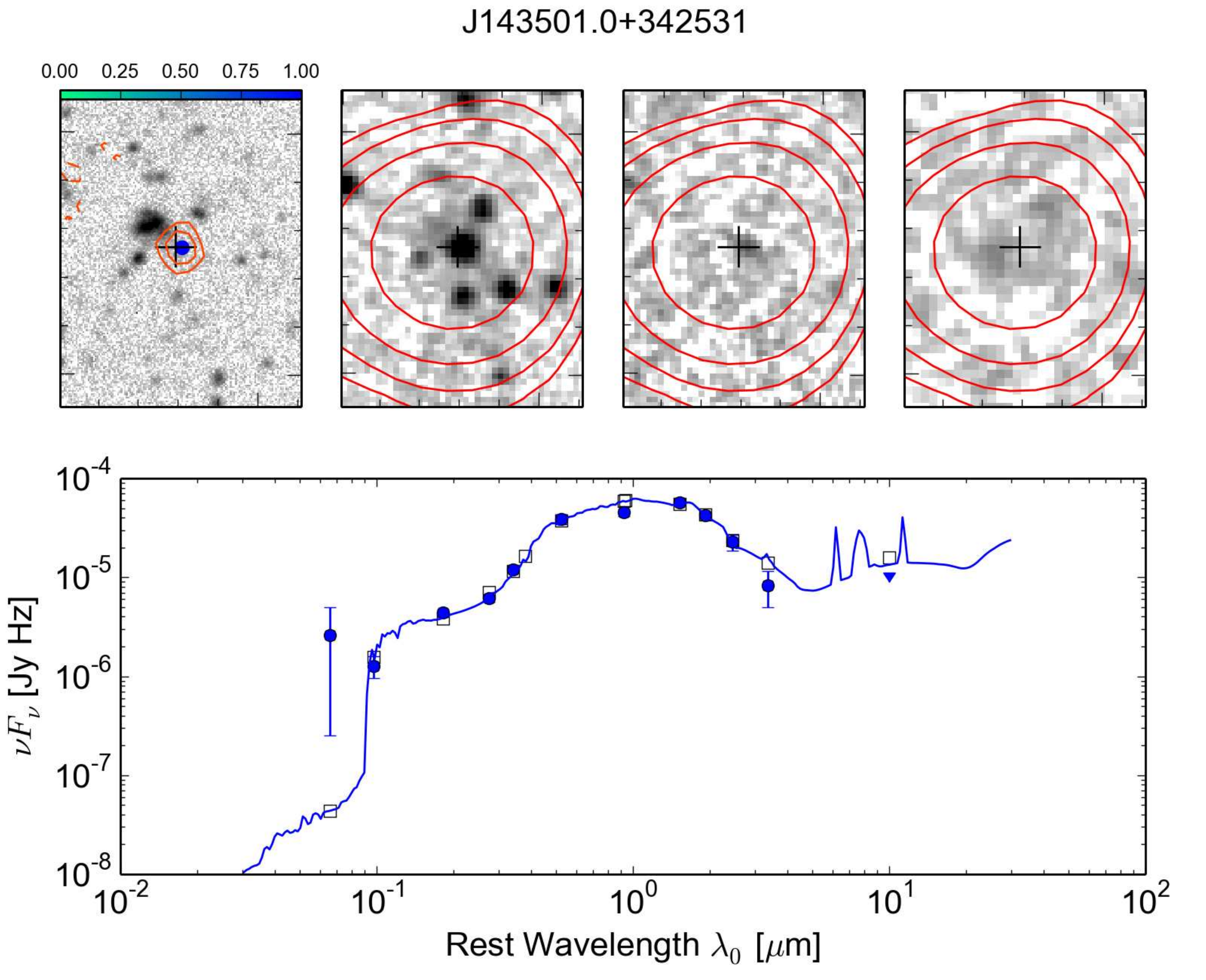}
\includegraphics[width=0.49\textwidth]{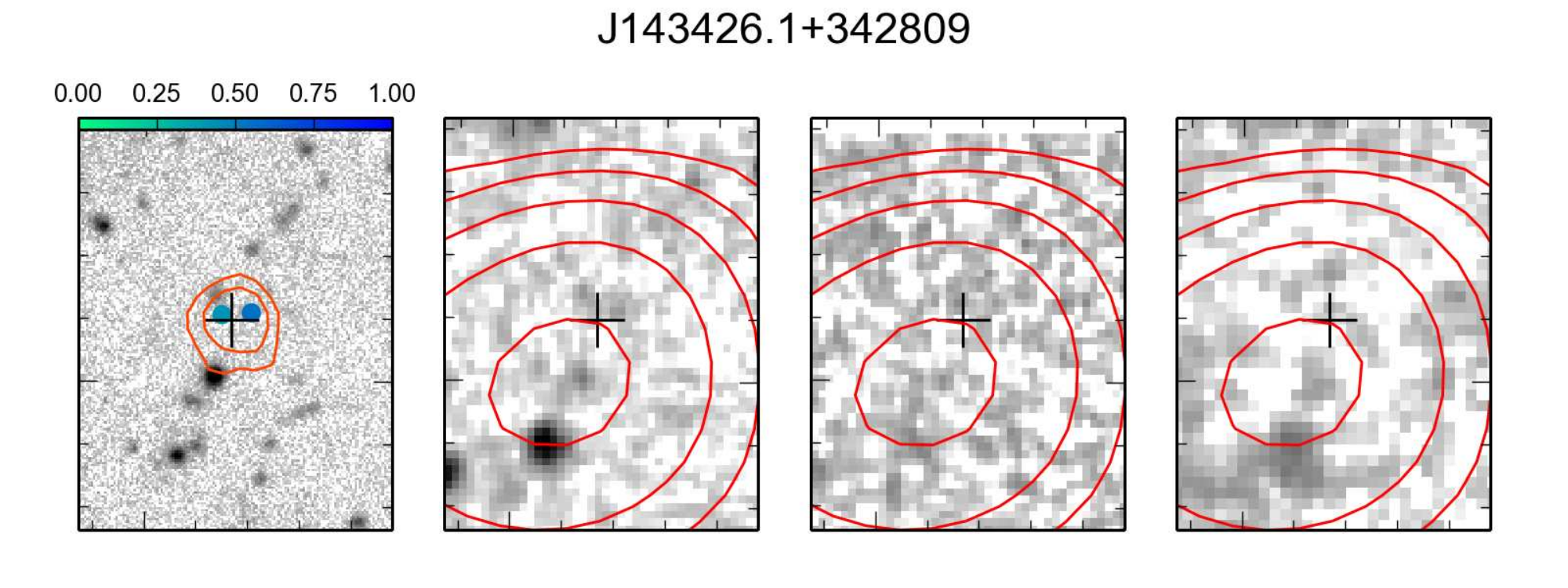}\\
\includegraphics[width=0.49\textwidth]{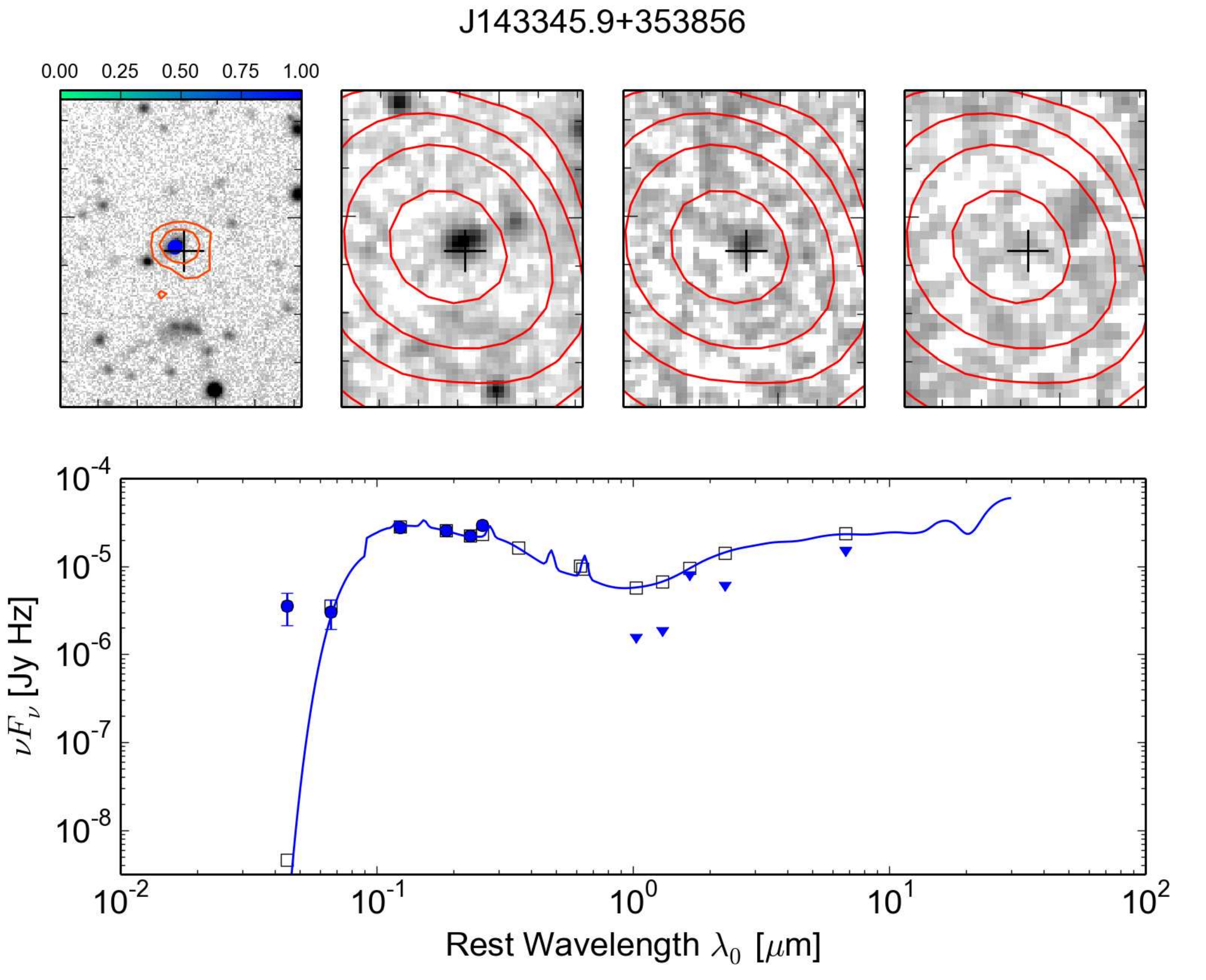}
\includegraphics[width=0.49\textwidth]{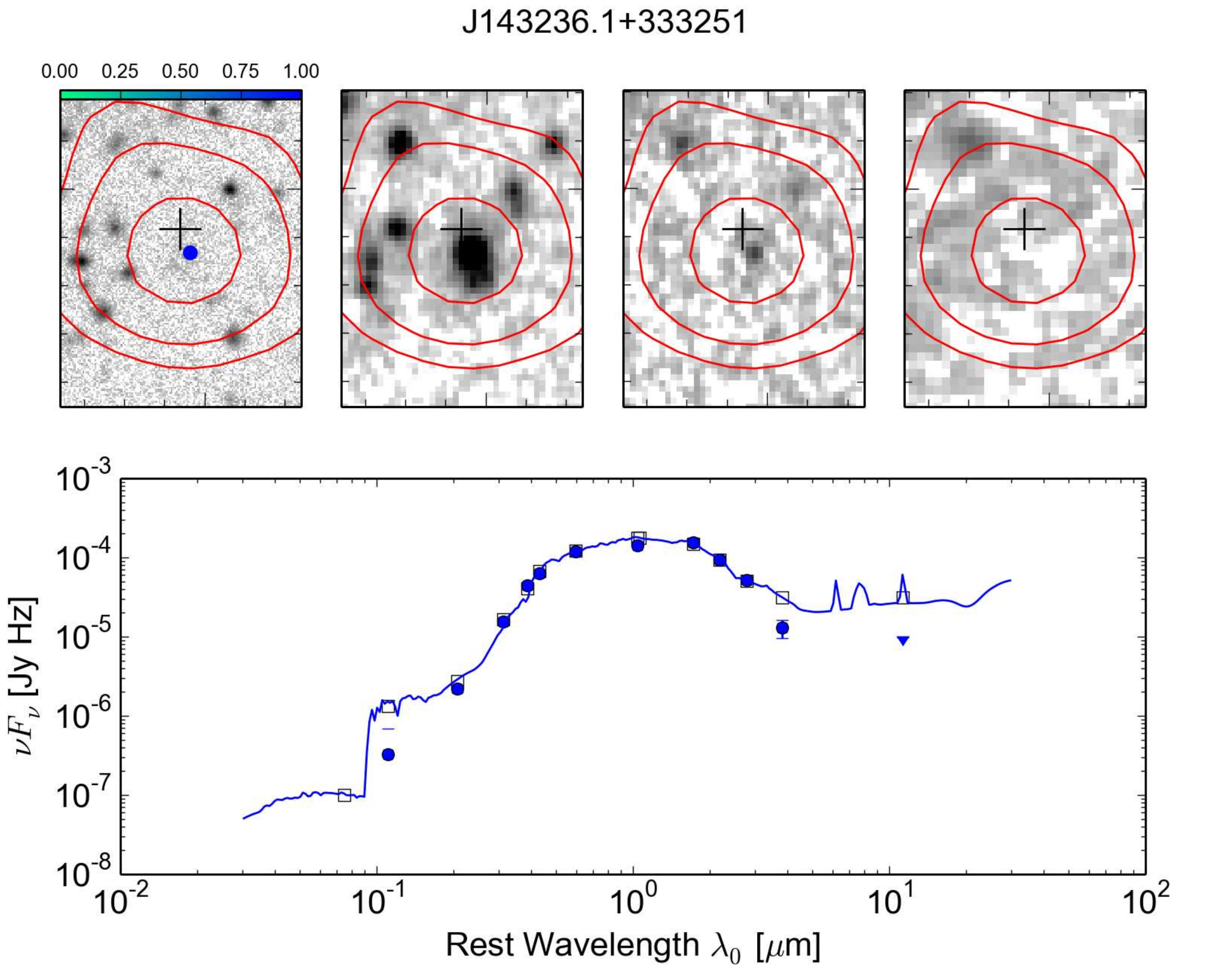}\\
\includegraphics[width=0.49\textwidth]{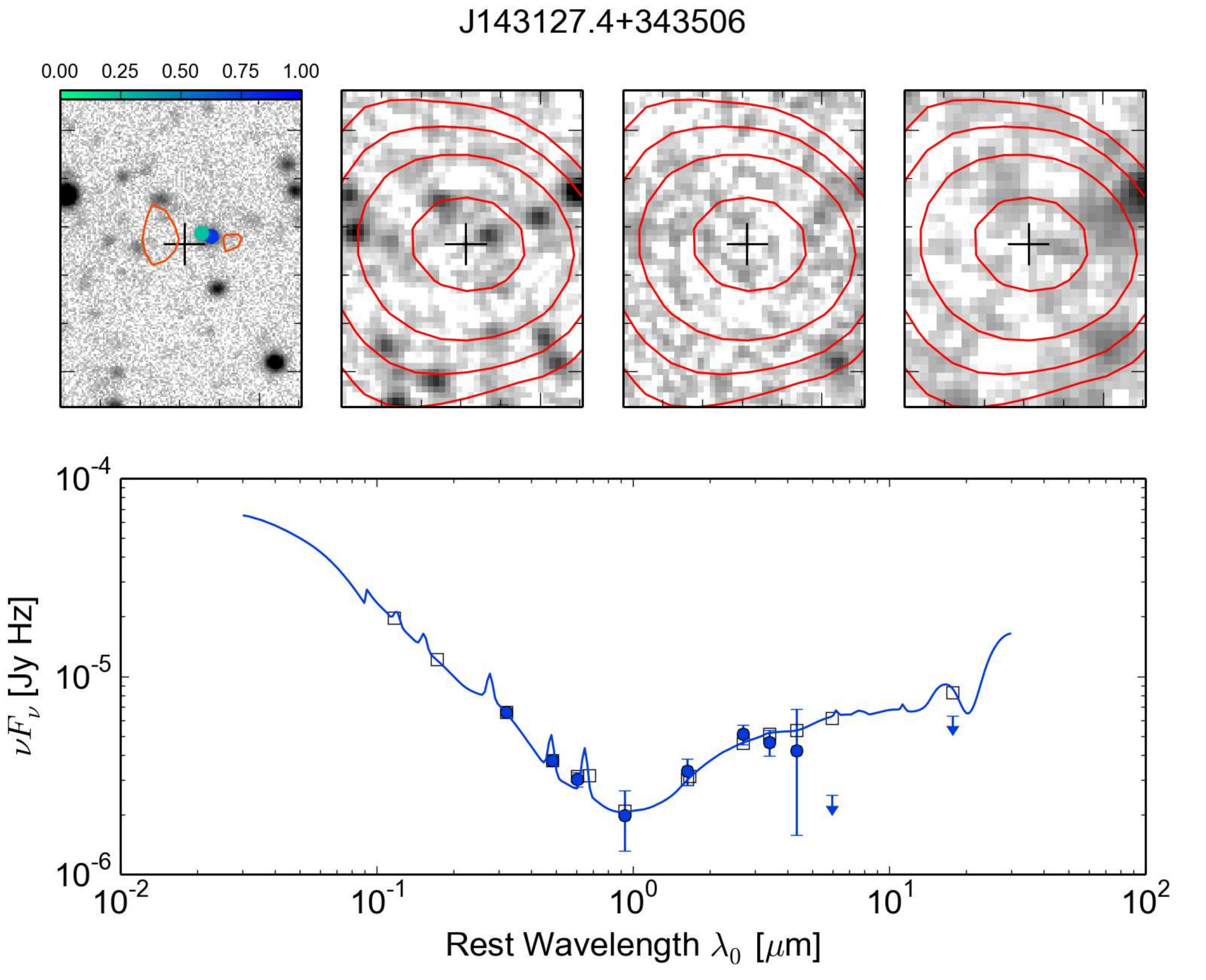}
\includegraphics[width=0.49\textwidth]{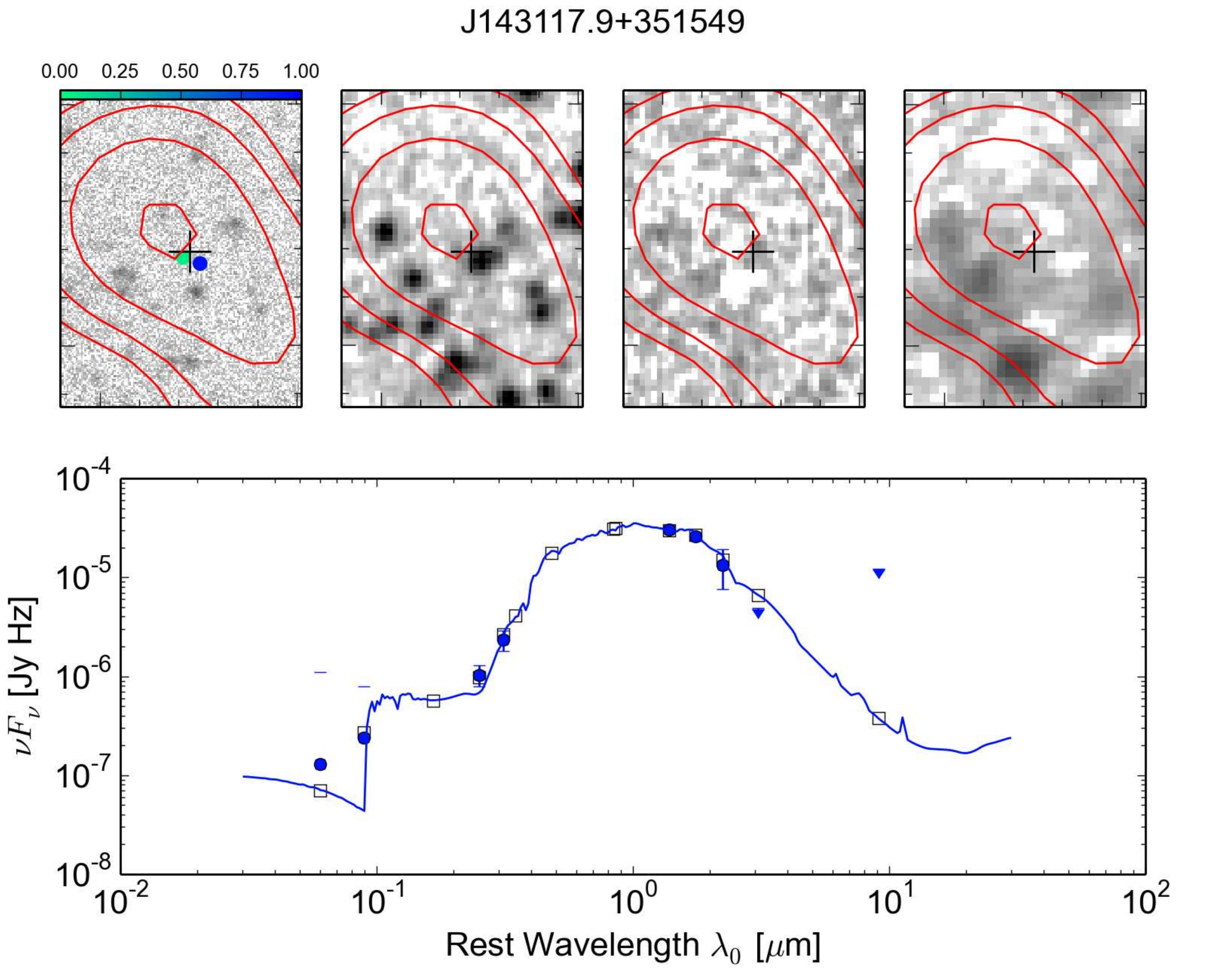}\\
 \caption{Continued.}
 \label{fig:sedss2}
\end{figure*}

\section{C. Source counts at 62~MH for the Bo\"otes field and 3C\,295 fields}

\begin{figure}
\begin{center}
\includegraphics[trim =0cm 0cm 0cm 0cm,angle=180, width=0.49\textwidth]{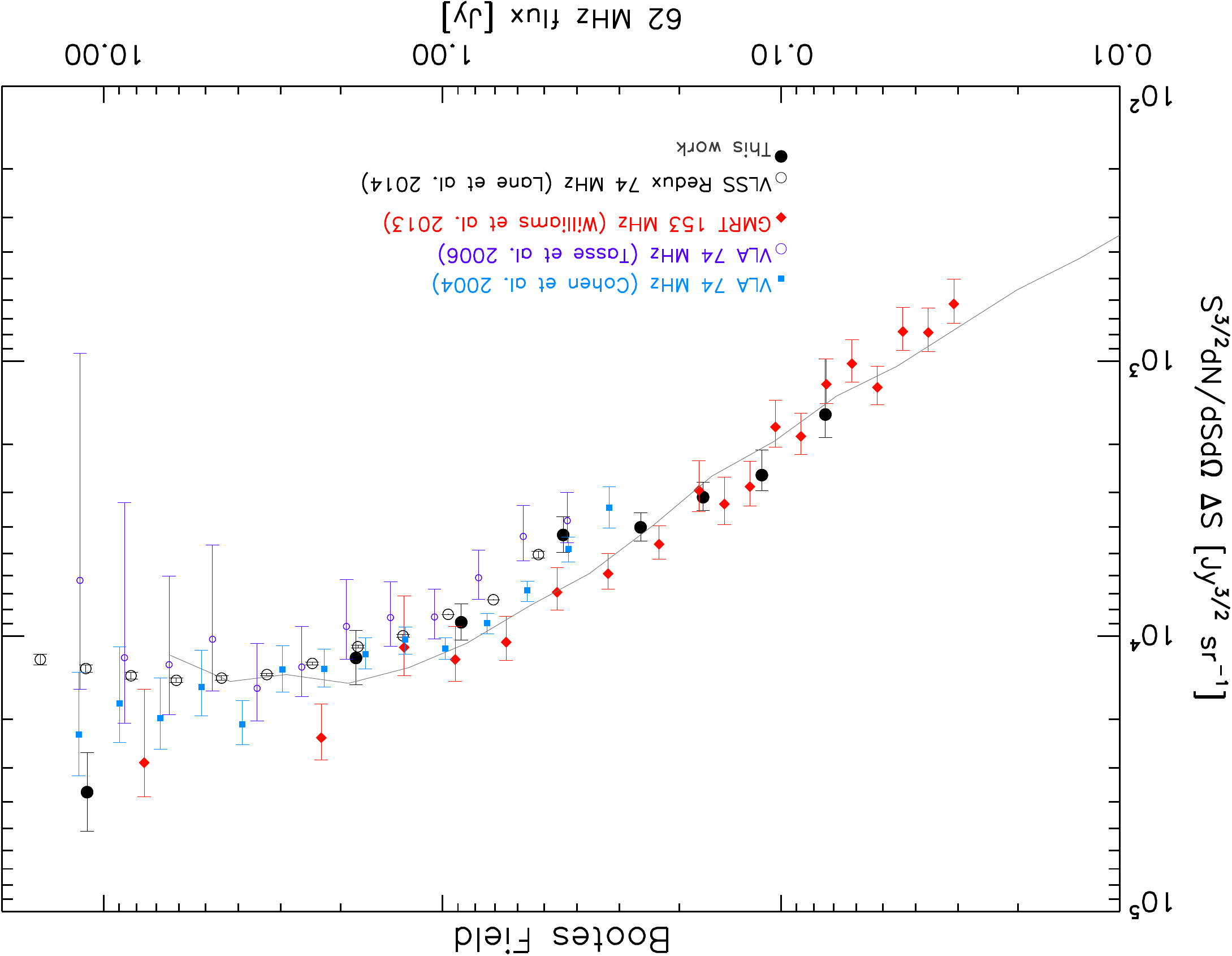}
\includegraphics[trim =0cm 0cm 0cm 0cm,angle=180, width=0.49\textwidth]{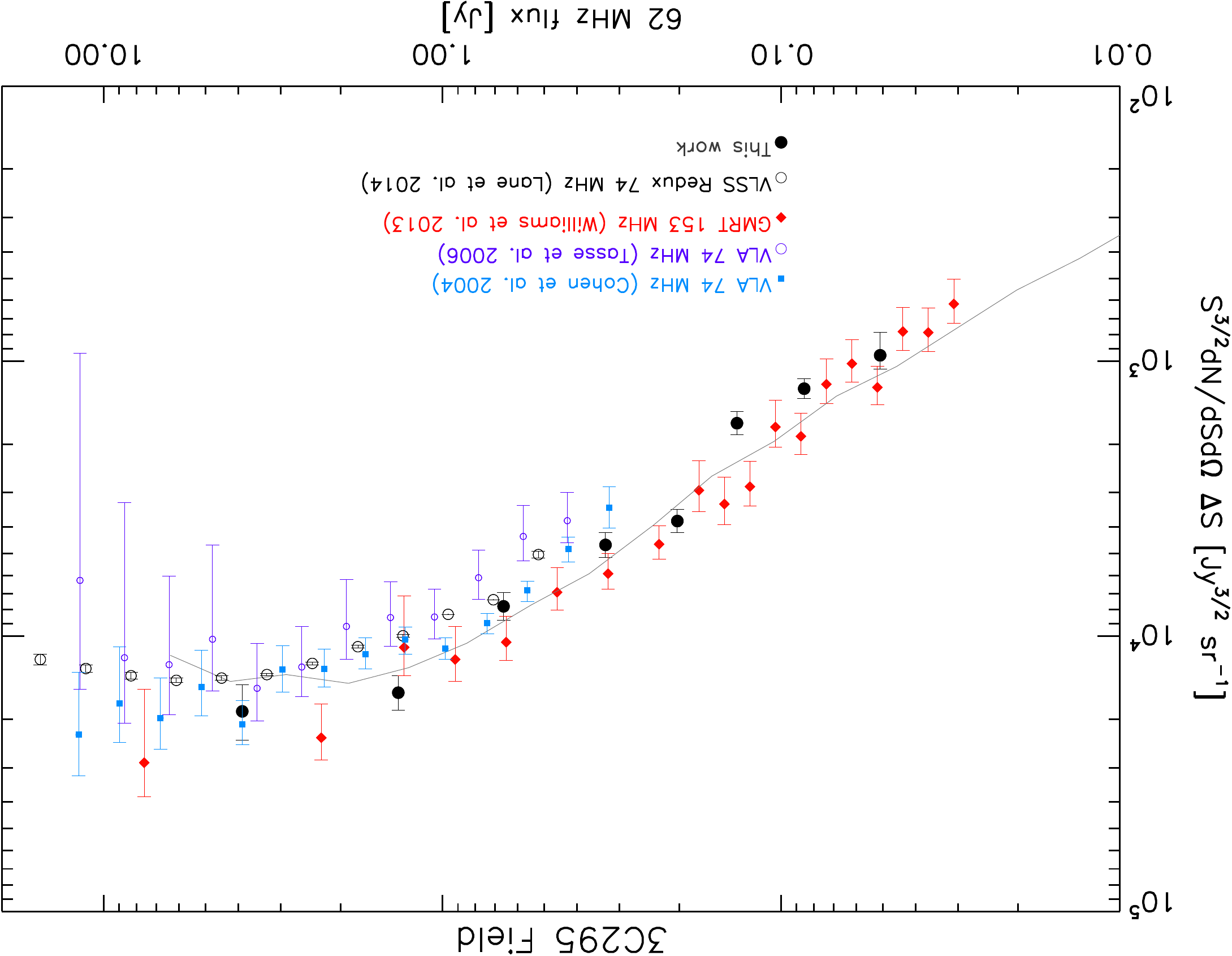}
\end{center}
\caption{Euclidean-normalized differential source counts at 62~MHz for the Bo\"otes and 3C\,295 fields. The LOFAR points are indicated by the black symbols. Red diamonds are Bo\"otes field source counts at 153~MHz, scaled to 62~MHz using $\alpha=-0.7$. 
Black open circles, blue squares and purple open circles are 74~MHz differential source counts from \cite{2006A&A...456..791T,2004ApJS..150..417C,2014MNRAS.440..327L} and the solid grey line displays the counts from the 151~MHz SKADS S$^{3}$-SEX simulation \citep{2008MNRAS.388.1335W}. These are all scaled to 62~MHz assuming $\alpha=-0.7$.}
\label{fig:ctshighsep}
\end{figure}

\bibliography{ref_filaments}

\end{document}